\documentclass[11pt,a4paper]{article}

\usepackage{jhepmacro}

\newcommand{\Hgl}{H_{\mathrm{gl}}}
\newcommand{\CHgl}{\CH_{\mathrm{gl}}}
\newcommand{\CHtgl}{\tilde{\CH}_{\mathrm{gl}}}
\newcommand{\Hgr}{H_{\mathrm{gr}}}
\newcommand{\CHgr}{\CH_{\mathrm{gr}}}
\newcommand{\CHtgr}{\tilde{\CH}_{\mathrm{gr}}}

\title{\textbf{Spinning correlators in $\boldsymbol{\mathcal{N}=2}$ SCFTs: Superspace and AdS amplitudes}}

\author[a]{Agnese Bissi,}
\author[a]{Giulia Fardelli,}
\author[a]{Andrea Manenti}
\author[b]{and Xinan Zhou}

\affiliation[a]{
\bigskip
Department of Physics and Astronomy\\
Uppsala University, Box 516, SE-751 20 Uppsala, Sweden
}

\affiliation[b]{
Kavli Institute for Theoretical Sciences,\\
University of Chinese Academy of Sciences, Beijing 100190, China.
\bigskip
}

\makeatletter
\renewcommand{\@email}[1]{#1}
\makeatother

\emailAdd{\{\href{mailto:agnese.bissi@physics.uu.se}{\tt agnese.bissi},
            \href{mailto:giulia.fardelli@physics.uu.se}{\tt giulia.fardelli},
            \href{mailto:andrea.manenti@physics.uu.se}{\tt andrea.manenti}\}\texttt{\textcolor{blue}{\,@physics.uu.se}}\lsp;\\
            \phantom{E-mail:\;\{}
            \href{mailto:xinan.zhou@ucas.ac.cn}{\tt xinan.zhou@ucas.ac.cn}
}

\abstract{We study four-point functions of spinning operators in the flavor current multiplet in four dimensional $\mathcal{N}=2$ SCFTs, using superspace techniques. In particular we explicitly construct the differential operators relating the different components of the super-correlator. As a byproduct of our analysis, we report the computation of the four-point amplitudes of gluons in bosonic Yang-Mills theories on $\mathrm{AdS}_5$ and we give evidence of an AdS double copy relation between the gluon amplitude and its gravitational counterpart.   }

\preprint{UUITP-36/22}

\begin{document}

\maketitle

\tableofcontents

\section{Introduction}
Recent years have seen significant progress in computing holographic correlators using bootstrap methods. A great deal of general results have been obtained in various full-fledged string theory/M-theory models with a varying amount of supersymmetry (see \cite{Bissi:2022mrs} for a recent review). In the bulk description, these holographic correlators correspond to scattering amplitudes in AdS. Consistent with this expectation, the modern holographic correlator program reveals much structural resemblance with the beautiful results in flat space. This makes one optimistic that one can export the techniques and notions in flat space and eventually establish a similar amplitude program in AdS.

Up until now, almost all the results in the literature are about correlators of scalar operators. This is because one usually focuses on correlators of the superprimaries of multiplets, which are scalars, in order to preserve the maximal amount of superconformal symmetry. The latter plays an important role in the bootstrap methods of computing correlators. Additionally, scalar correlators have a unique conformal tensor structure, which simplifies the kinematical analysis and allows one to use powerful tools such as the Mellin representation \cite{Mack:2009mi,Penedones:2010ue}. However, correlators of spinning operators are also of great importance. To see the necessity of studying spinning correlators, let us only mention two instances. The first example is the computation of higher-point correlators. Either as part of the algorithm~\cite{Goncalves:2019znr} or as a consistency check~\cite{Alday:2022lkk}, one considers the factorization of the AdS amplitudes at internal propagators~\cite{Goncalves:2014rfa}. Even though the external particles are restricted to be scalar superprimaries, spinning superconformal descendants are still exchanged. Factorization then leads one to consider correlators of spinning operators. The second example is to consider scattering amplitudes of purely bosonic theories in AdS, such as Yang-Mills theory (YM) and Einstein-Hilbert gravity. There are no scalar operators in the dual theories on the boundary and the basic operators are conserved currents. Note that in flat space, it is the amplitudes of spinning particles which exhibit the most remarkable properties. Similarly, in AdS we also need to study these spinning amplitudes in order to have a more direct comparison with the flat-space story. 

In this paper, we consider spinning correlators in 4d $\mathcal{N}=2$ SCFTs. More precisely, we will study four-point correlators of operators residing in the flavor current multiplet. In AdS, the flavor current is dual to the gluon field. We also revisit and rederive some of the results of \cite{Korchemsky:2015ssa,Belitsky:2014zha} for $\mathcal{N}=4$ theories, where the four-point correlator of the stress tensor multiplet is considered, and express these results in a more explicit form. 

Our motivation is threefold. The first is to obtain an array of explicit results for four-point correlators of superconformal descendants. Thanks to superconformal symmetry, all component correlators can be packaged into a ``super correlator'' using the superspace formalism. Correlators of superconformal descendants can be related to that of the superconformal primary via differential operators which we will derive. However, these expressions of correlators are rather implicit. We will evaluate the action of the differential operators and write the results in Mellin space and in terms of $D$-functions so that they can be readily used, for example, in the construction of higher-point correlators via factorization. 

The second motivation is to obtain explicit expressions for four-point amplitudes of gluons and gravitons for bosonic YM and gravity in $\mathrm{AdS}_5$. These amplitudes are extremely difficult to compute via diagrammatic expansions. This is especially true for the latter case as already in flat space one faces an excessive amount of diagrams. However, we can obtain them via supersymmetry even though these theories themselves are not supersymmetric. Here we are using the important property that the {\it tree-level} four-gluon and four-graviton amplitudes are the {\it same} in the supersymmetric and non-supersymmetric theories. This can be seen from the fact that other fields are forbidden in the exchanges due to the R-symmetry selection rule and the contributing diagrams are the same in both cases.

The third motivation is to explore possible generalizations of the double copy relation~\cite{Bern:2010ue} in AdS. The double copy relation in flat space is a remarkable structure which allows one to obtain graviton amplitudes from gluon ones. However, the curved space generalization of double copy is so far still elusive. In AdS, one such relation was found in Mellin space in \cite{Zhou:2021gnu} for four-point correlators of the bottom components of the multiplets. Since the component correlators are unified by the super correlator, the double copy property should also exist in some form in other correlators. Specially, thanks to the aforementioned identification with non-supersymmetric amplitudes at tree level, there should be a double copy relation which relates the bosonic YM theory and the bosonic gravity theory in AdS. This would constitute a direct analogy of the story in flat space. 

In this paper, we will achieve the first two goals via a thorough analysis of the superspace and the associated differential operators. While we did not manage to identify a precise double copy relation for bosonic YM and gravity four-point amplitudes, we found ample evidence for the existence of such a structure. For example, the differential operator relating the four-gluon amplitude to the superprimary amplitude is precisely half of that of the gravity one. Moreover, as a byproduct of our analysis, we will also introduce a useful kinematic configuration where all polarizations are orthogonal to the positions in the embedding space. This configuration has the advantage that spinning correlators effectively become scalar correlators.  We will write down explicitly the AdS four-gluon and four-graviton amplitudes in this configuration which take a quite simple form. The AdS double copy relation, which we leave to future work, should first manifest itself in this simplified limit. Although holographic correlators are our main focus, the usefulness of this configuration goes beyond this context and should be useful for the numerical bootstrap as well.

\subsection{Summary of results}

\paragraph{$\boldsymbol{\mathcal{N}=2}$ and $\boldsymbol{\mathcal{N}=4}$ Superspace} In order to obtain the four-point function of flavor currents we need to express the correlator of the entire multiplet in superspace. This can be done with the formalism of harmonic superspace~\cite{Galperin:1984av, gios_harmonic_superspace}. This is possible because in~\cite{Eden:2000qp}, it was shown that the entire superspace expression depends on only one scalar function of the cross ratios, which can be fixed by looking at the bottom component of the multiplet. Such correlator was recently computed at tree level~\cite{Alday:2021odx} and at one loop~\cite{Alday:2021ajh}.  In this paper, we find a way to relate the superspace formalism with the modern index-free formalism and with that, all superspace computations become very explicit. Using the results of the superspace literature, we obtain a differential representation of the supercharges $Q$ and $\Qb$ and the superconformal charges $S$ and $\Sb$, which we report here in full detail with our conventions. Then we follow the same logic that was adopted in~\cite{Belitsky:2014zha} to promote to superspace the correlator of four $[0,2,0]$ half-BPS operators in $\CN=4$ super Yang-Mills (SYM).  In this case,  the correlator of four stress tensor superfields can be written as some kinematical prefactor times  a superspace invariant function,  which must be annihilated by all supercharges.  By imposing that, the correlator takes the form
\eqn{
\langle \BBT(x_1,\theta_1,\thetab_1,t_1)\cdots \BBT(x_4,\theta_4,\thetab_4,t_4)\rangle =(\hat{g}_{12}\lsp \hat{g}_{34})^2 \,Q^8\Sb{}^8\, \theta_1^4\theta_2^4\theta_3^4\theta_4^4\,\hat{A}(x_i,t_i)\,, 
}[]
where $\hat{g}_{ij}$ is the superpropagator in~\eqref{NfourSuperprop} and $t_i$ denotes the SU(4)$_R$ polarizations.
The function $\hat{A}$ is proportional to the four-point function of the bottom component $\CO\sim \tr\varphi^{\{I}\varphi^{J\}}$. The entire superspace expression is thus fixed, provided we can actually compute those superspace differentials explicitly. We discovered that the situation for eight supercharges is almost identical: it is sufficient to merely halve the number of supercharges and Grassmann variables in every formula. Indeed, the correlator of four supergluons is
\eqn{
\langle \BBO_2(x_1,\theta_1,\thetab_1,\xi_1)\cdots \BBO_2(x_4,\theta_4,\thetab_4,\xi_4)\rangle =(g_{12}\lsp g_{34})^2 \,Q^4\Sb{}^4\, \theta_1^2\theta_2^2\theta_3^2\theta_4^2\,A(x_i,y_i)\,,
}[]
with $g_{ij}$ defined in~\eqref{gijdef} and $y_i$ is now the SU(2)$_R$ polarizations.
The function $A$, as before, is proportional to the four-point function of the lowest component $\CO_2$. In the body of the paper we work out this study in full detail and also provide some explicit results in position and Mellin space for both the $\CN=4$ and $\CN=2$ case.

\paragraph{Component correlators} Once we have the expression in superspace we can extract the various components by doing a Taylor expansion in the Grassmann variables $\theta$ and $\thetab$.  More precisely, one can write an appropriate superconformal differential operator, which selects a specific operator inside the multiplet. The one for  the stress tensor $\CT_{\mu\nu}$ was computed in~\cite{Korchemsky:2015ssa} and the one for the flavor current $\CJ_\mu$ is obtained here and is given in~\eqref{diffopDef}.
Moreover, in~\cite{Korchemsky:2015ssa} it was shown that the various components of $\langle \BBT \BBT \BBT \BBT\rangle$ can actually be expressed in a rather compact form in terms of a simple differential operator $\BBD_i$
\eqn{
\diff_i=\etab_{i}^\alphad \frac{\partial}{\partial x_i^{\alpha \alphad}} \frac{\partial}{\partial \eta_{i \alpha}}\;,
}[]
acting on a seed function.\footnote{The quantities $\eta$ and $\etab$ introduced above are polarizations used to contract the spin indices. } We find a similar structure for the components of~$\langle \BBO_2 \BBO_2 \BBO_2 \BBO_2 \rangle$. For example, the correlators of four flavor currents $\CJ$ in $\CN=2$ and of four stress tensors $\CT$ in $\CN=4$ have a remarkably simple structure and are, again, related by simply halving all the powers
\twoseqn{
\langle\CJ\CJ\CJ\CJ\rangle &= \diff_1\lsp\diff_2\lsp\diff_3\lsp\diff_4\,\Lambda(x_i, \eta_i)\, \Hgl \,,
}[]{
\langle\CT\CT\CT\CT\rangle &= \diff_1^2\lsp\diff_2^2\lsp\diff_3^2\lsp\diff_4^2\,\Lambda(x_i,  \eta_i)^2\, \Hgr \,.
}[][JandT]
Here $\Lambda(x_i,  \eta_i)$ is a kinematic function defined in~\eqref{mathcalMdef} and $\Hgl$ and $\Hgr$ are the four-point functions of the respective bottom components. When we input the tree-level correlators, \eqref{JandT} give AdS spinning gluon and graviton amplitudes which coincide with the tree-level amplitudes in bosonic theories.   

\paragraph{Correlators of top components} Among the various component correlators,  the ones of the top components of the multiplets are particularly interesting. These are the Lagrangians $\CL$, $\overbar\CL$ in $\CN=4$ and the superpotentials $\CW$, $\overbar\CW$ in $\CN=2$.  We compute their four-point function, which  take a very simple form
\twoseqn{
\langle\CW\overbar\CW\CW\overbar\CW\rangle &= \frac{1}{(x_{12}^2x_{34}^2)^3} \Delta^{(4)} \, (x_{12}^2x_{34}^2)^2 x_{13}^2x_{24}^2  \, \Hgl\,,
}[]{
\langle\CL\bar\CL\CL\bar\CL\rangle &= \frac{1}{(x_{12}^2x_{34}^2)^4}\,\Delta^{(8)}\, (x_{12}^2x_{34}^2x_{13}^2x_{24}^2)^2\, \Hgr \,,
}[][]
where $\Delta^{(4)}$ and $\Delta^{(8)}$ are differential operators in the cross ratios, defined in~\eqref{deltaFourDef} and~\eqref{deltaEightDef}, respectively. These differential operators are very interesting in their own right. The one of order eight first appeared in~\cite{Drummond:2006by}. Later they were observed to be relevant for a, still conjectural, hidden conformal symmetry in $\CN=4$ SYM~\cite{Caron-Huot:2018kta} and also in these $\CN=2$ theories~\cite{Drummond:2022dxd}.\footnote{The hidden conformal symmetry for $\CN=2$ theories was found in \cite{Alday:2021odx}.} Another interesting fact about them is that they are Casimir operators for the $u$-channel OPE (when the four-point function is expressed as above). More details about this are given in Sec.~\ref{Sec:topGluon}.

\paragraph{Orthogonal polarization} Spinning correlators with generic polarizations are quite complicated. In this paper, we propose to study and compute explicitly these correlators in a special kinematic configuration in which all the polarizations are orthogonal to the positions in embedding space.  This has the advantage of effectively turning the spinning correlator into a scalar one. We enforce this condition by choosing a particular frame~\eqref{xjCF} where, for each point, we fix a plane for $x_i$. This, consequently, leaves two possibilities for the polarizations in the orthogonal complement. We denote these different configurations as $\textttpm$, as they can be interpreted as the charge under a global U(1) group arising from the breaking of the conformal group SO(2,4) $\rightarrow$ U(1) $\times$ SO(2,2). This fact also makes it clear why only the correlators $\texttt{-++-}, \,  \texttt{++-{-}}, \, \texttt{-+-+}$ are nonzero, while the all $\texttt{+}$ and all $\texttt{-}$ configurations vanish.

\paragraph{Hints of a double copy} Finally, we turn our attention to the double copy relation. The results summarized above are already very indicative of the existence of such a structure, which even extends to the spinning components. These observations are so far only kinematical. However, we are able to write down the dynamical part, {\it i.e.}, the reduced correlator of four supergluons in position space, as follows
\eqn{
\Hgl  \propto \mathtt{c}_s\lsp\BBN_sW_s + \mathtt{c}_t\lsp\BBN_tW_t + \mathtt{c}_u\lsp\BBN_sW_u\,,
}[]
in terms of some seed functions $W_{s,t,u}$ and some differential operators $\BBN_{s,t,u}$, all defined in detail in Sec.~\ref{sec:bottomDC}. Here $W_{s,t,u}$ play the role of scalar propagator and $\BBN_{s,t,u}$ are the ``numerators'' in the  flat-space amplitude parlance.  Then, by performing a simple replacement
\eqn{
\mathtt{c}_{s,t,u}\;\longrightarrow\;\BBN_{s,t,u}\,,
}[]
we obtain the supergraviton four-point function, up to a rational function $\texttt{R}$
\eqn{
\mathbb{N}^2_s W_s+\mathbb{N}^2_tW_t+ \mathbb{N}^2_uW_u=\frac{9\pi^2N^2}{8}\Hgr +\mathtt{R}\,.
}[]
The existence of such an ``almost'' double copy for the bottom components naturally suggests an extension to spinning correlators when written in superspace as in~\eqref{JandT}.

\subsection{Structure of the paper}

The rest of the paper is organized as follows. In Sec.~\ref{Sec:superspace} we introduce the setup and review the basics of superspace. In Sec.~\ref{Sec:2ptfun}-\ref{sec:supergluons} we apply the superspace techniques to two-, three- and four-point functions and obtain correlators involving different $\mathcal{N}=2$ superconformal descendants. We then compare the results with the case of stress tensor multiplet four-point functions with $\mathcal{N}=4$ superconformal symmetry in Sec.~\ref{sec:supergravitons}. In Sec.~\ref{Sec:altpresentation} we present the spinning correlators in an alternative form by explicitly evaluating the action of the differential operators and introduce the orthogonal polarization configuration. So far, our discussions are purely kinematical and apply to any $\mathcal{N}=2$ SCFTs at any point in the moduli space. Starting from Sec.~\ref{Sec:holographic} we will specialize to the holographic limit and consider correlators from SYM and supergravity in AdS. We then present in Sec.~\ref{Sec:doublecopy} various structures which are indicative of a double copy structure. The paper also contains several appendices to which we relegate the technical details.  Some explicit results are made available in an ancillary file to the arXiv submission of this paper. 

\section{Superspaces for four dimensional \texorpdfstring{$\boldsymbol{\CN=2}$}{N=2}}\label{Sec:superspace}

\subsection{Setup}
In this paper we will study half-BPS operators in four dimensional $\CN=2$ superconformal field theory (SCFT).  There exist various holographic origins for these theories.  An $\CN=2$ theory can arise for instance from a stack of $N$ D3-branes near 7\nobreakdash-brane singularities in F-theory~\cite{Fayyazuddin:1998fb,	Aharony:1998xz}.  The near horizon metric is $\mathrm{AdS}_5$ times a compact five-dimensional space.  Only specific periodicities of this compact space, corresponding to different singularities, give rise to a SCFT on the D3-branes.  
The 7-brane,  which fills $\mathrm{AdS}$, wraps an $S^3$ of the compact space, which locally is an $S^5$.  
Hence its presence breaks the SO(6) isometry group to SU(2)$_R\times$U(1)$_r$ --- corresponding to the $\CN=2$ R-symmetry group --- times an additional global SU$(2)_L$.  Alternatively one can also start from the usual $\CN=4$ SYM setup and add probe flavor branes~\cite{Karch:2002sh}.  In particular,  adding $N_F$ D7-branes, $N_F\ll N$, filling $\mathrm{AdS}_5$ and wrapping an $S^3$ inside the  $S^5$, breaks half of the original supersymmetries thus giving a  $\CN=2$ SCFT.

In both cases, on the 7-brane there is an eight dimensional $\CN=1$ vector transforming in the adjoint representation of a gauge group $G_F$, which, from the CFT side,  constitutes a global flavor symmetry.  In the second example we have introduced, this group has to be identified with the  SU($N_F$). When reduced on the $S^3$, this vector multiplet gives rise to an infinite tower of Kaluza-Klein modes, which correspond to the $\CN=2$ half-BPS operators we are interested in.\footnote{
$B_1 \bar{B}_1 [0;0]^{(R;0)}_R$ in the notation of~\cite{Cordova:2016emh} and $\hat\CB_{R/2}$ in the notation of~\cite{Dolan:2002zh}.
}
They are protected scalar superprimaries of the form
\eqn{
\mathcal{O}^I_R(x, \xi, \xi^\prime)=\xi_{a_1} \cdots \xi_{a_R}\lsp \xi^\prime_{a^\prime_1} \cdots \xi^\prime_{a^\prime_{R-2}}\mathcal{O}_R^{I;\lsp a_1 \cdots a_R;\lsp a^\prime_1 \cdots a^\prime_{R-2} }(x)\, ,
}[half-BPS]
where $I=1, \ldots, \mathrm{dim}(G_F)$ is the color index in the adjoint representation of the flavor group $G_F$, $a_i$ is the $\SU(2)_R$ R-symmetry index while $a^\prime_i$ is an $\SU(2)_L$ global symmetry index.  We have contracted these indices with  auxiliary (commuting) polarization spinors $\xi$ and $\xi^\prime$ in order to impose the appropriate symmetrization properties. Their half-BPS nature fixes the conformal dimension $\Delta=R$.

\begin{figure}[t]
\centering
\begin{tikzpicture}
\node[rectangle, draw=black, very thick] (prim) at (0,0) {$\CO_2=[0;\lnsp0]^{(2;0)}_{2}$};

\node[rectangle, draw=black,inner sep=3pt] (glu) at (-2,-1.5) {$\lambda_\alpha=[1;\lnsp0]^{(1;-1)}_{5/2}$};
\node[rectangle, draw=black,inner sep=3pt] (glub) at (2,-1.5) {$\bar\lambda_\alphad=[0;\lnsp1]^{(1;1)}_{5/2}$};
\draw [->,>=stealth, semithick, shorten >=2pt] (prim) -- (glu) node[midway, fill=white, inner sep=1pt] {\footnotesize{$Q$}};
\draw [->,>=stealth, semithick, shorten >=2pt] (prim) -- (glub) node[midway, fill=white, inner sep=1pt] {\footnotesize{$\Qb$}};

\node[rectangle, draw=black] (T) at (-4,-3) {$\CW=[0;\lnsp0]^{(0;-2)}_3$};
\node[rectangle, draw=black] (Tb) at (4,-3) {$\overbar\CW=[0;\lnsp0]^{(0;2)}_3$};
\draw [->,>=stealth, semithick, shorten >=2pt] (glu) -- (T) node[midway, fill=white, inner sep=1pt] {\footnotesize{$Q$}};
\draw [->,>=stealth, semithick, shorten >=2pt] (glub) -- (Tb) node[midway, fill=white, inner sep=1pt] {\footnotesize{$\Qb$}};

\node[rectangle, draw=black] (J) at (0,-3) {$\CJ_\mu=[1;\lnsp1]^{(0;0)}_3$};
\draw [->,>=stealth, semithick, shorten >=2pt] (glub) -- (J) node[midway, fill=white, inner sep=1pt] {\footnotesize{$Q$}};
\draw [->,>=stealth, semithick, shorten >=2pt] (glu) -- (J) node[midway, fill=white, inner sep=1pt] {\footnotesize{$\Qb$}};
\end{tikzpicture} 
\caption{The \textit{supergluon} multiplet. We are denoting the operators with $[\ell, \bar{\ell}\lsp]^{(R;r)}_\Delta$, where $\ell$ and $\bar{\ell}$ stand for the Lorentz representation, $R$ corresponds to the SU(2)$_R$ symmetry, $r$ to the $U(1)_r$ and $\Delta$ represents the conformal dimension.}\label{Fig:multiplet}
\end{figure}
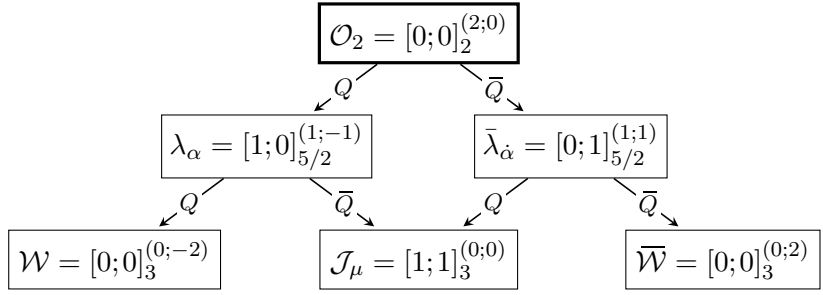

In this work, we will focus on the superprimary with the lowest possible R-charge, namely $R=2$. This is dual to the \textit{supergluon} in $\mathrm{AdS}$. The corresponding supermultiplet is in fact rather special since it contains the conserved flavor current for the group $G_F$.  As depicted in the diagram of Fig.~\ref{Fig:multiplet}, this supermultiplet starts with the scalar $\CO_2^{ab}$ and then it continues with two gluinos $\lambda^{a}_\alpha$ and $\lambdab^{a}_{\alphad}$, the flavor current $\CJ_\mu$ and two complex scalars of opposite $\rmU(1)$ R-charge $\CW$ and $\overbar{\CW}$. 
Let us conclude this section by mentioning that in both holographic setups we have described,  at large $N$ the self-couplings of the gluons are parametrically larger than their couplings to the gravitons~\cite{Alday:2021odx, Alday:2021ajh}, with the latter being suppressed by powers of $1/N$.  Consequently, at leading order in the large $N$ expansion, {\it i.e.}, the tree-level approximation we will consider, the contribution from gravitons to the connected correlator is absent.

\subsection[\texorpdfstring{$\CN=2$}{N=2} Superspaces]{$\boldsymbol{\CN=2}$ Superspaces}
An efficient approach to study different components of the same supermultiplet in a supersymmetry preserving way is to group them, possibly in a way that makes shortening conditions manifest.  The best suited language for these purposes is the one of superspace. The underlying idea is to enlarge the Minkowski space by adding Grassmann coordinates, and possibly a compact internal manifold, in such a way that the superconformal group is realized as the superspace isometries. In this work we will consider the $\CN=2$ harmonic superspace~\cite{Galperin:1984av, gios_harmonic_superspace} to derive the supercharges and then move to the analytic superspace to make computations easier~\cite{Eden:2000qp}. See also~\cite{Hartwell:1994rp} for a review.

Let us take a generic multiplet in an $\CN=2$ superconformal field theory whose primary is a scalar $(R+1)$-plet of $\SU(2)_R$.\footnote{From this moment on, in order to avoid cluttering, when referring to the half-BPS operators in~\eqref{half-BPS}, we will often suppress the flavor and SU(2)$_L$ indices, reinserting them only when necessary.} We write it as a rank-$R$ symmetric tensor which depends on a four-vector $x^\mu$ and two doublets of Grassmann spinors $\theta_a^\alpha$ and $\thetab^{a\alphad}$, transforming in the fundamental and anti-fundamental of $\SU(2)_R$
\eqn{
\mathbb{O}^{a_1\cdots a_R}(x,\theta,\thetab) = \CO^{a_1\cdots a_R}(x) + \theta_a^\alpha\lsp (Q^a_\alpha \CO)(x) + \cdots
}[]
A priori there are at most $2^8$ terms in the expansion, but the shortening conditions can reduce them.

We would like to consider the multiplets in~\eqref{half-BPS}, thus we need to impose the half-BPS shortening conditions which can be realized in superspace as the following differential equation
\eqn{
D_\alpha^{(a} \mathbb{O}^{a_1\cdots a_R)}(x,\theta,\thetab) = \Db_{\alphad b}\epsilon^{b(a} \mathbb{O}^{a_1\cdots a_R)}(x,\theta,\thetab) = 0\,,
}[shortening]
where the parentheses denote symmetrization and the differential operators are defined in~\eqref{DDbdef}.  As we did for the scalar component, in order to enforce the symmetrization of all the indices, we contract them with the same auxiliary variable $\xi_a$
\eqn{
\mathbb{O}(x,\theta,\thetab,\xi) = \xi_{a_1}\cdots\xi_{a_R}\lsp \mathbb{O}^{a_1\cdots a_R}(x,\theta,\thetab)\,.
}[]
With this definition, the differential equations above become just differentiation with respect to the operators $\xi_a D^a_\alpha$ and $\xi^b \Db_{b\alphad}$ --- where indices are raised and lowered with the $\epsilon_{ab}$ tensor.\footnote{In our conventions $\epsilon^{12}=\epsilon_{21}=1$.} The idea now is to rotate the  $\theta_a^\alpha$ and $\thetab^{a\alphad}$ to a different basis $\theta^\pm_\alpha$ and $\thetab^\pm_\alphad$ where the latter two operators become simple derivatives. This is achieved by
\threeseqn{
\theta^+_\alpha &= \xi_a \theta_\alpha^a\,,\qquad\;\,\hspace{.2pt} \theta^-_\alpha = \xib^a \theta_{a\alpha}\,,
}[]{
\thetab^+_\alphad &= \xi^a \thetab_{a\alphad}\,,\qquad \thetab^-_\alphad = \xib_a\thetab^a_{\alphad}\,,
}[]{
z^\mu &= x^\mu + i\lsp \theta^{a}\sigma^\mu\thetab^{b}\lsp (\xi_a\xib_b + \xi_b\xib_a)\,,
}[][thetapmWithxi]
with $\xib$ satisfying the property $\xi_a\xib^a = 1$. After performing the change of variables the covariant derivatives can be written as in~\eqref{Dnewvar}. From there one easily sees that the differential operators above become
\eqn{ 
\xi_a D^a_\alpha = \frac{\partial}{\partial \theta^{-\alpha}}\,,\qquad 
\xi^b \Db_{b\alphad} = \frac{\partial}{\partial \thetab^{-\alphad}}\,.
}[]
We conclude that the operator $\mathbb{O}$ depends only on the variables $\theta^+$, $\thetab^+$ and  $z$ 
\eqn{
\mbox{half-BPS}\lsp:\qquad \mathbb{O}(z,\theta^+,\thetab^+,\xi)\,.
}[]
This is precisely the set of coordinates that defines the harmonic superspace~\cite{Galperin:1984av, gios_harmonic_superspace}. Indeed, in harmonic superspace one has two Grassmann spinors $\theta^+_\alpha$ and $\thetab_\alphad^+$, a four-vector $z^\mu$ and an $\SU(2)$ matrix
\eqn{
\CU=\left(
\begin{array}{cc}
u^+_1 & u^-_1 \\
u^+_2 & u^-_2
\end{array}
\right) = 
\left(
\begin{array}{cc}
\cos \sqrt{\chi\chib} & i \chi \frac{\sin\sqrt{\chi\chib}}{\sqrt{\chi\chib}} \\
i \chib \frac{\sin\sqrt{\chi\chib}}{\sqrt{\chi\chib}} & \cos \sqrt{\chi\chib} 
\end{array}\right) = 
\frac{1}{\sqrt{1+y\yb}}\left(
\begin{array}{cc}
1  & -\yb \\ y & 1
\end{array}\right) \,.
}[Udef]
The parameters $\chi,\chib$ and $y,\yb$ are equivalent parametrizations of the matrix $\CU$. Unimodularity of $\CU$ implies $u_a^+ u^{-a} = 1$.

We have showed that harmonic superspace is the correct tool to accurately describe the half-BPS operators of interest, although it is still not clear how to make contact with the polarizations $\xi$. We will address this point shortly. Let us denote with $T^{\pm \pm}$, $T^0$ the generators of SU$(2)_R$, then from~\cite{gios_harmonic_superspace} we know that $\mathbb{O}(z,\theta^+,\thetab^+,\xi)$ can be written as a function on a supercoset as follows
\eqna{
\mathbb{O}(z,\theta^+,\thetab^+,\xi) &= \Omega(z,\theta^+,\thetab^+,\xi)\cdot \CO(0)\,,\\
\Omega(z,\theta^+,\thetab^+,\xi) &\equiv \exp\mleft[i\left(\chi T^{++}+\chib T^{--}\right)\mright]\,\exp\mleft[i\left(-z^\mu P_\mu - \theta^{+\alpha} Q^-_\alpha - \thetab^{+\alphad}\Qb{}^-_\alphad\right)\mright]\,,
}[OmegaDef]
where $\CO(0)$ is a superconformal primary operator which is annihilated by this set of generators
\eqn{
M_{\mu\nu},\; K_\nu,\;S^\alpha_a,\;\Sb{}^{\alphad a},\;Q^+_\alpha,\;\Qb{}^+_\alpha,\;T^{++},\;(T^{--})^{R+1}\,,
}[]
and has a definite eigenvalue under $D$, $r$ and $T^0$. The definitions of the generators together with their commutation relations can be found in Appendix~\ref{app:SuperconformalAlgebra}. The $\pm$ notation is related to the usual one as follows
\eqna{
Q_\alpha^1 &= Q_\alpha^+\,,&\qquad Q_\alpha^2 &= Q_\alpha^-\,,&\qquad \Qb_{\alphad1} &= \Qb^-_\alphad \,,&\qquad \Qb_{\alphad2} &= -\Qb_\alphad^+\,,\\
S^\alpha_1 &= S^{\alpha-}\,,&\qquad S^\alpha_2 &= -S^{\alpha+}\,,&\qquad \Sb^{\alphad1} &= \Sb^{\alphad+} \,,&\qquad \Sb^{\alphad2} &= \Sb^{\alphad-}\,.\label{plusminusNotation}
}[]
If we set $z=\theta^+=\thetab^-=0$ we only have the R-symmetry group element acting on a highest-weight state of weight $R/2$. This highest-weight state can be represented as a tensor with $R$ indices all in the ``1'' position $\CO^{11\cdots1}(0)$. It is possible to show that
\eqn{
\exp\mleft[i\left(\chi T^{++}+\chib T^{--}\right)\mright]\cdot \CO^{11\cdots1}(0) = \xi_{a_1}\cdots\xi_{a_R}\lsp \CO^{a_1\cdots a_R}(0)\,,
}[]
with
\eqn{
\xi_a = u_a^+\,,\qquad \xib_a = u_a^-\,,
}[]
where $u_a^\pm$ are functions of $\chi$ and $\chib$ as shown in the first parametrization of~\eqref{Udef}. This completes our identification between operators and elements in the supercoset.

An equivalent formulation of superspace goes under the name of analytic superspace~\cite{Eden:2000qp}. It is best seen by using the $y,\yb$ parametrization of $\CU$~\eqref{Udef}. It suffices to complexify the matrix $\CU$ regarding $y$ and $\yb$ as independent, and then setting $\yb = 0$. This in turn leads to a very simple form for the polarizations
\eqn{
\xi_a = \left(\begin{array}{c}
1 \\ y
\end{array}\right)\, , \qquad 
\bar{\xi}_a = \left(\begin{array}{c}
0 \\ 1
\end{array}\right)\,.
}[xitoymap]
In the context of analytic superspace, the variables $\theta^+_\alpha$ and $\thetab^+_\alphad$ are often renamed as $\lambda_\alpha$ and $\pi_\alphad$. Note that, if we forget about superspace, $\xi$ is defined up to a rescaling $\xi\to\lambda \xi$, where $\lambda$ can always be reabsorbed in the normalization of $\CO_2$. Furthermore, $\xib$ is determined by $\xi_a\xib^a=1$, which fixes it only up to a shift $\xib_a \to \xib_a + \alpha \xi_a$ for any $\alpha$. With this point of view in mind, \xitoymap can be seen as merely a gauge fixing of the definitions of $\xi$ and $\xib$.

Finally, let us remark that the topological twist,  achieved with the chiral algebra construction of~\cite{Beem:2013sza},  takes a rather nice form in these variables
\eqn{
y = \zb\,.
}[]

\subsection{Computing supercharges}

In order to compute the supercharges we can use the coset element $\Omega(z,\theta^+,\thetab^+,\xi)$ defined in~\eqref{OmegaDef}. We act on a state $|0\rangle$ which is annihilated by all generators that we have taken at the denominator of the coset, namely
\eqn{
\big\{M_{\mu\nu},K_\nu,D,S^\alpha_a,\Sb{}^{\alphad a},r,Q^+_\alpha,\Qb{}^+_\alpha,T^0\big\} |0\rangle = 0\,.
}[]
Let us now consider a generator $X^A\in \SU(2,2|2)$ associated with a parameter $\omega_A$. This generates a transformation of the form
\eqn{
\exp \mleft(i \omega_A X^A\mright)\cdot (z^\mu,\theta^+_\alpha,\thetab^+_\alphad)  = (z^\mu+ \delta z^\mu,\theta^+_\alpha + \delta\theta^+_\alpha,\thetab^+_\alphad+\delta \thetab^+_\alphad)\,.
}[]
We need to find those $\delta z$, $\delta \theta^+$ and $\delta \thetab^+$. This can be done by equating two quantities: the first is the action of the generator on $\Omega$, while the second is a general variation of $\Omega$ with respect to its parameters. For simplicity we can multiply on the left by the inverse of $\Omega$ so that we obtain quantities that live in the algebra. Let us denote the superspace coordinates collectively by  $\bfz=(z^\mu,\theta^+_\alpha,\thetab^+_\alphad)$. We arrive to the equation
\eqn{
\Omega(\bfz)^{-1} \lsp \exp \mleft(i \omega_AX^A\mright)\lsp \Omega(\bfz)\,|0\rangle = \Omega(\bfz)^{-1}\lsp\Omega(\bfz+\delta\bfz)\,|0\rangle\,.
}[]
Solving this equation gives the variations $\delta\bfz$ for an arbitrary generator $X^A$, which could be $Q^a_\alpha$, $\Sb{}^{a\alphad}$, etc. Then with the variations we can compute the operator by doing simply the chain rule
\eqn{
X^A \equiv \delta_{\omega_A} = \frac{\partial (\delta \theta^{+\alpha})}{\partial\omega_A}\frac{\partial}{\partial\theta^{+\alpha}} + \frac{\partial (\delta \thetab^{+\alphad})}{\partial\omega_A}\frac{\partial}{\partial\thetab^{+\alphad}} + \frac{\partial (\delta z^\mu)}{\partial\omega_A} \frac{\partial}{\partial z^\mu}+ \frac{\partial(\delta u^+_a)}{\partial \omega_A}\frac{\partial}{\partial u_a^+}\,.
}[]
Note that in the above equation we did not put the variation $\delta u^-_a$ because, in harmonic superspace, this variation is always vanishing. This is a consequence of the fact that the harmonic superspace is closed under the superconformal algebra.\footnote{
Indeed we want that setting to zero $\xib = u^-$ in~\eqref{thetapmWithxi} is consistent. Namely we want to make sure that transformations within the harmonic subspace will not reintroduce any $\theta^-$'s.
}$^{,}$\footnote{
This also means that the variations $\delta u^+_a$ must be orthogonal to $u^{-a}$, otherwise the constraint $u_a^+u^{-a}=1$ would be spoiled. It is easy to check that this is indeed the case.
} Furthermore, if we make the change of variables to the $y$ and $\yb$ parametrization, we also observe that derivatives with respect to $\yb$ never appear. It is thus consistent to set $\yb=0$ as we remarked previously.

In harmonic superspace the generators are given in~\eqref{Qharmonic} for the $Q$'s and in~\eqref{Sharmonic} for the $S$'s. The same generators in analytic superspace can instead be found in~\eqref{Qanalytic} and~\eqref{Sanalytic} respectively.

\section{Correlator of two supergluons (and higher Kaluza-Klein modes)}\label{Sec:2ptfun}

Now that we have introduced all the necessary ingredients to treat half-BPS operators in superspace, we can start constructing correlators out of them. Let us start from the easiest one, namely the two-point function.
We denote the superfield corresponding to the operators in~\eqref{half-BPS} as follows
\eqn{
\mathbb{O}^I_R(z,\theta^+,\thetab^+,\xi,\xi') \qquad \text{such that}\qquad \mathbb{O}^I_R(\bfz,\xi')|_{\theta^+=\thetab^+=0}=\mathcal{O}^I_R(x, \xi,\xi')\,.
}[]
Here $\bfz$ collectively denotes $z$, $\theta^+$, $\thetab^+$ and $\xi$.  The two-point function can be written in terms of the ``superpropagator'' $g_{ij}$ defined as follows
\eqn{
g_{ij} \equiv \frac{1}{z_{ij}^2}\left(\xi_{ij}-4 i\frac{\theta_{ij}^{+\alpha}  (\rmz_{ij})_{\alpha\alphad}\thetab_{ij}^{+\alphad} }{z_{ij}^2} \right)\,,
}[gijdef]
where we have defined $\xi_{ij}=\xi_i^a\xi_{ja}$, $\theta^+_{ij}=\theta^+_i-\theta^+_j$ and the same for $\thetab^+_{ij}$. With this definition the two-point function reads simply
\eqn{
\langle \mathbb{O}^{I}_R(\bfz_1,\xi'_1) \, \mathbb{O}^{J}_R(\bfz_2,\xi'_2)\rangle=\delta^{I J}\,(\xi'_{12})^{R-2}\,(g_{12})^R\,.
}[twopfDef]
One can show that this superpropagator is indeed covariant under SU$(2,2|2)$ transformations.\footnote{In particular one can check that
\eqna{
(Q^\pm_i +Q^\pm_j) g_{ij}&=(\Qb^\pm_i +\Qb^\pm_j) g_{ij} = 0\,,\qquad &(S^-_i +S^-_j)g_{ij}&=(\Sb^-_i +\Sb^-_j)g_{ij}=0
\,,\\
(S^{\alpha +}_i +S^{\alpha +} _j)g_{ij}&=-4 (\theta^\alpha_i+\theta^\alpha_j)g_{ij}\, , \qquad &
(\Sb^{\alphad +}_i +\Sb^{\alphad +} _j)g_{ij}&=4 (\thetab^\alphad_i+\thetab^\alphad_j)g_{ij}\,.
}\vspace{-\baselineskip}} Also note that if the Grassmann variables are set to zero $g_{ij}$ reduces to the scalar propagator
\eqn{
g_{ij}\big|_{\theta^+=\thetab^+=0} = \frac{\xi_{ij}}{x_{ij}^2}\,.
}[]

The two-point function, other than being useful for the disconnected correlator, can be used to compute the differential operator that extracts the spin-one component in the multiplet $\CV_{R}^\mu\equiv[1;1]^{(R-2;0)}_{R+1}$.
Indeed we are looking for an operator $\CD_{\alpha\alphad}$ of the form
\eqn{
\CD_{k,\alpha\alphad} = i \frac{\partial}{\partial \theta_k^{+\alpha}}\frac{\partial}{\partial \thetab_k^{+\alphad}} + \CA(R_k)\, \frac{\partial}{\partial y_k} \, \sigma^\mu_{\alpha\alphad} \frac{\partial}{\partial z_k^\mu}\,,
}[diffopDef]
for some coefficient $\CA(R)$ which might, in principle, depend on the $\SU(2)_R$ representation of the operator it acts on.  In theory, one should explicitly verify that this operator commutes with $K_\mu$ (see Appendix~\ref{app:SuperconformalAlgebra}), to guarantee that it yields a conformal primary.  However, as anticipated before, it is much easier to start from the correlators of two superfields and impose that 
\eqn{
\CD_{1,\alpha\alphad}\CD_{2,\beta\betad}\,\langle \mathbb{O}_R(\bfz_1)\lsp \mathbb{O}_R(\bfz_2)\rangle\big|_{\theta^+_i = \thetab_i^+=0} = c_{\CV_R}\,\langle \CV_{R\,\alpha\alphad}(x_1)\,\CV_{R\,\beta\betad}(x_2) \rangle
}
is a conformally covariant two-point function of a spin-one operator.\footnote{The same problem was solved in general for $\CN=1$ supersymmetry in~\cite{Manenti:2019jds}, where there are also $\CN=2$ results but those are different from what we need. The differential operators defined there extract $\CN=1$ superprimaries from $\CN=2$ superprimaries rather than the conformal primaries directly. They are also expressed in full superspace instead of harmonic superspace.} This fixes the value of $\CA(R)$ to be
\eqn{
\CA(R) = \frac{2}{R}\,.
}[]
Furthermore, the normalization $c_{\CV_R}$ turns out to be
\eqn{
c_{\CV_R} = 16\lsp(R^2-1)\,.
}[]
From now on we will focus on the operator with smallest dimension, namely $\mathbb{O}_2^I(\bfz)$, whose superdescendant $\CV_2^\mu$ is precisely the flavor current $\CJ^\mu$.

\section{Correlator of three supergluons}\label{Sec:3ptfun}

The three-point function of supergluons was computed in $\CN=2$ superspace in~\cite{Kuzenko:1999pi}. Using~\eqref{thetapmWithxi} we can map their notation to ours. In particular, one should observe that many quantities drastically simplify as they should only depend on $\theta^+$ and $\thetab^+$. As an example, we have the identity
\eqn{
\xi^a_i\frac{\hat{u}_a^{\phantom{a}b}(z_{ij})}{(x_{\bar\imath j}^2x_{\bar\jmath\llsp i}^2)^{1/2}}\xi_{j\lsp b} = g_{ij}\,,
}
with $g_{ij}$ given by~\eqref{gijdef}. This necessarily follows from the general form of the two-point function~\eqref{twopfDef}.

The other invariants, such as $\mathbf{u}_a^{\phantom{a}b}(\mathbf{Z}_i)$, $\mathbf{X}_i$ and $\overbarUp{\mathbf{X}}_i$, can be computed explicitly and expressed in terms of the analytic superspace variables. We can directly drop the $\theta^-$'s since we know they will have to disappear in the end.

The various components can be extracted using the differential operator defined in~\eqref{diffopDef}. We can then compare the results with some known basis of three-point functions. For concreteness, we choose the one implemented in the package appeared in~\cite{Cuomo:2017wme}, whose notations are reviewed in Appendix~\ref{app:tensors}. We find
\eqn{
\langle \CO_2^I(x_1)\CO_2^J(x_2)\CO_2^K(x_3)\rangle = f^{IJK} \frac{\xi_{12}\lsp\xi_{13}\lsp\xi_{23}}{x_{12}^2\lsp x_{13}^2\lsp x_{23}^2}\,,
}[]
as expected. Applying the operator $\CD_{\alpha\alphad}$ on the first point leads to
\eqn{
\langle \CJ_{\alpha\alphad}^I(x_1)\CO_2^J(x_2)\CO_2^K(x_3)\rangle =  -2f^{IJK}\lsp(\xi_{23})^2 \frac{(\hat{\mathbb{J}}^1_{23})_{\alpha\alphad}}{(x_{12}^2x_{13}^2)^2}\,,
}[]
where $\hat{\mathbb{J}}^1_{23}$ is defined in~\eqref{defTensor}.
Applying two operators at the first two points yields zero. This is expected since the flavor symmetry Ward identities relate this to a two-point function $\langle \CJ_\mu\CO_2\rangle$. Finally, applying the operators on all three fields we get
\eqn{
\langle \CJ_{\alpha\alphad}^I(x_1)\CJ_{\beta\betad}^J(x_2)\CJ_{\gamma\gammad}^K(x_3)\rangle = f^{IJK} \sum_{i=1}^4 \lambda_i^+\,(\mathbb{T}^+_i)_{\alpha\alphad,\beta\betad,\gamma\gammad}\,,
}[tensors3]
where the $\mathbb{T}_i^+$ are the basis of three-point structures in~\eqref{tensorthreeE}. The coefficients that multiply them are given by
\eqn{
\lambda_1^+ = 48\,,\qquad \lambda_2^+ = -\lambda_3^+ = \lambda_4^+ = -80\,.
}[]
The precise linear relations satisfied by them are necessary to ensure conservation of $\CJ_{\alpha\alphad}^I(x)$ at all three points. Furthermore, conformal invariance allows in principle for another structure $\mathbb{T}^-$ which is parity odd. This structure does not appear in our example and this is expected because the parity odd structure is associated to a 't Hooft anomaly~\cite{Lin:2019vgi}, which cannot be present in $\CN=2$.

Note that in all these cases $\CA(2)=1$ was necessary to even have a combination of conformally covariant structures at all. Any other value of $\CA$ would have led to something that is not a three-point function of a primary.

\section{Correlator of four supergluons}\label{sec:supergluons}
The first correlator containing nontrivial dynamical information is the four-point function.
In~\cite{Eden:2000qp} it has been shown that the superspace correlator of four analytic superfields can be entirely determined in terms of a single bosonic scalar function of the superconformal cross ratios $u$ and $v$, which are defined as
\eqn{
u=\frac{x_{12}^2 x_{34}^2}{x_{13}^2 x_{24}^2}=z \zb \, , \qquad v=\frac{x_{14}^2 x_{23}^2}{x_{13}^2 x_{24}^2}=(1-z)(1- \zb) \, ,
}[]
Such function can in turn be fixed by the correlator of the lowest component of the multiplet, {\it i.e.}  $\langle \CO_2\CO_2\CO_2\CO_2 \rangle$. To perform the uplift from the bottom component to the full superspace answer, we will follow closely the analysis in~\cite{Belitsky:2014zha,Korchemsky:2015ssa} for $\CN=4$ SYM.  We will also briefly summarize their results in Sec.~\ref{sec:supergravitons}.

From now on we will drop the ``$+$'' from the Grassmann variables and denote $z^\mu$ as $x^\mu$. Furthermore, we will trade $\xi_i$ for $y_i$ using~\eqref{xitoymap}.\footnote{
Note that $y_i - y_j \equiv y_{ij} =\xi_{ij} \equiv \xi_i^a\xi_{ja}$.\label{footyxi}
} The correlator of four superfields, in full analytic superspace,  at leading order in the large $N$ expansion can be written as 
\eqn{
\langle \mathbb{O}_2^{I_1} (\bfz_1)\mathbb{O}_2^{I_2} (\bfz_2)\mathbb{O}_2^{I_3} (\bfz_3) \mathbb{O}_2^{I_4} (\bfz_4) \rangle=g_{12}^2 \lsp g_{34}^2 \left( \mathbb{G}_{\mathrm{rational}}^{I_1I_2I_3I_4} + \mathbb{G}_{\mathrm{anom}}^{I_1I_2I_3I_4}\right)\, .
}[]
The rational part can be simply obtained by starting from the rational part of the correlator of four $\CO_2^I$, as  in~\cite{Alday:2021ajh}, and promoting the scalar propagators to the superpropagator $g_{ij}$ in~\eqref{gijdef}. 
\eqna{
\mathbb{G}_{\mathrm{rational}}^{I_1I_2I_3I_4}&= \delta^{I_1I_2}\delta^{I_3I_4} +  \delta^{I_1I_3}\delta^{I_2I_4}\,\CU^2+ \delta^{I_1I_4}\delta^{I_2I_3}\lsp\left(\frac\CU\CV\right)^2 \\
& \quad + \frac{(C_{2,2,2})^2}{3}\left(
(\mathtt{c}_t-\mathtt{c}_s)\, \frac\CU\CV +(\mathtt{c}_s-\mathtt{c}_u)\,  \CU + (\mathtt{c}_t-\mathtt{c}_u)\,  \frac{\CU^2}\CV
\right)\,,
}[GratGl]
with
\eqn{
\CU \equiv\frac{g_{13}\lsp g_{24}}{g_{12} \lsp g_{34}}\; \xrightarrow{\theta,\thetab=0} \;\alpha \lsp u\,,\qquad \CV \equiv \frac{g_{13}\lsp g_{24}}{g_{14} \lsp g_{23}}\;\xrightarrow{\theta,\thetab=0}\;\frac{\alpha\lsp v}{1-\alpha}\,,
}[]
and
\eqn{
\alpha= \frac{y_{13}y_{24}}{y_{12}y_{34}}\, , \qquad \alpha-1=  \frac{y_{14}y_{23}}{y_{12}y_{34}}\, .
}
If we denote the structure constants of $G_F$ as $f^{I J K}$, then we can express the color structures as
\eqn{
\mathtt{c}_s=f^{I_1 I_2 J}f^{J I_3 I_4}\,,\qquad \mathtt{c}_t=f^{I_1 I_4 J}f^{J I_2 I_3}\, , \qquad \mathtt{c}_u=f^{I_1 I_3 J}f^{J I_4 I_2}\, .
}[css]
Finally the constant $C_{2,2,2}$ represents the $\CO_2$'s three-point coefficient and it is related to the flavor central charge through
\eqn{
C_{2,2,2}^2=\frac{6}{C_{\CJ}}\, .
}[]
The interesting part, however, is the anomalous one. Similarly to the $\CN=4$ case, we would like to write an ansatz that satisfies the superconformal Ward identities, namely that it is annihilated by all  supercharges and reproduces the lowest component correlator once we set all the $\theta$'s and $\thetab$'s to zero.  A natural solution is\footnote{Notice that the order of $Q^{\pm}$ and $\Sb^{\pm}$ does not matter since they  are all mutually anticommuting.}$^{,}$\footnote{What may sound confusing at first is that we just said that all $Q$'s and $\Qb$'s acting on the correlator give zero. In order to extract the $Q$, resp. $\Qb$, superdescendant from a superspace expression, however, one has to act with the superspace covariant derivative $D$, resp. $\Db$, defined in Appendix~\ref{app:covariantDeriv}. The same is true for imposing shortening conditions: the equation is $D\CO=\Db\CO=0$. The difference is in the order in which they compose: $D_1 D_2 f(\theta) = D_1 (D_2 f(\theta))$ whereas $[Q_1,[Q_2,\CO\}\} = (Q_2(Q_1\CO))$. Other than that, the algebra is the same.}
\eqna{
\mathbb{G}_{\mathrm{anom}}^{I_1I_2I_3I_4}= (Q^-)^2(Q^+)^2(\Sb^-)^2(\Sb^+)^2\left[\theta_1^2\theta_2^2\theta_3^2\theta_4^2\lsp \frac{F(x)}{g_{12}^2g_{34}^2}\right]\, ,
}[ansatz]
where $x$ denotes collectively $(x_1,x_2,x_3,x_4)$ and $F(x)$ implicitly contains the flavor indices.  Denoting with the label ``$i$''  the point in the correlator, we  define
\threeseqn{
\theta_i^2 &= \theta_i^{+\alpha}\theta_{i\alpha}^{+}\,,\qquad
\thetab_i^2 = \thetab_{i\alphad}^+\thetab_i^{+\alphad}\,.
}[]{
(Q^{\pm})^2 &= \Big(\sum\nolimits_{i=1}^4 Q_i^{\alpha\pm}\Big)\Big(\sum\nolimits_{i=1}^4 Q_{i\alpha}^{\pm}\Big)\,,
}[]{
(\Sb^{\pm})^2 &= \Big(\sum\nolimits_{i=1}^4 \Sb_{i\alphad}^\pm\Big)\Big(\sum\nolimits_{i=1}^4 \Sb_i^{\alphad\pm}\Big)\,.
}[][]
Written in the form~\eqref{ansatz}, and given the fact that all the $Q$'s and the $S$'s are nilpotent, it is clear that the correlator satisfies half of the Ward Identities. It is less trivial to check that this satisfies the other half, but this is indeed the case.  Alternatively, we could have taken the barred version of~\eqref{ansatz} and the same reasoning would have applied.

Now that we have this formula we would like to find an explicit expression for $F(x)$ and we will do so in the next subsection by making a connection with the four-point function of the lowest component.  But before doing that, we need to simplify the expression in order to make it more manageable.  This is achievable by resorting to some tricks and clever observations similar to the ones in~\cite{Korchemsky:2015ssa}. We will proceed in  two steps: first we will use the fact that $g_{ij}$ is annihilated by $Q^\pm$ and $\Sb^-$,  thus we can pass it through them.  At the same time $g_{ij}$ commutes with $\theta^2_i$, so that~\eqref{ansatz} becomes
\eqna{
g_{12}^2 g_{34}^2 \,\mathbb{G}_{\mathrm{anom}}^{I_1I_2I_3I_4}&=(Q^-)^2(Q^+)^2(\Sb^-)^2 \tilde{S}^2\left[\theta_1^2\theta_2^2\theta_3^2\theta_4^2\lsp F(x)\right]\,,\\
\tilde{S}^\alphad &\equiv g_{12}^2 g_{34}^2\Big(\sum\nolimits_i\Sb_i^{\alphad +}\Big) g_{12}^{-2}g_{34}^{-2}=\sum_{i=1}^4\Sb_i^{\alphad +} -8 \lsp  \thetab_i^{\alphad}\, .
}[ansatz1]
Notice now that $Q^-$ and $\Sb^-$ do not involve any derivative with respect to $x$ and the $y$ derivative in $\Sb^-$ gives trivially zero since there is no dependence from $y$ in~\eqref{ansatz1}. So they can only act on the product  of $\theta_i^2$ returning 
\eqna{
g_{12}^2 g_{34}^2\,\mathbb{G}_{\mathrm{anom}}^{I_1I_2I_3I_4}&=(Q^+)^2 \tilde{S}^2 \times\\&\;\quad\times\left[ x_{12}^2 x_{13}^2 x_{14}^2\left( \frac{\theta_{13}x_{13}}{x_{13}^2}-\frac{\theta_{12}x_{12}}{x_{12}^2} \right)^2 \left( \frac{\theta_{14}x_{14}}{x_{14}^2}-\frac{\theta_{12}x_{12}}{x_{12}^2} \right)^2 F(x)\right]\, ,
}[ansatz2]
where by $X^2$ we mean $X_\alphad \epsilon^{\alphad \betad} X_\betad$.

In the following subsections we will present the results for the various components of the superconformal four-point function. We mainly focus on insertions of the flavor current but we also present some results for the top components $\CW$ and $\overbar\CW$. The computations presented here were performed in Mathematica with the package \texttt{Superspace4d}.\footnote{The source code is available at \href{https://gitlab.com/maneandrea/superspace4d}{\texttt{gitlab.com/maneandrea/superspace4d}}.}

\subsection{Lowest component}

The lowest component can be obtained from~\eqref{ansatz2} by keeping only the terms that remove a $\theta$. This is because all other terms will either generate $\thetab$'s that will not be needed at this order, or do not remove enough $\theta$'s. In this way we get for the anomalous part 
\eqn{
\langle  \CO_2^{I_1}(x_1) \CO_2^{I_2}(x_2) \CO_2^{I_3}(x_3) \CO_2^{I_4}(x_4) \rangle = \frac{y_{12}^2 \lsp y_{34}^2}{(x_{12}^2 \lsp x_{34}^2)^2}(z \alpha-1)(\zb \alpha-1)\lsp\CHgl(z,\zb)\, ,
}[lowestComp]
where we finally introduced the function of the cross ratios $\CF(z,\zb)$ which is related to $F(x)$ as follows
\eqn{
F(x) \equiv \Hgl(x)= \frac{C_{2,2,2}^2}{2^4}\frac{\CHgl(z,\zb)}{(x_{12}^2x_{34}^2)^2x_{13}^2x_{24}^2}\,.
}[FxLowest]
Notice that this result is consistent with the Ward identities~\cite{Nirschl:2004pa}. Comparing with~\cite{Alday:2021ajh}, the reduced correlator at the holographic limit is given by 
\eqna{
\CHgl(z,\zb) &\equiv  \CHgl^{I_1I_2I_3I_4}(z,\zb)=\mathsf{c_s}\CH_s+\mathsf{c_t}\CH_t+\mathsf{c_u}\CH_u\,,\\
\CH_s&=\frac{1}{3} (z \zb)^2 \left(\Db_{2321}-\Db_{3221}\right)\, , \\
\CH_t&= \frac{1}{3} (z \zb)^2 \left(\Db_{2231}-\Db_{2321}\right)\, , \\
\CH_u&= \frac{1}{3} (z \zb)^2 \left(\Db_{3221}-\Db_{2231}\right)\, .
}[]
\subsection{Current insertions}
Let us start by considering the four-point function of one current and three $\CO_2$, by definition this can be obtained as 
\eqn{
\CD_{1, \alpha\alphad}\langle \mathbb{O}_2^{I_1} (\bfz_1)\mathbb{O}_2^{I_2} (\bfz_2)\mathbb{O}_2^{I_3} (\bfz_3) \mathbb{O}_2^{I_4} (\bfz_4) \rangle\big|_{\theta_i=\thetab_i=0} \equiv \langle \CJ_{\alpha\alphad}^{I_1}(x_1)\CO_2^{I_2}(x_2) \CO_2^{I_3}(x_3) \CO_2^{I_4}(x_4)  \rangle 
}[]
where we have used the differential operator in~\eqref{diffopDef} with $\CA(2)=1$. An explicit computation gives us\footnote{The complete result together with the ones for the other correlators can be find in an ancillary Mathematica file.}
\eqna{
\eta_1^{\alpha} \etab_1^{\alphad} \langle \CJ^{I_1}_{\alpha \alphad} \CO_2^{I_2}\CO_2^{I_3}\CO_2^{I_4}\rangle&=2^4 \lsp y_{23} y_{34} y_{24}\lsp  \diff_1 \left[ \left( \eta_1 \rmx_{12} \rmx_{23} \rmx_{34} \rmx_{41} \eta_1 \right) \Hgl(x) \right]\\
&=-2^4 \lsp y_{23} y_{34} y_{24}\lsp  \diff_1 \big[ \BBL_{234}^1 \Hgl(x) \big]\, ,
}[JOOO]
where in the second line we have used~\eqref{defTensor} and similarly to~\cite{Korchemsky:2015ssa} we have defined
\eqn{
\diff_i=\etab_{i}^\alphad \frac{\partial}{\partial x_i^{\alpha \alphad}} \frac{\partial}{\partial \eta_{i \alpha}}\, .
}[diffDef]
Here and in the following we will contract the $\alpha, \alphad$ indices with auxiliary commuting spinor variables, respectively $\eta^\alpha$ and $\etab^{\alphad}$.
Now passing to two insertions, we find 
\eqna{
\eta_1^{\alpha} \etab_1^{\alphad} \eta_2^{\beta} \etab_2^{\betad} \langle \CJ^{I_1}_{\alpha \alphad} \CJ^{I_2}_{\beta \overset{\phantom{1}}{\betad}}\CO_2^{I_3}\CO_2^{I_4}\rangle& =-2^4 \lsp y_{34}^2 \lsp  \diff_1 \diff_2 \left[ \left( \eta_1 \rmx_{13} \rmx_{32} \eta_2 \right) \left( \eta_1 \rmx_{14} \rmx_{42} \eta_2 \right) \Hgl(x) \right]\\
& =-2^4 \lsp y_{34}^2 \lsp  \diff_1 \diff_2 \left[ \BBK_3^{12} \BBK_4^{12} \Hgl(x) \right]
\, ,
}[JJOO]
where the $\mathbb{K}$ is the invariant tensor in~\eqref{defTensor}.

The four-point function of three currents and one $\CO_2$ simply vanishes because of R-symmetry conservation. In fact $\mathcal{J}$ is neutral under the $SU(2)$ R-symmetry while $\mathcal{O}_2$ has $SU(2)$ spin 1. 

The next and last non-trivial correlator is the one with all flavor currents. This reads
\eqn{
\eta_1^{\alpha} \etab_1^{\alphad} \eta_2^{\beta} \etab_2^{\betad} \eta_3^{\delta} \etab_3^{\deltad} \eta_4^{\gamma} \etab_4^{\gammad} \langle \CJ^{I_1}_{\alpha \alphad} \CJ^{I_2}_{\beta \overset{\phantom{1}}{\betad}}\CJ^{I_3}_{\delta \deltad}\CJ^{I_4}_{\gamma \gammad}\rangle=2^4  \lsp  \diff_1 \diff_2 \diff_3 \diff_4 \left[ \Lambda(x, \eta)\Hgl(x) \right]\, ,
}[JJJJ]
where we have introduced
\eqna{
\Lambda(x, \eta)&=\left(\frac{(\eta_1 \rmx_{12} \rmx_{23} \eta_3)(\eta_4 \rmx_{41} \rmx_{12} \eta_2)-(\eta_1 \rmx_{12} \rmx_{24} \eta_4)(\eta_3 \rmx_{31} \rmx_{12} \eta_2)}{x_{12}^2}\right)^2\\
 &=\frac{1}{x_{12}^4}\left( \BBK_2^{13}\BBK_1^{42}-\BBK^{14}_2 \BBK^{32}_1\right)^2\, .
}[mathcalMdef]

\subsection{Correlator of the top components}\label{Sec:topGluon}
As one can see from Fig.~\ref{Fig:multiplet},  together with the conserved current,  at the same level of the supermultiplet and with the same conformal dimension, we find another operator, the superpotential $\CW$ and its $r$-charge conjugate $\overbar\CW$. 
From the diagram in Fig.~\ref{Fig:multiplet}, it is also clear that the combination of $Q$'s or $\Qb$'s that give the right operator does not mix with derivatives $\partial_\mu$, unlike~\eqref{diffopDef} for example. Therefore we immediately get
\twoseqn{
\CW(x) &= \left(\frac{\partial}{\partial\xi_a}\frac{\partial}{\partial\theta^a_\alpha}\right)^2\mathbb{O}(x,\theta,\thetab,\xi)\big|_{\theta,\thetab=0}\,,
}[]{
\overbar\CW(x) &= \left(\frac{\partial}{\partial\xi_a}\frac{\partial}{\partial\thetab^a_\alphad}\right)^2\mathbb{O}(x,\theta,\thetab,\xi)\big|_{\theta,\thetab=0}\,.
}[][]
We want to translate this result into analytic superspace. This can be done by a simple change of variables using~\eqref{thetapmWithxi} and
\eqn{
y = \frac{\xi_2}{\xi_1} = -\frac{\xi^1}{\xi^2}\,.
}[]
When acting on $\BBO_2$ we do not need to consider the action of the Grassmann derivatives on $z^\mu$ since this is going to produce factors of $\thetab$ (for $\CW$) and $\theta$ (for $\overbar\CW$) that will be set to zero anyway. The only derivatives we need to be careful about are those in $y$. A simple computation shows
\eqn{
\frac{\partial}{\partial\xi_a}\frac{\partial}{\partial\theta^a_\alpha}\BBO_2 = \bigg(2+ \theta_\beta^+\frac{\partial}{\partial\theta_\beta^+}+\thetab_\betad^+\frac{\partial}{\partial\thetab_\betad^+}\bigg)\frac{\partial}{\partial\theta_\alpha^+}\BBO_2\,,
}[]
and similarly for the barred analogue. In particular the $\partial/\partial y$ term drops out due to \eqn{\xi_a\frac{\partial y}{\partial\xi^a}=0\,.}
The terms in the parentheses give an overall prefactor since they simply count the homogeneity degree of the $\theta$'s and $\thetab$'s. All in all, quite unsurprisingly, the differential operators that we need are
\twoseqn{
\CW(x) &= \frac{\partial}{\partial\theta^+_\alpha}\frac{\partial}{\partial\theta^{+\alpha}}\mathbb{O}_2(z,\theta^+,\thetab^+,y)\big|_{\theta^+,\thetab^+=0}\,,
}[]{
\overbar\CW(x) &= \frac{\partial}{\partial\thetab^{+\alphad}}\frac{\partial}{\partial\thetab^+_\alphad}\mathbb{O}_2(z,\theta^+,\thetab^+,y)\big|_{\theta^+,\thetab^+=0}\,.
}[][]
By an explicit computation one can also confirm that $\CW$ and $\overbar\CW$ do not depend on $y$, which is not manifest from the above equation.

Since both these operators are charged, the only nonzero four-point function with them is
\eqn{
\langle \CW(x_1)\overbar\CW(x_2)\CW(x_3)\overbar\CW(x_4)\rangle = \frac{1}{(x_{12}^2 x_{34}^2)^3}\CG_\CW(z,\zb)\,,
}[WWb]
and permutations thereof. In order to obtain it, we can compute the derivatives in~\eqref{ansatz2} and retain only the terms $\theta_1^2\thetab_2^2\theta_3^2\thetab_4^2$. The result is a differential operator acting on the function $\CHgl$ defined in~\eqref{FxLowest}.  Let us denote $\CHtgl(z\zb,(1-z)(1-\zb)) \equiv \CHgl(z,\zb)$ and $\tilde\CG_\CW$ in the same way. The result takes a form reminiscent of~\cite{Drummond:2006by}, indeed it is a ``square root'' of their result
\eqn{
\tilde\CG_\CW(u,v) = u^3\,\Delta^{(2)}\lsp u v\lsp \Delta^{(2)}\lsp u^{-2} \CHtgl(u,v) \equiv \Delta^{(4)} \CHtgl(u,v)\,,
}[deltaFourDef]
with
\eqn{
\Delta^{(2)} \equiv u\partial_u^2+v\partial_v^2+(u+v-1)\partial_u\partial_v+2(\partial_u+\partial_v)\,.
}[delta2]
Another way to present this result is through the Casimir differential operator\footnote{The hat emphasizes that this Casimir operator does not act naturally on the blocks of the $(12)$ OPE but rather on the $2\leftrightarrow3$ crossed blocks
\[
\hat{\CD}_z\,\kappa_h(1/z) = -h(h-1) \,\kappa_h(1/z)
\,,\qquad \kappa_h(x)\equiv z^h\lsp{}_2F_1(h,h;2h;x)\,.
\]\vspace{-\baselineskip}\label{foot:crossedCasimr}}
\eqn{
\hat{\CD}_x \equiv \partial_x\,x(1-x)\,\partial_x\,.
}[]
We obtain
\eqn{
\CG_\CW(z,\zb) = \frac{(z\zb)^3}{z-\zb}\hat{\CD}_z\hat{\CD}_\zb\,(z-\zb)(z\zb)^{-2}\CHgl(z,\zb)\,.
}[]
As footnote~\ref{foot:crossedCasimr} emphasizes, the Casimir that we used here acts naturally on the OPE of $\CW\times\CW$. If we considered the correlator
\eqn{
\langle \CW(x_1)\CW(x_2)\overbar\CW(x_3)\overbar\CW(x_4)\rangle = \frac{\CG_\CW^{u}(z,\zb)}{(x_{12}^2 x_{34}^2)^3}\,,\qquad \CG_\CW^{u}(z,\zb) \equiv (z\zb)^3\CG_\CW\mleft(\frac1z,\frac1\zb\mright)\,,
}[]
then we would have had\footnote{Here we used the crossing relation $\CHgl(z,\zb) = z\zb\,\CHgl(1/z,1/\zb)$ which is easy to derive from~\eqref{lowestComp}.}
\eqn{
\CG_\CW^u(z,\zb) = \frac{z\zb}{z-\zb}\CD_z\CD_\zb\,(z-\zb)\CHgl(z,\zb)\,,
}[]
which now is expressed in terms of the usual Casimir operator
\eqn{
\CD_x \equiv x^2\partial_x(1-x)\partial_x\,.
}[]
It seems that the combination of Casimir operators wants to act on the chiral channel, namely the one which is annihilated by the highest number of supercharges. A possible, handwavy, explanation of why this is true is the following: the unprotected operators that are exchanged need to be half-BPS themselves, in order to respect the shortening condition in the $\BBO_2\times\BBO_2$ OPE. Let us denote the respective superprimaries as $\CO_\mathrm{L}$. In the neutral $\CW\times\overbar\CW$ channel $Q^2\Qb{}^2\CO_\mathrm{L}$ is allowed to appear together with some other superdescendants as well, up to $Q^4\Qb{}^4\CO_\mathrm{L}$. So the superconformal blocks are nontrivial linear combinations of ordinary bosonic blocks. This means that the expansions of the seed function $\CHgl$ in \emph{super}conformal blocks is not related in a simple way to the expansion of $\CG_\CW$ in ordinary conformal blocks. If we consider the chiral channel, on the other hand, the R\nobreakdash-charge conservation allows only $Q^4\CO_\mathrm{L}$ to be exchanged. Therefore the conformal blocks expansion of $\CF^u_\CW$ is going to be the same as the superconformal block expansion of $\CHgl$, up to prefactors in each single block. Since the operator $(z-\zb)^{-1}\CD_z\CD_\zb(z-\zb)$ has the blocks as eigenfunctions, its job is precisely to provide these prefactors. The physical interpretation of such prefactors is simply the linear relation between OPE coefficients
\eqn{
\lambda_{\CO_2\CO_2(Q^4\CO_{\mathrm{L}})} = \CP(\Delta,\ell)\, \lambda_{\CW\CW (Q^4\CO_{\mathrm{L}})}\,.
}[]
Notice that there is an alternative way of rewriting \eqref{WWb} in terms of $\Hgl(x)$ in \eqref{FxLowest}
\eqn{
\langle \CW(x_1)\overbar\CW(x_2)\CW(x_3)\overbar\CW(x_4)\rangle=\square_2  \square_4\lsp (x_{24}^2)^2 \Hgl(x)=\square_1  \square_3\lsp (x_{13}^2)^2 \Hgl(x)\, ,
}[WWbInF]
where $\square_i=\frac{\partial}{\partial x_i^\mu}\frac{\partial}{\partial x_{i\mu}}$.
\section{Correlator of four supergravitons} \label{sec:supergravitons}
This section is devoted to a very brief review of the results in~\cite{Belitsky:2014zha,Korchemsky:2015ssa}. In doing that, our main intention is to show how the ``squaring'' of the results for supergluons in a $\CN=2$ SCFT recovers their results.

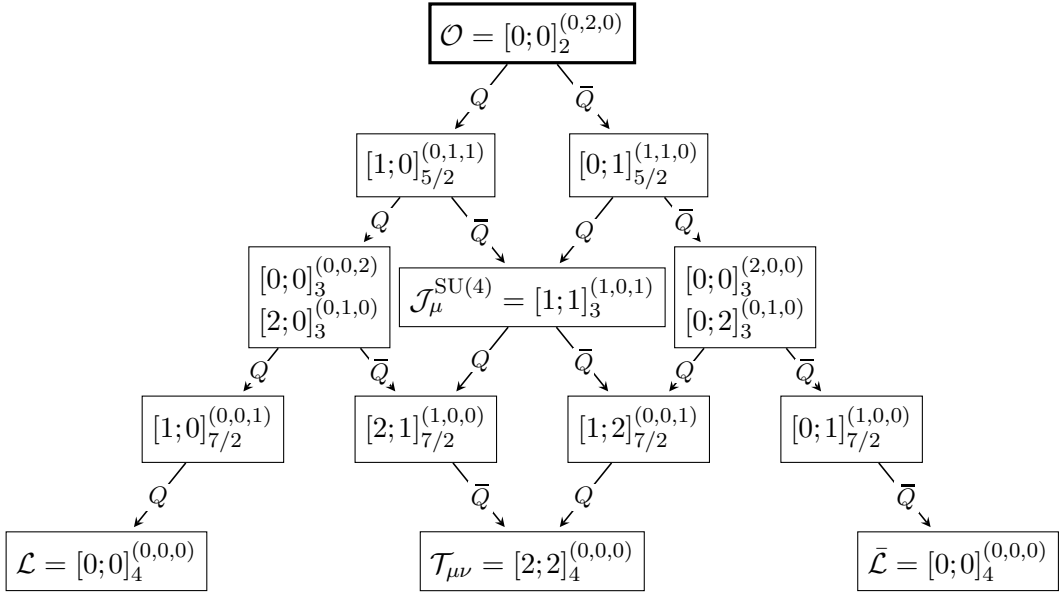
\begin{figure}[t]
\centering
\begin{tikzpicture}
\node[rectangle, draw=black, very thick] (prim) at (0,0) {$\CO=[0;\lnsp0]^{(0,2,0)}_{2}$};

\node[rectangle, draw=black,inner sep=3pt] (glu) at (-1.4,-1.75) {$[1;\lnsp0]^{(0,1,1)}_{5/2}$};
\node[rectangle, draw=black,inner sep=3pt] (glub) at (1.4,-1.75) {$[0;\lnsp1]^{(1,1,0)}_{5/2}$};
\draw [->,>=stealth, semithick, shorten >=2pt] (prim) -- (glu) node[midway, fill=white, inner sep=1pt] {\footnotesize{$Q$}};
\draw [->,>=stealth, semithick, shorten >=2pt] (prim) -- (glub) node[midway, fill=white, inner sep=1pt] {\footnotesize{$\Qb$}};

\node[rectangle, draw=black] (W) at (-2.8,-3.5) {\parbox{1.6cm}{$[0;\lnsp0]^{(0,0,2)}_3$\\$[2;\lnsp0]^{(0,1,0)}_3$}};
\node[rectangle, draw=black] (Wb) at (2.8,-3.5) {\parbox{1.6cm}{$[0;\lnsp0]^{(2,0,0)}_3$\\$[0;\lnsp2]^{(0,1,0)}_3$}};
\draw [->,>=stealth, semithick, shorten >=2pt] (glu) -- (W) node[midway, fill=white, inner sep=1pt] {\footnotesize{$Q$}};
\draw [->,>=stealth, semithick, shorten >=2pt] (glub) -- (Wb) node[midway, fill=white, inner sep=1pt] {\footnotesize{$\Qb$}};

\node[rectangle, draw=black] (J) at (0,-3.5) {$\CJ_\mu^{\SU(4)}=[1;\lnsp1]^{(1,0,1)}_3$};
\draw [->,>=stealth, semithick, shorten >=2pt] (glub) -- (J) node[midway, fill=white, inner sep=1pt] {\footnotesize{$Q$}};
\draw [->,>=stealth, semithick, shorten >=2pt] (glu) -- (J) node[midway, fill=white, inner sep=1pt] {\footnotesize{$\Qb$}};

\node[rectangle, draw=black] (P) at (-4.2,-5.25) {$[1;\lnsp0]^{(0,0,1)}_{7/2}$};
\node[rectangle, draw=black] (Pb) at (4.2,-5.25) {$[0;\lnsp1]^{(1,0,0)}_{7/2}$};
\draw [->,>=stealth, semithick, shorten >=2pt] (W) -- (P) node[midway, fill=white, inner sep=.2pt] {\footnotesize{$Q$}};
\draw [->,>=stealth, semithick, shorten >=2pt] (Wb) -- (Pb) node[midway, fill=white, inner sep=0pt] {\footnotesize{$\Qb$}};

\node[rectangle, draw=black] (M) at (-1.4,-5.25) {$[2;\lnsp1]^{(1,0,0)}_{7/2}$};
\node[rectangle, draw=black] (Mb) at (1.4,-5.25) {$[1;\lnsp2]^{(0,0,1)}_{7/2}$};
\draw [->,>=stealth, semithick, shorten >=2pt] (W) -- (M) node[midway, fill=white, inner sep=.1pt] {\footnotesize{$\Qb$}};
\draw [->,>=stealth, semithick, shorten >=2pt] (Wb) -- (Mb) node[midway, fill=white, inner sep=.1pt] {\footnotesize{$Q$}};
\draw [->,>=stealth, semithick, shorten >=2pt] (J) -- (M) node[midway, fill=white, inner sep=.1pt] {\footnotesize{$Q$}};
\draw [->,>=stealth, semithick, shorten >=2pt] (J) -- (Mb) node[midway, fill=white, inner sep=.1pt] {\footnotesize{$\Qb$}};

\node[rectangle, draw=black] (L) at (-5.6,-7) {$\CL=[0;\lnsp0]^{(0,0,0)}_{4}$};
\node[rectangle, draw=black] (Lb) at (5.6,-7) {$\bar\CL=[0;\lnsp0]^{(0,0,0)}_{4}$};
\draw [->,>=stealth, semithick, shorten >=2pt] (P) -- (L) node[midway, fill=white, inner sep=1pt] {\footnotesize{$Q$}};
\draw [->,>=stealth, semithick, shorten >=2pt] (Pb) -- (Lb) node[midway, fill=white, inner sep=1pt] {\footnotesize{$\Qb$}};

\node[rectangle, draw=black] (T) at (0,-7) {$\CT_{\mu\nu}=[2;\lnsp2]^{(0,0,0)}_4$};
\draw [->,>=stealth, semithick, shorten >=2pt] (Mb) -- (T) node[midway, fill=white, inner sep=1pt] {\footnotesize{$Q$}};
\draw [->,>=stealth, semithick, shorten >=2pt] (M) -- (T) node[midway, fill=white, inner sep=1pt] {\footnotesize{$\Qb$}};

\end{tikzpicture} 
\caption{The \textit{supergraviton} multiplet. We are denoting the operators as $[\ell, \bar{\ell}\lsp]^{(p_1,p_2,p_3)}_\Delta$, where $[\ell, \bar{\ell}]$ stands for the Lorentz representation, $(p_1,p_2,p_3)$ are the SU(4)$_R$ Dynkin labels and  $\Delta$ represents the conformal dimension.}\label{Fig:Tmultiplet}
\end{figure}

The above-mentioned works investigate four-point functions of different components of the energy-momentum supermultiplet in four dimensional $\CN=4$ SYM.  Among them, it is given an explicit and compact form for the one involving one, two, three or four insertions of the stress tensor  $\CT_{\alpha\beta,\alphad\betad}$ itself. These are the ones that we wish to compare to the results discussed in the previous section.  Let us recall that $\CN=4$ is the holographic counterpart of a theory of gravity and the stress tensor multiplet is mapped to the \textit{supergraviton} in $\mathrm{AdS}_5 \times S^5$. 

In $\CN=4$ superspace, given its half-BPS nature, the stress tensor superfield $\mathbb{T}$ depends only on half of the original Grassmann variables $\theta^A_\alpha$ and $\thetab_A^\alphad$, where $A$ is the $\SU(4)$ R\nobreakdash-symmetry group index.
The lowest component is a scalar operator of conformal dimension two, transforming in the $(0,2,0)$ of $\SU(4)$ and in the adjoint of the $\SU(N)$ gauge group.  It takes the schematic form
\eqn{
\CO(x, t)=(t_{M_1} \Sigma^{M_1}_{AB})(t_{M_2} \Sigma^{M_2}_{CD})\, \CO^{AB, CD}(x) \, ,
}[]
where the matrix $\Sigma^M_{AB}$ maps from $\SO(6)$ fundamental index $M$ to antisymmetric $\SU(4)$ ones and $t_M$ are some additional null polarization vectors. At the opposite side of the multiplet, instead, we have the stress tensor  $\CT_{\alpha\beta, \alphad\betad}$ and the self-dual and anti-self-dual part of the SYM Lagrangians $\CL$ and $\overbar{\CL}$, which have opposite charge under the bonus $\rmU(1)_Y$ symmetry~\cite{Intriligator:1999ff}.
In order to better parametrize this multiplet, similarly to what we have done before, we can introduce the harmonic variables, defining this time an SU(4) matrix as
\eqn{
u^{A}_B=\left( \begin{array}{c:c}u^{+a}_B \;\;&\; u^{-a^\prime}_B \end{array}\right)\, .
}[]
The upper index $A$ gets split in two fundamental $\SU(2)$ and $\SU(2)'$  indices $a$ and $a^\prime$, carrying respectively  $\pm 1$ $\rmU(1)$ charge. In this way one realizes the coset space 
\eqn{
\frac{\SU(4)}{\SU(2) \times \SU(2)' \times \rmU(1)}\,.
}[]
As a consequence, the Grassmann variables $\theta^A_\alpha$ also split as
\eqn{
\theta^{a}_\alpha=u^{+a}_A \theta^A_\alpha \, , \qquad \theta^{a^\prime}_\alpha=u^{-a^\prime}_A\theta^A_\alpha \,,
}[]
and similarly for $\thetab_A^\alphad.$  An equivalent description can be given in analytic superspace, where the analytic variables are defined through the identification
\eqn{
\left( \begin{array}{c:c}u^{+a}_B\;\;&\; u^{-a^\prime}_B \end{array}\right)=\left(\begin{array}{c:c}
\,\delta^a_b & 0\, \\\hdashline
\,y^{a}_{b^\prime} & \delta^{a^\prime}_{b^\prime} \,
\end{array}\right)\,.
}[]
Accordingly, the $\SU(4)$ polarization vectors can be written as
\eqn{
t_M \Sigma^M_{AB}=u_A^{+c} \epsilon_{cd} u^{+d}_B =\left(\begin{array}{c:c}
\,\epsilon_{ab} & -y_{a b^\prime}\,\\\hdashline
\, y_{b a^\prime}& \epsilon_{a^\prime b^\prime} y^2
\end{array}\right)\, ,
}
where $y^2=\det y_{a a^\prime}$.
With these coordinates we can write the four-point function of $\mathbb{T}(x_i, \theta_i^a, \thetab_i^{a^\prime}, t_i)\equiv \mathbb{T}(\hat{\bfz}_i)$ as
\eqna{
\langle\mathbb{T}(\hat{\bfz}_1) \mathbb{T}(\hat{\bfz}_2) \mathbb{T}(\hat{\bfz}_3)\mathbb{T}(\hat{\bfz}_4)  \rangle 
&=\hat{g}_{12}^2\hat{g}_{34}^2\left(\hat{\BBG}^{\mathrm{rational}}+\hat{\BBG}^{\mathrm{anom}}\right)\, .
}[Ggrav]
The rational terms can be written in terms of superpropagators
\eqna{
\hat{g}_{ij}= \frac{\hat{y}_{ij}^2}{x_{ij}^2}\, , \qquad \hat{y}_{ij}^{a a^\prime}=y_{ij}^{a a^\prime}-4i \frac{\theta^{\alpha a}_{ij}(x_{ij})_{\alpha\alphad}\thetab^{\alphad a^\prime}_{ij}}{x_{ij}^2}\, .
}[NfourSuperprop]
This superpropagator is defined in such a way that when we set all the $\theta$'s and $\thetab$'s to zero we recover the usual free propagator
\eqn{
\hat{g}_{ij}\big|_{\theta^a=\thetab^{a'}=0} = \frac{y_{ij}^2}{x_{ij}^2}=\frac{t_{ij}}{x_{ij}^2}\, ,
}[]
with $y_{ij}^2\equiv(y_i-y_j)^2=t_i \cdot t_j \equiv t_{ij}$. The explicit expression for the rational part is
\eqn{
\hat{\BBG}^{\text{rational}}=1+\widehat{\CU}^2+\left(\frac{\widehat{\CU}}{\widehat{\CV}}\right)^2+\frac{4}{N^2}\left(\frac{\widehat{\CU}}{\widehat{\CV}}+\widehat{\CU}+\frac{\widehat{\CU}^2}{\widehat{\CV}}\right)\, ,
}[GfreeGr]
with
\eqna{
\widehat{\CU} \equiv\frac{\hat{g}_{13}\lsp \hat{g}_{24}}{\hat{g}_{12} \lsp \hat{g}_{34}}\; \xrightarrow{\theta,\thetab=0} \;\alpha \alphab \lsp u\,,\qquad \widehat{\CV} \equiv \frac{\hat{g}_{13}\lsp \hat{g}_{24}}{\hat{g}_{14} \lsp \hat{g}_{23}}\;\xrightarrow{\theta,\thetab=0}\;\frac{\alpha \alphab\lsp v}{(1-\alpha)(1-\alphab)}\,,
}
and  the $\SO(6)$ cross ratios
\eqn{
\alpha \alphab = \frac{t_{13}\lsp t_{24}}{t_{12}\lsp t_{34}}\,,\qquad
(\alpha -1)(\alphab -1)= \frac{t_{14}\lsp t_{23}}{t_{12}\lsp t_{34}}\,.
}[]
The anomalous correlation function can be written as the action of supercharges 
\eqna{
Q_{a}^\alpha&=\bar{u}_{+a}^A Q_{A}^\alpha \, ,  \qquad &&Q_{a^\prime}^\alpha=\bar{u}_{-a^\prime}^A Q_{A}^\alpha\, ,  \\ 
\bar{S}_{\alphad\lsp a}&=\bar{u}_{+a}^A \bar{S}_{\alphad\lsp A}\,, \qquad && \bar{S}_{\alphad\lsp a^\prime}=\bar{u}_{-a^\prime}^A \bar{S}_{\alphad\lsp A}\, ,
}[]
on a single function
\eqn{
\hat{\BBG}^{\mathrm{anom}}=Q^4 Q^{\prime 4 } \bar{S}^4 \bar{S}^{\prime 4 }\left[ \theta_1^4 \theta_2^4 \theta_3^4 \theta_4^4 \lsp \frac{F(x)}{\hat{g}_{12}^2\hat{g}_{34}^2}\right]\, .
}[Ganomgrav]
Here we have defined $Q=\frac{1}{12} Q^\alpha_a Q^b_\alpha Q^\beta_b Q^a_\beta$ and $Q^\prime=\frac{1}{12} Q^{\alpha \lsp a^\prime}Q_{\alpha\lsp b^\prime}Q^{\beta \lsp b^\prime}Q_{\beta\lsp a^\prime}$ and similarly for $\theta, \, \bar{S}, \, \bar{S}^\prime$.  The function $F$ now has to be identified with 
\eqn{
F(x)\equiv \Hgr(x) =  \frac{4}{N^2}\frac{\CHgr(z,\zb)}{(x_{12}^2x_{34}^2x_{13}^2x_{24}^2)^2}\,, 
}[Fgrav]
in such a way that for the lowest component, {\it i.e.}, the one at $\theta=0=\thetab$, the anomalous part of the correlator takes the familiar form 
\eqna{
\langle \CO\CO\CO\CO\rangle& = \frac{(t_{12}\lsp t_{34})^2}{(x_{12}^2\lsp x_{34}^2)^2}(z\alpha\lnsp-\lnsp1)(\zb\alpha\lnsp-\lnsp1)(z\alphab\lnsp-\lnsp1)(\zb\alphab\lnsp-\lnsp1)\lsp \CHgr(z,\zb)\, .
}[Hgrav]
In particular, in the supergravity limit we are considering, one can identify
\eqna{
\CHgr(z,\zb)&=-(z \zb)^2 \Db_{2422}\, .
}[]
With an expression for $\hat{\BBG}^{\mathrm{anom}}$, one can obtain correlators for the different components in the supermultiplet by acting on \eqref{Ggrav} with an appropriate differential operators, similar to \eqref{diffopDef}, which selects the correct operator.\footnote{This is the operator (2.16) in~\cite{Belitsky:2014zha}.} 
\subsection{Stress tensor insertions}
In this section we report the results of~\cite{Korchemsky:2015ssa} for the correlators involving the stress tensor and the scalar $\CO$. 
Let us denote for  brevity  $\eta^{\alpha_i} \etab^{\alphad_i} \eta^{\beta_i} \etab^{\betad_i} \CT_{\alpha_i\beta_i,\alphad_i\betad_i}$ as $\CT$.
With one single stress tensor insertion we find
\eqna{
 \langle \CT\CO\CO\CO\rangle&=t_{23} t_{34} t_{24}\,  \diff_1^2 \,\big[ ( \eta_1 \rmx_{12} \rmx_{23} \rmx_{34} \rmx_{41} \eta_1)^2 \Hgr(x) \big]\,,
}[TOOO]
where $\diff_i$ is defined in~\eqref{diffDef}. The two stress tensor insertions reads
\eqna{
\langle \CT \CT\CO\CO\rangle&=(t_{34})^2\,  (\diff_1\diff_2)^2 \,\big[  \left( \eta_1 \rmx_{13} \rmx_{32} \eta_2 \right)^2 \left( \eta_1 \rmx_{14} \rmx_{42} \eta_2 \right)^2 \Hgr(x) \big]\,.
}[TTOO]
Also in this case, the correlator with three $\CT$ insertions and one $\CO$ is trivially zero due to R-symmetry conservation: $\CT$ is a singlet under SU$(4)_R$, while $\CO$ belongs to the $(0,2,0)$. 

Finally, the correlator with all four stress tensors is given by
\eqn{
\langle \CT\CT\CT\CT\rangle = 4^4 \lsp (\diff_1\diff_2\diff_3\diff_4)^2\big[
\Lambda(x, \eta)^2 \Hgr(x)
\big]\,,
}[TTTT]
with $\Lambda(x, \eta)$ defined in~\eqref{mathcalMdef}.

\subsection{Correlator of the top components}\label{Sec:topGraviton}
Let us finish this section with the correlators of the Lagrangian. Since $\CL$ and $\overbar{\CL}$ carry opposite $\rmU(1)_Y$ charge,  the only non vanishing four-point functions are the ones with two $\CL$ and two $\overbar{\CL}$.  These are~\cite{Drummond:2006by}
\eqn{
\langle \CL({x_1}) \overbar{\CL}({x_2})  \CL({x_3}) \overbar{\CL}({x_4})\rangle =\frac{1}{(x_{12}^2 x_{34}^2)^4}\Delta^{(8)} \CHtgr(u,v)\equiv\frac{1}{(x_{12}^2 x_{34}^2)^4} {\CG}_\CL
(z,\zb)\, ,
}[LLb]
together with its crossing symmetric version.  Here $\CHtgr(z \zb,(1-z)(1-\zb))\equiv \CHgr(z, \zb)$ is the reduced correlator defined in \eqref{Hgrav} and the eight-derivative operator $\Delta^{(8)}$ is constructed from $\Delta^{(2)}$ in~\eqref{delta2} as
\eqn{
\Delta^{(8)}=u^4( \Delta^{(2)})^2 u^2 v^2 ( \Delta^{(2)})^2 u^{-2}\, .
}[deltaEightDef]
An alternative representation of the Lagrangian correlator is
\eqn{
\langle \CL({x_1}) \overbar{\CL}({x_2})  \CL({x_3}) \overbar{\CL}({x_4})\rangle=\square_2^2 \square_4^2\lsp (x_{24}^2)^4 \Hgr(x)=\square_1^2 \square_3^2\lsp (x_{13}^2)^4 \Hgr(x)\, ,
}[]
with $\Hgr(x)$ defined in~\eqref{Fgrav}. This expression can be viewed as the ``square'' of the similar representation of the supergluon case~\eqref{WWbInF}.

\section{Evaluating the differential representation and the orthogonal frame}\label{Sec:altpresentation}

\subsection{Quick overview of formalisms for four-point tensor structures}

It can be convenient to write the correlator in a different basis of tensor structures. In the previous section we have seen a representation in terms of $\diff_i$ acting on some seed function. Here we want to use the embedding formalism --- see Appendix~\ref{app:embedding} for notations. The four-dimensional formalism for general spin representations allows for tensor structures of both even and odd parity. Here we want to instead use the dimension-agnostic formalism, which deals with parity-even structures and can only be used for symmetric traceless representations in parity-preserving theories, which is all we need. There are only two building blocks
\eqn{
H_{ij} = \eta_i \rmx_{ij}\etab_j\; \eta_j \rmx_{ji} \etab_i\,,\qquad
V_{i,jk} = \frac{1}{\sqrt{2}} \frac{x_{ij}^2\,x_{ik}^2}{x_{jk}^2}\bigg(\frac{\eta_i \rmx_{ik} \etab_i}{x_{ik}^2} - \frac{\eta_i \rmx_{ij} \etab_i}{x_{ij}^2}\bigg)
\,.
}[HandVdefs]
We also need to define the kinematic prefactor
\eqn{
\CK_{(\Delta_1,j_1,\jb_1)\ldots(\Delta_4,j_4,\jb_4)} = \left(\frac{x_{24}^2}{x_{14}^2}\right)^{\frac{\kappa_1-\kappa_2}2}
\left(\frac{x_{14}^2}{x_{13}^2}\right)^{\frac{\kappa_3-\kappa_4}2}
\frac{1}{(x_{12}^2)^{\frac{\kappa_1+\kappa_2}{2}}(x_{34}^2)^{\frac{\kappa_3+\kappa_4}{2}}}\,,
}[kinPrefDef]
where $\kappa_i \equiv \Delta_i + (j_i+\jb_i)/2$. For brevity, the labels $(\Delta,j,\jb)$ will be denoted as $(\Delta,\ell)$ is $j=\jb=\ell$ and as $\Delta$ is $j=\jb=0$.

In order to translate from the seed representation to the embedding one, we evaluate both correlators in a conformal frame and then compare them. Since the little group for four points in four dimensions is $\SO(2)$, the structures are manifestly linearly independent and therefore the comparison becomes trivial. The conventional conformal frame configuration sets the point $x_4$ at infinity~\cite{Kravchuk:2016qvl}. While in theory this is as good a frame as another, in practice it becomes very computationally expensive to evaluate the limit $x_4\to\infty$. For this reason we chose a different frame defined as follows
\eqna{
x_1^\mu &=(0,0,0,0)\,,\\
x_2^\mu &=\left(\frac{z-\zb}{(1+z)(1+\zb)},0,0,\frac{z+\zb+2z\zb}{(1+z)(1+\zb)}\right)\,,\\
x_3^\mu &=(0,0,0,1)\,,\\
x_4^\mu &=(0,0,0,2)\,.
}[CFalt]
The structures in this frame are related to those in the usual frame in the way explained in~\cite{Cuomo:2017wme}. However, we will not need this result since we never compare structures in different frames. As one can easily check, the frame~\eqref{CFalt} has been chosen in such a way that the cross ratios $z$ and $\zb$ follow the usual definition.

\subsection{Orthogonal configuration}\label{subsec:ortho}

Later we will also consider a special kinematic limit in which the structures simplify. Let us introduce it now: it consists in taking the vector polarizations to be perpendicular to the positions. Namely, if we contract Lorentz indices $\mu$ with a null vector $h^\mu$, the configuration we are looking for is $h^\mu_i x_{ij,\mu} = 0$ (without summing over $i$ and for all $j$). In terms of our spinor notation this is
\eqn{
\eta_i \rmx_{ij}\etab_i = 0\,,\qquad \mbox{with}\qquad h_i^\mu = -\frac12\sigma^\mu_{\alpha\alphad}\etab_i^\alphad\eta_i^\alpha\,,
}[hDef]
In embedding space, reviewed in Appendix~\ref{app:embedding}, this reads
\eqn{
P_i\cdot Z_j=0\,,
}[]
which immediately implies that $V_{i,jk}$ is sent to zero. 

When we have a result written directly in terms of embedding structures $H$ and $V$, taking this configuration is very easy. However, the correlators with four insertions are rather involved and it will not be possible to write them in embedding (at least without a lot of work). Therefore it will be necessary to evaluate them in an explicit frame where this condition is automatically enforced. Note that the linear conditions that we must impose are actually more than it seems because
\eqn{
\eta_i\rmx_j\etab_i = \eta_i\rmx_i\etab_i \quad\Longrightarrow\quad
\eta_i\rmx_{jk}\etab_i=0\quad \forall\;i,j,k\,.
}[constr]
Let us now try to understand the most general solution to this system. Without loss of generality we can put the first point at the origin. The remaining three points can be arranged via conformal transformations so that they span a space of dimension $p\leq 2$. If $p=1$ then we can choose $h_i$'s without issues but the cross ratios become dependent, namely $z=\zb$. The only possible choice is $p=2$.\footnote{Since the metric is not definite, one can in principle have the planes of the $h$'s and the $x$'s overlapping. However, the only way to solve the constraints~\eqref{constr} is to have all $x$'s to be null and one $h_i$ proportional to them. We want to avoid that because we need two independent cross ratios.} In this case, since the $h_i$'s are null, they span a one-dimensional space
\twoseqn{
h_j^\mu &= (0,\;h_j^{\mathbf{1}},\; h_j^{\mathbf{2}}=\pm i\lsp h_j^{\mathbf{1}},\;0)^\mu\,,
}[]{
x_j^\mu &= \left(\frac{z_j-\zb_j}2,\;0,\;0,\;\frac{z_j+\zb_j}2\right)^\mu\,.
}[xjCF][]
There are in total 16 different choices that give non-vanishing polarizations (the boldface index is an $\alpha$ or $\alphad$ index)\footnote{Incidentally, this choice is equivalent to taking $q_i=-\bar{q}_i$ and $q_i=\pm \ell_i$ in the formalism and notations of~\cite{Cuomo:2017wme}.}
\eqn{
h^{\texttt+,\lsp\mu}_j \equiv-\frac12\,\big(
0,\;\eta_j^{\mathbf2}\etab_j^{\mathbf1},\;i\lsp\eta_j^{\mathbf2}\etab_j^{\mathbf1},\;0
\big)^\mu\qquad\mbox{or}\qquad
h^{\texttt-,\lsp\mu}_j \equiv-\frac12\,\big(
0,\;\eta_j^{\mathbf1}\etab_j^{\mathbf2},\;-i\lsp\eta_j^{\mathbf1}\etab_j^{\mathbf2},\;0
\big)^\mu\,.
}[choices]
In these configurations all the $V_{i,jk}$'s will vanish by construction, but, unfortunately, some $H_{ij}$'s will vanish as well. Out of these 16, there are 6 configurations that keep nonzero the maximal amount of $H_{ij}$, which is four. If we call \texttt{+} the first choice in~\eqref{choices} and \texttt{-} the second choice, then choosing as an example \texttt{-++-} for $h_1$, $h_2$, $h_3$ and $h_4$ leads to
\eqna{
H_{12} &= \frac12\lsp z_{12}\,\zb_{12}\,\eta_1^{\mathbf1}\eta_2^{\mathbf2}
\etab_1^{\mathbf2}\etab_2^{\mathbf1}\,,\qquad &
H_{13} &= \frac12\lsp z_{13}\,\zb_{13}\,\eta_1^{\mathbf1}\eta_3^{\mathbf2}\etab_1^{\mathbf2}\etab_3^{\mathbf2}\,,\qquad &
H_{14} &=0\,,\\
H_{23} &=0\,,\qquad &
H_{24} &= \frac12\lsp z_{24}\,\zb_{24}\,\eta_2^{\mathbf2}\eta_4^{\mathbf1}
\etab_2^{\mathbf1}\etab_4^{\mathbf2}\,,\qquad\\
H_{34} &= \frac12\lsp z_{34}\,\zb_{34}\,\eta_3^{\mathbf2}\eta_4^{\mathbf1}
\etab_3^{\mathbf1}\etab_4^{\mathbf2}\,.
}[]
In general $H_{ij}$ is nonzero if the signs at the points $i$ and $j$ are opposite. Thus, the choices with an equal number of \texttt{+} and \texttt{-} have four nonzero $H_{ij}$'s, those with three equal choices have only three nonzero $H_{ij}$'s and the two \texttt{++++} and \texttt{-{-}-{-}} have all $H_{ij}$'s set to zero.

The positions on the plane can be taken in the standard conformal frame, or in the frame that we introduced before. Taking $z$ and $\zb$ to be the usual cross ratios, we choose
\eqna{
&z_1 = \zb_1 = 0\,,\qquad
z_2 = \frac{2\lsp z}{z+1}\,,\qquad
\zb_2 = \frac{2\lsp\zb}{\zb+1}\,,\\
&z_3 = \zb_3 = 1\,,\qquad
z_4 = \zb_4 = 2\,.
}[ziDef]

If we evaluate a correlator in the frame \texttt{++{-}-} and in the frame \texttt{-++-} we can obtain the functions multiplying all structures which are made out of $H_{ij}$'s, except for those that are proportional to the product of five or six different $H_{ij}$'s: those will always be vanishing.

Finding an explicit form for the correlators~\eqref{JJJJ} and~\eqref{TTTT} in terms of derivatives of $\CHgl$ and $\CHgr$ is very hard and the final expressions would be too cumbersome. If however one is interested in the orthogonal frame only, it is possible to greatly simplify the computations by expressing the differential operator $\diff_i$, in~\eqref{diffDef},  directly in that frame.\footnote{Normally going to the conformal frame does not commute with the action of generic differential operators such as $\partial/\partial x_i^\mu$ or $\partial/\partial\eta^\alpha_i$. However, for our purposes, we only care about the differential operator~$\diff_i$. } To this aim, let us consider separately the two possible cases: point $i$ is in the \texttt{+} configuration or point $i$ in the \texttt{-} configuration.

In the first case we have that $\etab_i^{\mathbf{2}} = \eta_i^\mathbf{1} = 0$, thus we get
\eqn{
\diff_i = 
\etab_i^{\mathbf{1}}\lsp\left(\frac{\partial}{\partial x_i^1} + i\frac{\partial}{\partial x_i^2}\right)\lsp \frac{\partial}{\partial \eta_i^{\mathbf{1}}}+
\etab_i^{\mathbf{1}}\lsp\left(\frac{\partial}{\partial x_i^0} - \frac{\partial}{\partial x_i^3}\right)\lsp \frac{\partial}{\partial \eta_i^{\mathbf{2}}}\,.
}[]
The same situation arises for the second case which reads
\eqn{
\diff_i =
-\etab_i^{\mathbf{2}}\lsp\left(\frac{\partial}{\partial x_i^1} - i\frac{\partial}{\partial x_i^2}\right)\lsp \frac{\partial}{\partial \eta_i^{\mathbf{2}}}
-\etab_i^{\mathbf{2}}\lsp\left(\frac{\partial}{\partial x_i^0} + \frac{\partial}{\partial x_i^3}\right)\lsp \frac{\partial}{\partial \eta_i^{\mathbf{1}}}\,.
}[]
In order to deal with the spacetime derivatives we can enlarge the conformal frame by adding a single coordinate $\zeta$ for the \texttt{-} case and a coordinate $\zetab$ for the \texttt{+} case. More precisely, we should define
\eqn{
x_i^\mu = \left(\frac{z_i-\zb_i}2,\frac{\zeta_i+\zetab_i}{2},\;\frac{\zeta_i-\zetab_i}{2i},\;\frac{z_i+\zb_i}2\right)^\mu\,.
}[]
So far this is just a relabeling of the four components, but the differential operators now take an easier form
\twoseqn{
\diff_i &= 
2\lsp \etab_i^{\mathbf{1}}\lsp\frac{\partial}{\partial\zetab_i}\lsp \frac{\partial}{\partial \eta_i^{\mathbf{1}}}-
2\lsp \etab_i^{\mathbf{1}}\lsp\frac{\partial}{\partial\zb_i}\lsp \frac{\partial}{\partial \eta_i^{\mathbf{2}}}\,,
&\mbox{\texttt{+} case}\,,
}[]{
\diff_i &= 
-2\lsp \etab_i^{\mathbf{2}}\lsp\frac{\partial}{\partial\zeta_i}\lsp \frac{\partial}{\partial \eta_i^{\mathbf{2}}}-
2\lsp \etab_i^{\mathbf{2}}\lsp\frac{\partial}{\partial z_i}\lsp \frac{\partial}{\partial \eta_i^{\mathbf{1}}}\,,
&\mbox{\texttt{-} case}\,.
}[][diffinOrthoFrame]
Thanks to this we can set to zero the $\zeta_i$'s for the points in the \texttt{+} case as well as the $\zetab_i$'s in the \texttt{-} case. Furthermore we can set to~\eqref{ziDef} the $z_i$'s or $\zb_i$'s that are not involved in the derivative. Then after each derivative we can also set the remaining $\zeta$'s and $\zetab$'s to zero as well as the $\eta_i^\alpha$ that is vanishing for the \texttt{+}/\texttt{-} case of point $i$. It is also without loss of generality that one can set $\etab_i^\alphad = 1$.

With this trick we can obtain all current and stress tensor insertions in the orthogonal frame without the need of evaluating the full expression first. The results are presented in the next section but are not shown explicitly. We refer the reader to the attached Mathematica file for the explicit expressions.

\subsection{Current and stress tensor correlators}

Now we will list all four-point functions involving the current and the $\CO_2$ scalar or  the stress tensor and the $\CO$ scalar .  Let us start with the single current insertion. It is given by
\eqn{
\langle \CJ^{I_1}\CO_2^{I_2}\CO_2^{I_3}\CO_2^{I_4}\rangle = \CK_{(3,1)222} \;y_{23}\lsp y_{24}\lsp y_{34}\,\big(
V_{1,23}\, a_1(z,\zb) + V_{1,24}\, a_2(z,\zb)
\big)\,,
}[JOOOE]
where $\CK_{(3,1)222}$ is the kinematic prefactor defined in~\eqref{kinPrefDef} and
\twoseqn{
a_1 &= 2\sqrt{2} \lsp v \,\big((1+z+\zb)\lsp \CHgl+\zb(\zb-1) \lsp \partial_{\zb}\lsp \CHgl+z(z-1) \lsp \partial_{z}\CHgl\big)\,,
}[]{
a_2 &= 2\sqrt{2} \lsp v \,\big(-\CHgl+\zb \lsp \partial_\zb \CHgl+z\lsp \partial_z\CHgl\big)\,.
}[][]
The analog for $\CN=4$ reads
\eqn{
\langle \CT\CO\CO\CO\rangle = \CK_{(4,2)222}\; t_{23} t_{34} t_{24}\,\big(
(V_{1,23})^2\, b_1 + V_{1,24} V_{1,23}\, b_2 + (V_{1,24})^2\, b_3
\big)\,,
}[TOOOE]
with the definitions 
\threeseqn{
b_1 &=16 \frac{v^2}{z-\zb}((z-1)^2 z^2 (z-\zb)
 \partial_{z}^2  \CHgr+4 (z-1) z^2
   (\zb-1) \zb \partial_z\partial_{\zb}\CHgr\\& \nonumber \, \quad +2 (z-1) z \left(3 z^2-2 z \zb+z+2 (1-2 \zb)
   \zb\right) \partial_z \CHgr\\&\nonumber\, \quad+2 z \left(3 z^2+2 z (\zb+1)+3 \zb^2+2 \zb-2\right)\CHgr- (z\leftrightarrow\zb))\,,
}[]{
b_2 &=32 \frac{v^2}{z-\zb}((z-1) z^2 (z-\zb)\partial^2_z \CHgr+2 z^2 \zb (z+\zb-2)\partial_z\partial_\zb\CHgr\\& \nonumber \, \quad +2 z \left(z^2-2 z \zb+z-2 (\zb-1) \zb\right)\partial_z\CHgr-2 z (z+\zb-2) \CHgr-(z \leftrightarrow\zb))\,,
}[]{
b_3 &=16 \frac{v^2}{z-\zb}(z^2 (z-\zb) \partial_z^2\CHgr	+4 z^2 \zb\partial_z\partial_\zb\CHgr-2 z (z+2 \zb) \partial_z\CHgr-4 z \CHgr-(z\leftrightarrow\zb))\,.
}[][]
Note that these two correlators are identically zero in the kinematic limit $\eta\perp x$.

The first nontrivial correlators in this kinematic configuration are the ones involving at least two spinning operators insertions. For currents, we can rewrite the correlators in embedding formalism as
\eqn{
\langle \CJ^{I_1}\CJ^{I_2}\CO_2^{I_3}\CO_2^{I_4}\rangle = \CK_{(3,1)(3,1)22} \;y_{34}^2\,\big(
W_1 W_2\lsp c_1+\overbar{W}_1W_2\lsp c_2+
W_1 \overbar{W}_2\lsp c_3+\overbar{W}_1\overbar{W}_2\lsp c_4+
H_{12}\lsp c_5
\big)\,,
}[JJOOOr]
with the definitions
\eqna{
W_1 &= V_{1,23}+V_{1,24}\,,\quad
W_2 = V_{2,13}+V_{2,14}\,,\\
\overbar{W}_1 &= V_{1,23}-V_{1,24}\,,\quad
\overbar{W}_2 = V_{2,13}-V_{2,14}\,,
}[]
and $c_i\equiv c_i(z, \zb)$ some polynomials involving up to second derivative in $z$ and $\zb$ of $\CHgl(z, \zb)$.\footnote{The explicit expressions for these correlators, together with all the others that will follow, can be found in an ancillary Mathematica file.} The function $c_5$ is the only one that survives in the orthogonal frame and it reads
\eqna{
c_5(z,\zb) &= \frac{4 (\zb-1) \zb (z^2-3 z \zb-6 z+8 \zb) }{z-\zb}\partial_\zb\CHgl(z,\zb)+
\\&\;\quad\frac{4 (z-1) z (3 z \zb-8 z-\zb^2+6 \zb)}{z-\zb}\partial_z\CHgl(z,\zb)
\\&\;\quad-16 (z-1) z (\zb-1) \zb \partial_z\partial_\zb\CHgl(z,\zb)-4 (z \zb-2 z-2 \zb+12) \CHgl(z,\zb)
}[]

As for the stress tensor, we use the trick explained in Sec.~\ref{subsec:ortho} to extract only the $H_{ij}$ component. The correlator restricted to the $H$'s reads
\eqn{
\langle \CT\CT\CO\CO\rangle = \CK_{(4,2)(4,2)22}\,(t_{34})^2\, H_{12}^2 \,d_1(z,\zb) + O(V_{i,jk})\,.
}[TTOOOr]
The coefficient function $d_1$ is an expression of the form
\eqn{
d_1= \sum_{i=0}^2\sum_{j=0}^2 q_{i,j}(z,\zb)\,\partial_z^i\partial_\zb^j\CHgr\,.
}[]
We relegate the explicit result to the ancillary file.

Finally let us show the correlators with four insertions. Also in this case we will sidestep the computation of the full correlator and go directly to the orthogonal frame configuration.\footnote{The full expression for $\langle\CJ\CJ\CJ\CJ\rangle$ is however available upon request.} There are three orthogonal frames which give a nonzero answer: \texttt{-++-}, \texttt{-+-+} and \texttt{-{-}++}. Of course there are also the conjugated versions of those, but they are redundant. In the current case the correlator can be written as
\eqn{
\langle \CJ^{I_1}\CJ^{I_2}\CJ^{I_3}\CJ^{I_4}\rangle = \CK_{(3,1)(3,1)(3,1)(3,1)}\big(H_s \, e_1 + H_t \, e_2 + H_u \, e_3 + \cdots\big)\,,
}[JJJJort1]
with $e_i$ being functions of the cross ratios and with the definition
\eqn{
H_s = H_{12}H_{34}\,,\qquad
H_t = H_{14}H_{23}\,,\qquad
H_u = H_{13}H_{24}\,.
}[]
The ellipses represent tensor structures that vanish in any orthogonal frame. Going to the three frames shown above allows us to solve for all the $e_i$. Note that in the \texttt{-++-} frame only $H_t$ vanishes, while in \texttt{-+-+} $H_u$ vanishes and finally in \texttt{-{-}++} $H_s$ vanishes. The specific linear combinations appearing in each \textttpm~basis are
\eqna{
\CE_1(u,v)&= e_1+\frac{1}{u}\lsp e_3\qquad &\mbox{from \texttt{-++-},}\\
\CE_2(u,v)&= e_1+{\frac{v}{u}}\lsp e_2\qquad &\mbox{from \texttt{-+-+},}\\
\CE_3(u,v)&=e_2+\frac{1}{{v}}\lsp e_3\qquad &\mbox{from \texttt{-{-}++}.}\\
}[JJJJort2]
Let us now for a moment forget the original correlator and let us reinterpret~\eqref{JJJJort1}, without the ellipses, as a new four-point function of identical scalar operators of conformal dimension~3. This is a consistent truncation of the crossing equations,\footnote{Notice that one can argue that the tensor structures appearing represent a complete set just by noticing that they are the same ones appearing in generalized free theory.} so we can study how this new correlator behaves under $1\leftrightarrow 3$ exchange and write the resulting crossing equations for the $\CE_i$'s. With the normalization in~\eqref{JJJJort2}, we find
\eqna{
\CE_1(u,v)&=\left(\frac{u}{v}\right)^3(- \CE_3(v, u)|_{\mathtt{c}_s \leftrightarrow \mathtt{c}_t})\,,\\
\CE_2(u,v)&=\left(\frac{u}{v}\right)^3(- \CE_2(v, u)|_{\mathtt{c}_s \leftrightarrow \mathtt{c}_t})\,.
}[preCrossEq]
Notice that the minus sign comes from the fact that, as can be seen from~\eqref{css}, under the $1\leftrightarrow 3$ exchange, $\mathtt{c}_s \rightarrow -\mathtt{c}_t$,  and vice-versa,   $\mathtt{c}_u \rightarrow -\mathtt{c}_u$. For any given choice of flavor group, one can compute the crossing matrix of the $\mathtt{c}_{s,t,u}$ and derive a system of crossing equations for the $\CE_i$'s from~\eqref{preCrossEq}.

The situation is different in the stress tensor case. The correlator can be written as a combination of six structures now, modulo vanishing ones
\eqn{
\langle \CT\CT\CT\CT\rangle = \CK_{(4,2)\cdots(4,2)}\big(H_s^2 \, f_1 + H_t^2 \, f_2 + H_u^2 \, f_3
+ H_s H_t\, f_4 + H_sH_u\, f_5 + H_tH_u\,f_6
 + \cdots\big)\,.
}[TTTTort]
Now we can only compute three independent combination of functions $f_i$, so not all of them. In particular we can only compute
\eqna{
\CF_1&= f_1 + u^{-2}f_3 + u^{-1}f_5\qquad &\mbox{from \texttt{-++-},}\\
\CF_2&=f_1 + v^2u^{-2}f_2 + vu^{-1}f_4\qquad &\mbox{from \texttt{-+-+},}\\
\CF_3&=v^2u^{-2}f_2 +u^{-2} f_3 + vu^{-2}f_6\qquad &\mbox{from \texttt{-{-}++}.}\\
}[CFs]
These functions satisfy the following crossing equations
\eqna{
\CF_3(u,v)=\left(\frac{u}{v}\right)^4 \CF_1(v,u)\, , \qquad \CF_2(u,v)= \left(\frac{u}{v}\right)^4\CF_2(v,u) =\CF_1\left(\frac{u}{v}, \frac{1}{v}\right)\, ,
}[crossT]
where we have used the fact that
\eqna{
\CHtgr(u,v)=\left(\frac{u}{v}\right)^2 \CHtgr(v,u)=\frac{1}{v^2}\CHtgr\mleft(\frac{u}{v}, \frac{1}{v}\mright)\, .
}[]

\subsection{A brief comment on the interpretation of the orthogonal frame}
In the previous section we took the correlator in a convenient frame in order to obtain simpler and more manageable expressions.  However, a closer look reveals a deeper physical interpretation of this configuration.  
Choosing a specific frame breaks part of the conformal group, but we can still consider the part which is left unbroken and try to reinterpret the $\textttpm$ polarizations we have introduced before. In particular, the choice of a plane for $x_i$ and its orthogonal complement for $h_i$ breaks the conformal group into $\SO(2,4) \,\to\, \rmU(1) \times \SO(2,2)$. We choose the plane $[03]$ for the coordinates and the plane $[12]$ for the spin polarizations
\twoseqn{
h_j^{\textttpm,\lsp\mu} &= (0,\;h_j^{\textttpm},\; \pm i\lsp h_j^{\textttpm},\;0)\lsp^\mu\,,
}[]{
x_j^\mu &= \left(\frac{z_j-\zb_j}2,\;0,\;0,\;\frac{z_j+\zb_j}2\right)^\mu\,.
}[][]
The $\rmU(1)$ factor is a global symmetry from the point of view of the two-dimensional theory on the plane. The coordinates are indeed left unchanged and the spin polarizations rotate with a sign that depends on whether we choose the \texttt{+} or the \texttt{-} frame. Vice versa, conformal transformations on the $[03]$-plane have no effect on the polarizations and  act on $z$ and $\zb$ as an holomorphic and an anti-holomorphic transformation respectively. To summarize, a rotation of an angle $\varphi$ in the $[12]$-plane and of an angle $\psi$ in the $[03]$-plane have the following effect
\eqna{
\eta^{\mathbf{1}} &\to e^{\frac12(i\varphi-\psi)}\,\eta^{\mathbf{1}}\,,\qquad &
\eta^{\mathbf{2}} &\to e^{\frac12(-i\varphi+\psi)}\,\eta^{\mathbf{2}}\,,\\
\etab^{\mathbf{1}} &\to e^{\frac12(-i\varphi-\psi)}\,\etab^{\mathbf{1}}\,,\qquad &
\etab^{\mathbf{2}} &\to e^{\frac12(i\varphi+\psi)}\,\etab^{\mathbf{2}}\,,\\
h^{\texttt{+}} &\to e^{-i\varphi}\,h^{\texttt{+}}\,, &
h^{\texttt{-}} &\to e^{i\varphi}\,h^{\texttt{-}}\,, \\
z &\to e^{\psi}\,z\,, &
\zb &\to e^{-\psi}\,\zb\,.
}[]

Consider now an operator $V_{\mu_1\cdots\mu_s}$ in four dimensions. The cases we are interested in are $s=1$ and $s=2$, but for generality we keep $s$ arbitrary. When we go to an orthogonal frame we contract all indices of $V$ either with $h^{\texttt{+},\lsp\mu}$ or with $h^{\texttt{-},\lsp\mu}$. Let us denote then
\eqn{
V^{\pm}(z,\zb) \equiv h^{\textttpm,\lsp\mu_1}\cdots h^{\textttpm,\lsp\mu_s}\,V_{\mu_1\cdots\mu_s}\,.
}[]
In the two-dimensional theory this is a scalar operator with global $\rmU(1)$ charge $\mp s$. From this interpretation it is obvious that there are only three possible frames: they correspond to the correlators of $V^\pm$ which are nonzero by charge conservation and these are exactly the ones we have seen before. The crossing equations in this orthogonal frame then reduce to those of a charged scalar operator of dimension $s+2$ in two dimensions.

An operator with dimension $\Delta$ and spin $(j,\jb)$ will contribute as a sum of operators with dimension $\Delta+n$ and different spins, as explained in detail in~\cite{Hogervorst:2016hal}. Let us denote the exchanged operator as
\eqn{
\CO^{(j,\jb)}_{\alpha_1\cdots\alpha_j,\lsp \alphad_1\cdots\alphad_\jb}(x)\,.
}[]
When we contract the indices with $\eta$ and $\etab$ we can put some indices along the plane and some along the orthogonal direction. The ones on the plane will contribute to the two-dimensional spin while the others will contribute to the global $\rmU(1)$ charge, $q$. Furthermore we can have some derivatives along the orthogonal direction which will still make the operator a primary in two dimensions. A generic component will look like
\eqn{
(\partial_1-i\partial_2)^m(\partial_1+i\partial_2)^n\CO^{(j,\jb)}_{\mathbf{1\ldots12\ldots2},\lsp\mathbf{1\ldots12\ldots2}}\,(\eta^{\mathbf{1}})^{a}(\eta^{\mathbf{2}})^{j-a}(\etab^{\mathbf{1}})^{b}(\etab^{\mathbf{2}})^{\jb-b}\,.
}[]
It then has the following charges under the parallel and orthogonal rotations
\eqna{
\rmU(1)\colon&\qquad& q &=\mathrlap{\frac{\jb-j}{2}+a-b+n-m\,,}\\
\SO(2,2)\colon&\qquad& \Delta &= \Delta+n+m\,,&\qquad l &=\frac{j+\jb}{2}-a-b\,.}[]
Each of this components is exchanged through the usual Dolan-Osborn conformal block~\cite{Dolan:2003hv}
\eqn{
g_{\Delta,l} = z^{\frac{\Delta-l}{2}}\zb^{\frac{\Delta+l}{2}}\,{}_2F_1\mleft(\tfrac{\Delta-l}2,\tfrac{\Delta-l}2;\Delta-l,z\mright)\,
\,{}_2F_1\mleft(\tfrac{\Delta+l}2,\tfrac{\Delta+l}2;\Delta+l,\zb\mright)+(z\leftrightarrow\zb)\,.
}[DOBlocks]

\section{AdS amplitudes}\label{Sec:holographic}
\subsection{Mellin space representation}
In the orthogonal configuration, all the correlators described so far effectively appear as scalar four-point functions.  For scalar correlators an alternative to position space representation is the Mellin space formalism~\cite{Mack:2009mi,Penedones:2010ue}, which is particularly illuminating in the holographic limit.  Consider four generic scalar operators of dimension $\Delta_i$. Their correlator can be written as
\eqn{
\langle \CO_{\Delta_1}(x_1)  \CO_{\Delta_2}(x_2)  \CO_{\Delta_3} (x_3) \CO_{\Delta_4}(x_4)\rangle =\CK_{\Delta_1\Delta_2\Delta_3\Delta_4}\CG_{\Delta_1\Delta_2\Delta_3\Delta_4}(u,v)\, .
}[]
The corresponding Mellin amplitude is defined through the integral
\eqna{
\CG_{\Delta_1\Delta_2\Delta_3\Delta_4}(u,v)&=\int_{-i \infty}^{i \infty} \frac{d\Ms d\Mt}{(4 \pi i)^2}u^{\frac{\Ms}{2}}v^{\frac{\Mt-\Delta_2-\Delta_3}{2}}  \CM_{\D{1}\D{2}\D{3}\D{4}}(\Ms, \Mt)\Gamma_{\D{1}\D{2}\D{3}\D{4}}(\Ms, \Mt, \Mu)\,,\\
\Gamma_{\D{1}\D{2}\D{3}\D{4}}(\Ms, \Mt, \Mu)&=\Gamma\mleft(\frac{\D{1}+\D{2}-\Ms}{2} \mright)\,\Gamma\mleft(\frac{\D{3}+\D{4}-\Ms}{2} \mright)\,\Gamma\mleft(\frac{\D{1}+\D{4}-\Mt}{2} \mright)\\
&\quad \, \lsp\lsp \Gamma\mleft(\frac{\D{2}+\D{3}-\Mt}{2} \mright)\,\Gamma\mleft(\frac{\D{1}+\D{3}-\Mu}{2} \mright)\,\Gamma\mleft(\frac{\D{2}+\D{4}-\Mu}{2} \mright)\, ,
}[]
where $\Ms, \, \Mt,\, \Mu$ are the Mellin-Mandelstam variables. They are not independent,  but are constrained to satisfy $\Ms+\Mt+\Mu=\sum_{i=1}^4 \Delta_i$.  The expression above allows us to define the Mellin amplitude $\CM(\Ms, \Mt)$ for a generic scalar correlator.  However, for the four-point function of the scalar supergluons $\CO_2^{I}$ and supergravitons $\CO$, we would like to include the constraints coming from the superconformal Ward Identities as well.  Superconformal symmetry dictates that the four-point function of all supergluons and all supergravitons takes respectively the form in~\eqref{FxLowest} and  in~\eqref{Hgrav}.\footnote{More precisely, we are only Mellin transforming the anomalous part. This is allowed because the protected part does not contribute to the Mellin amplitude \cite{Rastelli:2017udc}.}  This is realized in Mellin space by  considering as our defining amplitude the \textit{reduced} Mellin amplitude~\cite{Rastelli:2016nze,Rastelli:2017udc,Alday:2021odx},  which is nothing but the Mellin transform of $\CHtgl(u,v)$ and $\CHtgr(u,v)$
\twoseqn{
\CHtgl(u,v)&=\int_{-i \infty}^{i \infty} \frac{d\Ms d\Mt}{(4 \pi i)^2}u^{\frac{\Ms}{2}}v^{\frac{\Mt-4}{2}} \overtilde{\CM}_{\mathrm{gl}}(\Ms, \Mt)\Gamma_{2222}(\Ms, \Mt, \tilde{\Mu})  &&\,\lsp\tilde{\Mu}=\Mu-2=6-\Ms-\Mt\, ,
}[]
{\CHtgr(u,v)&=\int_{-i \infty}^{i \infty} \frac{d\Ms d\Mt}{(4 \pi i)^2}u^{\frac{\Ms}{2}}v^{\frac{\Mt-4}{2}} \overtilde{\CM}_{\mathrm{gr}}(\Ms, \Mt)\Gamma_{2222}(\Ms, \Mt, \tilde{\Mu})  && \,\lsp \tilde{\Mu}=\Mu-4=4-\Ms-\Mt\, .
}[][]
The expressions for the supergluon and supergraviton reduced Mellin amplitude are respectively~\cite{Alday:2021odx, Rastelli:2016nze}
\twoseqn{
&\begin{aligned}
\overtilde{\CM}_{\mathrm{gl}}(\Ms, \Mt)&=\frac{4\lsp \mathtt{c}_s}{3(\Ms-2)}\left( \frac{1}{\Mt-2}-\frac{1}{\tilde{\Mu}-2}\right)+\frac{4\lsp \mathtt{c}_t}{3(\Mt-2)}\left( \frac{1}{\tilde{\Mu}-2}-\frac{1}{{\Ms}-2}\right)\\
&\quad\, +\frac{4\lsp \mathtt{c}_u}{3(\tilde{\Mu}-2)}\left( \frac{1}{\Ms-2}-\frac{1}{\Mt-2}\right)\,,\end{aligned}
}[]
{&\overtilde{\CM}_{\mathrm{gr}}(\Ms, \Mt)=\frac{8}{(\Ms-2)(\Mt-2)(\tilde{\Mu}-2)}\, .
}[][]

On the other hand, we have shown that in the orthogonal frame, the four-point function with two or four insertions are simply given in terms of polynomials in $u$ and $v$ multiplying derivatives of the lowest-component correlator $\CHtgl(u,v)$ and  $\CHtgr(u,v)$.  This action is easily realized in Mellin space where any differential operator becomes a difference operator.\footnote{In particular, the Mellin amplitudes of the bottom components are related to the reduced Mellin amplitudes via the action of difference operators \cite{Rastelli:2016nze,Rastelli:2017udc,Alday:2021odx}.}  Consider a generic function $f(\Ms, \Mt)$, then
\eqna{
u \partial_u &: f(\Ms, \Mt) \to \frac{\Ms}{2}\times f(\Ms, \Mt)\, ,\\
v \partial_v &: f(\Ms, \Mt) \to \frac{\Mt-4}{2}\times f(\Ms, \Mt)\, ,\\
u^m v^n &: f(\Ms, \Mt) \to \frac{\left(f(\Ms, \Mt) \Gamma_{2222}(\Ms,\Mt, \tilde{\Mu})\right)_{\Ms\to \Ms-2m, \Mt\to \Mt-2n+4-\D{2}-\D{3}}}{\Gamma_{\D{1}\D{2}\D{3}\D{4}}(\Ms,\Mt, \Mu)}\, ,
}[]  
where $\tilde{\Mu}=-\Ms-\Mt+6$ or $\tilde{\Mu}=-\Ms-\Mt+4$ depending whether we are acting on $\overtilde{\CM}_{\mathrm{gl}}$ or~$\overtilde{\CM}_{\mathrm{gr}}$. 
Through the action of this difference operator\footnote{We provide the expressions for all the difference operators in an ancillary Mathematica file.} we are able to determine the Mellin amplitude for all the correlators we have described before: namely the ones with two or four current/tensor insertions in the orthogonal frame and also $\CM_{\CW\overbar{\CW}\CW\overbar{\CW}}$ and $\CM_{\CL\overbar{\CL}\CL\overbar{\CL}}$ which are naturally scalars.  From these results, we can then go back to position space by guessing an expression in terms of $\Db$-function through~\eqref{DbMellin}. We expect that the correlators can be written as a sum of $\Db$-functions because, if we were to perform an honest diagrammatic calculation, all the exchange Witten diagram have the correct quantum numbers to truncate into a finite sum of contact diagrams~\cite{DHoker:1999mqo}. It is important to notice that the position space result obtained as such correspond to the \textit{full} correlator, namely it consists of both the anomalous part and the protected one. In the following, we will show explicitly that if we compare the results from the inverse Mellin transform with the one obtained by the direct action of the differential operator on $\CHtgl$ and $\CHtgr$, their difference is exactly the connected component of the protected part in~\eqref{GratGl} and~\eqref{GfreeGr}.

\subsection{One spinning insertion}

Let us start by briefly discussing the correlators with just one current or stress-tensor insertion in the supergravity limit.  These examples fall outside the above discussion of scalar Mellin amplitudes since they vanish in the orthogonal configuration. It is nonetheless possible to describe them in Mellin space following~\cite{Goncalves:2014rfa}. More precisely, in~\cite{Goncalves:2014rfa} they adopt the Mellin formalism to describe a correlator of scalars plus a spin-$J$ insertion. We are not reviewing this method here since all the details for $\langle \CT \CO \CO \CO \rangle$ can already be found in~\cite[Appendix C]{Goncalves:2019znr} and the insertion of one $\CJ$ is very similar.  We instead just report for completeness the position space results for the \textit{full} correlators. Following the notation in~\eqref{JOOOE} and~\eqref{TOOOE}, for $\langle \CJ \CO_2\CO_2\CO_2 \rangle$ we find
\eqna{
a_1^{\mathrm{full}}(u,v)&=-2 \sqrt{2}u^2 v(\mathtt{c}_s \Db_{3221}+\mathtt{c}_t \Db_{2231}-\mathtt{c}_u \Db_{3131})\, ,\\
a_2^{\mathrm{full}}(u,v)&=-2 \sqrt{2}u^2 \left(\mathtt{c}_s\Db_{3212}-\mathtt{c}_t \frac{1}{v}\Db_{3113}+\mathtt{c}_u \Db_{3122}\right)\, 
}[]
with the rational part 
\eqna{
a_1^{\mathrm{full}}-a_1^{\mathrm{anom}}&=-\frac{2 \sqrt{2} u}{3}(v(\mathtt{c}_s-\mathtt{c}_u)+u (\mathtt{c}_t-\mathtt{c}_u))\, ,\\
a_2^{\mathrm{full}}-a_2^{\mathrm{anom}}&=\frac{2 \sqrt{2} u}{3 v}(-\mathtt{c}_s+\mathtt{c}_t+u(\mathtt{c}_t-\mathtt{c}_u))\, .
}[]
By substituting $V_{1,23}= v^{-1}(V_{1,24}-u \lsp V_{1, 34})$, in (C.29) of~\cite{Goncalves:2019znr}, for the stress-tensor insertion we find
\eqna{
b_1^{\mathrm{full}}(u,v)&=32 u^2 v \left(\bar{{D}}_{4132}+v
   \left(\bar{{D}}_{4141}+\bar{{D}}_{4231}+\bar{{D}}_{4242}\right)\right)\, ,\\
   b_2^{\mathrm{full}}(u,v)& = -64 u^2 (\bar{{D}}_{4132}+u \bar{{D}}_{4222}+v
   \left(\bar{{D}}_{4141}+\bar{{D}}_{4242}+(1-2 u) \bar{{D}}_{4231}-2 u
   \bar{{D}}_{4321}\right)
   \\& \quad \, +u v \bar{{D}}_{4332})\, , \\
   b_3^{\mathrm{full}}(u,v)&=v^2\lsp  b_1^{\mathrm{full}} \left( \frac{u}{v}, \frac{1}{v}\right)\, , 
}[]
with
\eqna{
b_1^{\mathrm{full}}-b_1^{\mathrm{anom}}&=32 u v (u+v) \, , \\ 
b_2^{\mathrm{full}}-b_2^{\mathrm{anom}}&=-64 u^2 \, .
}[]
Notice that both these results are consistent with (and can be extracted from) the factorization of the five-point function  of supergluons in~\cite{Alday:2022lkk} and supergravitons in~\cite{Goncalves:2019znr} respectively.

\subsection{Two spinning insertions}
Let us now consider the case with two spinning operators which are inserted at $x_1$ and $x_2$. In the orthogonal frame, only the terms in~\eqref{JJOOOr} and~\eqref{TTOOOr} proportional to $H_{12}$ survive.  By using the Schouten identities, this tensor can be rewritten in terms of $h_i$ in~\eqref{hDef}
\eqn{
H_{ij}=- \eta_i \rmx_{ij}\etab_i\; \eta_j \rmx_{ji} \etab_j+x_{ij}^2 \lsp \eta_i\eta_j \; \etab_i \etab_j \xrightarrow[\mathrm{frame}]{\mathrm{orthogonal}} \frac{1}{2} x_{ij}^2 \; h_i \cdot h_j\, .
}[]
For two current insertions, the four point function in the orthogonal frame can be written as
\eqna{
\langle \CJ \CJ \CO_2 \CO_2 \rangle_{\mathrm{ortho}}=\frac{1}{2}h_1\cdot h_2 \lsp y_{34}^2 \lsp \CK_{3322} \lsp \CG_{\CJ\CJ\CO_2\CO_2}(u,v)\,.
}[]
The corresponding Mellin amplitude then reads
\eqn{
\CM_{\CJ\CJ\CO_2\CO_2}(\Ms, \Mt)=\mathtt{c}_s \frac{10(\Mt-\Mu)}{\Ms-2}-\mathtt{c}_t\frac{4}{\Mt-3}+\mathtt{c}_u \frac{4}{\Mu-3}-10(\mathtt{c}_t-\mathtt{c}_u)\, .
}[]
The interpretation of these terms is quite clear. The coefficient of $\mathtt{c}_s$ is from the exchange of the spin-1 gluon field. The singular parts of $\mathtt{c}_t$ and $\mathtt{c}_u$ are from exchanging the scalar supergluon.  Yet,  their effective dimensions are shifted because of the external spinning operators. The last two terms represent additional contact interactions.  Now let us transform this Mellin amplitude back into position space,\footnote{It is important to note that the translation into position space is in principle not unique due to the existence of $\bar{D}$-function identities which add up to rational terms of the cross ratios. In Mellin space, the rational terms are invisible and therefore gives rise to ambiguities. One can avoid the ambiguities by writing down an ansatz which includes only the $\bar{D}$-functions expected from Witten diagram calculations.} 
\eqna{
\CG_{\CJ\CJ\CO_2\CO_2}^{\mathrm{full}}=4 u^3\left(5\mathtt{c}_s \left(\Db_{4321}-\Db_{4312}\right)-\mathtt{c}_t \Db_{2321}+\mathtt{c}_u \Db_{3221}+5 (\mathtt{c}_t-\mathtt{c}_u) \Db_{3322} \right)\, ,
}[]
where we have introduced the superscript ``full'' to distinguish between the correlator obtained from the Mellin amplitude and the correlator $\CG^{\mathrm{anom}}$ obtained by the direct action of the differential operator on $\tilde{\CF}$ as in~\eqref{JJOOOr}. If we compare the two, we find the two quantities differ by a rational function of the cross ratios
\eqna{
\CG_{\CJ\CJ\CO_2\CO_2}^{\mathrm{full}}-\CG_{\CJ\CJ\CO_2\CO_2}^{\mathrm{anom}}=\frac{4}{3}u \left(\mathtt{c}_s \left(\frac{1}{v}-1\right)+\frac{\mathtt{c}_t}{v}+\mathtt{c}_u\right)\, .
}[]
As mentioned earlier, this term should be identified with the protected part of the correlator which we can also compute independently by acting on the protected part of the super primary correlator with the superconformal differential operators. We find that we arrive at the same answer
\eqna{
\CD_{1}\CD_{2} \left(g_{12}^2 g_{34}^2 \mathbb{G}_{\mathrm{rational}}^{I_1I_2I_3I_4}\right)\xrightarrow[\mathrm{frame}]{\mathrm{ortho}} \frac{1}{2} x_{12}^2 \; h_1 \cdot h_2\lsp \CK_{3322}\frac{4}{3}u \left(\mathtt{c}_s \left(\frac{1}{v}-1\right)+\frac{\mathtt{c}_t}{v}+\mathtt{c}_u\right)\, 
}[]
where $\CD_i$ is defined in~\eqref{diffopDef}.

For two stress tensor insertions we have
\eqna{
\langle \CT \CT \CO \CO \rangle_{\mathrm{ortho}}=\frac{1}{4}(h_1\cdot h_2)^2\lsp t_{34}^2 \lsp \CK_{4422} \lsp \CG_{\CT\CT\CO\CO}(u,v)\, ,
}[]
with the corresponding Mellin amplitude
\eqna{
\CM_{\CT\CT\CO\CO}=
2^8  \left(\frac{23}{\Ms-2}(3\lsp \Mt\lsp \Mu-76)-\frac{2}{\Mt-4}-\frac{2}{\Mu-4}+250\right)\, .
}[]
The form of the amplitude also agrees with the expectation: the singular part in $\Ms$ corresponds to an exchange diagram of a spin-2  graviton, while the poles in $\Mt$ and $\Mu$ come from the exchange of the supergraviton, with an effective dimension shifted by 2.  From this, we can find the corresponding expression in position space
\eqna{
\CG_{\CT\CT\CO\CO}^{\mathrm{full}}=2^8 u^4 \left(\Db_{3412}+\Db_{3421}+46 \Db_{4411}-164 \Db_{4422}-138
   \Db_{6422}\right)\, .
}[]
The difference with the anomalous part of the correlator gives the following protected part
\eqna{
  \CG_{\CT\CT\CO\CO}^{\mathrm{full}}- \CG_{\CT\CT\CO\CO}^{\mathrm{anom}}=\frac{2^8 u (v+1)}{v}\,.
}[]
\subsection{Four spinning insertions: bosonic YM and gravity}
In this subsection, we consider correlators in the orthogonal frame with four spinning operator insertions. In the holographic limit, they correspond to gluon and graviton scattering amplitudes in AdS. At tree level, these correlators obtained from the supersymmetric theories coincide with those of the bosonic theories. 

For the correlators of four currents,  we have three  components, each one corresponding to a different $e_i$.  Let us define $h_{ij}=h_i\cdot h_i$,
\eqna{
\langle \CJ \CJ \CJ \CJ \rangle_{\mathrm{ortho}}=\frac{ \CK_{3333} }{4}\left(h_{12}h_{34} e_1+h_{14}h_{23} \frac{v}{u}e_2 +h_{13}h_{24} \frac{1}{u}e_3 \right)\, ,
}[]
In Mellin space, we find
\eqna{
\CM_\CJ^{(1)}=
2^6 \left( \mathtt{c}_s\frac{25
   (\Mt-\Mu)}{2(\Ms-2)}+\mathtt{c}_s\frac{4 (\Mt-\Mu)}{ (s-4)}- \mathtt{c}_t\frac{ (\Ms+21)}{(\Mt-4)}+ \mathtt{c}_u\frac{ (\Ms+21) }{ (\Mu-4)}-33 (\mathtt{c}_t-\mathtt{c}_u)\right)\, ,
}
\eqna{
\CM_\CJ^{(2)}=-\CM_\CJ^{(1)}\Big|_{\substack{ \,\Ms\lsp\leftrightarrow \lsp\Mt,\\  \mathtt{c}_s\leftrightarrow  \lsp\mathtt{c}_t}}\, , \qquad \CM_\CJ^{(3)}=-\CM_\CJ^{(1)}\Big|_{\substack{ \,\Ms\lsp \leftrightarrow\lsp\Mu,\\  \mathtt{c}_s\leftrightarrow \lsp \mathtt{c}_u}}
}[]
Transforming back into position space, we find the full correlator of the first component is given by
\eqna{
 e_1^{\mathrm{full}}&=
2^5 u^3 \big\lbrace\mathtt{c}_s \left(25( \Db_{4312}- \Db_{3412})+33
 (  \Db_{4323}- \Db_{3423})\right)\\&\;\quad+\mathtt{c}_t (-39 \Db_{2332}
+31 \Db_{2343}+33
   \Db_{3342})\\&\;\quad +\mathtt{c}_u \left(39 \Db_{2323}-31 \Db_{3243}-33 \Db_{3324}\right)\lnsp\big\rbrace\;,
}[]
and the protected part is 
\eqna{
e_1^{\mathrm{full}}- e_1^{\mathrm{anom}}&=\frac{2^5 u}{3v^2} \big\lbrace  \mathtt{c}_s 
 (v-1)\left(-4 u (v+1)+4 v^2+v+4\right)\\&  \;\quad +\mathtt{c}_t \left(4 u \left(u^2-u (v+1)-1\right)-3
   v+4\right)\\&\;\quad + \mathtt{c}_u \left(-4 u^3+4 u^2 (v+1)+4 u v^2+(3-4 v) v^2\right)\lnsp\big\rbrace\, .
}[]
For completeness, we report also the expressions corresponding to $e_2$ and $e_3$
\twoseqn{&\begin{aligned}
e_2^{\mathrm{full}}&=2^5 \frac{u^4}{v} \big\lbrace\mathtt{c}_s \left(39 \Db_{3322}-31 \Db_{4323}-33
   \Db_{4332}\right)\\&\quad\;+\mathtt{c}_t (25( \Db_{1432}- \Db_{1342})
 +33 (\Db_{2433}  - \Db_{2343}))\\&\quad\;+\mathtt{c}_u \left(-39 \Db_{3232}+33
   \Db_{3243}+31 \Db_{4233}\right)\lnsp\big\rbrace\, ,\end{aligned}}[]{
&\begin{aligned}
   e_3^{\mathrm{full}}&=
2^5 u^4 \big\lbrace\mathtt{c}_s \left(-39 \Db_{3322}+31 \Db_{3423}+33
   \Db_{4323}\right)\\&\quad\;+\mathtt{c}_t (39 \Db_{2332}-33 \Db_{2343}-31
   \Db_{2433})\\&\quad\;+\mathtt{c}_u \left(25 (\Db_{3142}-
   \Db_{4132})+33( \Db_{3243}- \Db_{4233})\right)\lnsp\big\rbrace\,,\end{aligned}
}[]
which are related to $ e_1^{\mathrm{full}}$ by crossing symmetry.

For the correlators of four stress tensors we will consider the Mellin transform $\CM_{\CT}^{(i)}$of the three independent component $\CF_i$ in~\eqref{CFs}, which are effectively correlators of scalars of dimension four. The result reads
\eqna{
\CM_{\CT}^{(1)}&=
3\cdot 2^{13}\Bigg\lbrace \Bigg(-\frac{529 \left(3 \Mt^2-42 t+148\right)}{\Ms-2}-\frac{
   \left(1731 \Mt^2-23676 t+77896\right)}{\Ms-4}-\\ &\;\quad \frac{ \left(183 \Mt^2-2526 \Mt+8254\right)}{\Ms-6}-\frac{3 (\Ms (75 \Ms+72 \Mu-1238)+2520)}{4(\Mt-6)}\Bigg)+ \Ms\leftrightarrow \Mu \Bigg\rbrace\,.
}[]
The other two components of the Mellin amplitude are related to this one through crossing
\eqn{
\CM_\CT^{(2)}=\CM_\CT^{(1)}\Big|_{ \Mt\leftrightarrow \Mu}\, , \qquad \CM_\CT^{(3)}=\CM_\CT^{(1)}\Big|_{ \Ms\leftrightarrow \Mt}\, .
}[]
In position space, this becomes
\eqna{
\CF_1^{\mathrm{full}}&=3 \cdot 2^{16} u^4 (48 \Db_{3443}+4232 \Db_{4141}-6 (3 \Db_{3555}+968
   \Db_{4253}+1200 \Db_{4354}\\&\,\quad
   +2116 \Db_{5153}+3270 \Db_{5254}+2334
   \Db_{5355})+3 (5040 \Db_{4242}+5646 \Db_{4343}\\&\,\quad
   -4768
   \Db_{4444}-9261 \Db_{5445})+4232 \left(\Db_{4411}-3
   \Db_{5513}\right)+12 (1260 \Db_{4422}\\&\,\quad 
   -484 \Db_{4523}-1635 \Db_{5524})+18 \left(941 \Db_{4433}-400 \Db_{4534}-778
   \Db_{5535}\right))\;,
}[]
and the protected part reads
\eqna{
\CF_1^{\mathrm{full}}-\CF_1^{\mathrm{anom}}&=3\cdot 2^{18}u\Bigg(\frac{ \left(u^3+1\right)}{2v}+\frac{11}{4} \left(37( u^2+1)-2 u\right)-\frac{837}{4} (u+1) v+108 v^2\Bigg)\, .
}[]
The other $\CF_i$ can be easily obtained using their crossing properties as in~\eqref{crossT}.
\subsection{Correlators of the top components}
Finally we study the four-point function  of the top components introduced in Sec.~\ref{Sec:topGluon} and~\ref{Sec:topGraviton}. These are scalars by themselves, so there is no need to go to the orthogonal frame and we can perform directly a Mellin transform. 
Let us start from the gluon case. Consider $
\CG_\CW$ as in~\eqref{WWb}, the corresponding Mellin amplitude reads
\eqna{
\CM_\CW=\mathtt{c}_s\left(\frac{2 (\Mt-\Mu)}{\Ms-2}+\frac{ (\Mt-\Mu)}{\Ms-4}\right)+\mathtt{c}_t\left(\frac{2
   (\Mu-\Ms)}{\Mt-2}+\frac{(\Mu-\Ms) }{\Mt-4}\right)+3 (\mathtt{c}_s-\mathtt{c}_t)\, ,
}[]
where the poles correspond to the $\Ms$- and $\Mt$-channel exchange of the gluon field and we have additional contact terms. The fact that $\mathtt{c}_u$ does not appear is due to the nontrivial and opposite U(1)$_r$ charge carried by $\CW$ and $\overbar{\CW}$ which allows exchange Witten diagrams in the neutral channels.  The position space expression in terms of $\Db$ function reads
\eqna{
\CG_{\CW}^{\mathrm{full}}&=
-2 u^3 \mathtt{c}_s \left(\Db_{3311}-4 \Db_{3322}-2
   \Db_{4312}+3 \Db_{4332}\right)\\
&\;\quad +2 u^3 \mathtt{c}_t \left(\Db_{1331}-4 \Db_{2332}-2 \Db_{1342}+3
   \Db_{3342}\right)\, ,
}[]
with the protected connected part given by
\eqna{
\CG_{\CW}^{\mathrm{full}}-\CG_{\CW}^{\mathrm{anom}}=\frac{u (u+v-1) (\mathtt{c}_s-\mathtt{c}_t)}{3 v^2}\, .
}[Wrational]
For the correlator of four Lagrangians in~\eqref{LLb}, the Mellin amplitude is\footnote{Similar expressions can also be found in~\cite{Goncalves:2014ffa}.}
\eqna{
\CM_\CL&=\Bigg(\!-\frac{12 \left(3 \Mt^2-42 \Mt+3 \Mu^2-42
   \Mu+296\right)}{\Ms-2}-\frac{48 \left(\Mt^2-12 \Mt+\Mu^2-12 \Mu+72\right)}{\Ms-4}\\&\;\quad -\frac{6 \left(\Mt^2-10 \Mt+\Mu^2-10 \Mu+48\right)}{\Ms-6}-90 (3 \Ms-16)\Bigg)+ \left(\Ms\leftrightarrow \Mt\right)\,.
}[]
Similar to the previous case,  the poles represent the $\Ms$- and $\Mt$-channel exchange Witten diagrams of gravitons and there are additional contact terms.  In position space, we find
\eqna{
\CG_{\CL}^{\mathrm{full}}&=24 u^4 \lbrace (4 \Db_{1441}-24 \Db_{1452}+12 \Db_{1463}+12 \Db_{2442}-48
   \Db_{2453}+20 \Db_{2464}\\&\;\quad +12 \Db_{3443}-40 \Db_{3454}+15
   \Db_{3465}-60 \Db_{4444}+45 \Db_{4455})+(\Delta_1\leftrightarrow \Delta_3)\rbrace\,,
}[]
where by $(\Delta_1\leftrightarrow \Delta_3)$ we simply  mean to take the expression among parentheses and send $\Db_{\D{1}\D{2}\D{3}\D{4}}\to\Db_{\D{3}\D{2}\D{1}\D{4}}$. In this way the expected $u \leftrightarrow v$ symmetry is manifest.
Finally, by comparing this expression with the one obtained by the direct action of $\Delta^{(8)}$ operator on $\CHtgr$, we recover the rational part 
\eqna{
\CG_{\CL}^{\mathrm{full}}-\CG_{\CL}^{\mathrm{anom}}=\frac{16 u \left(u^2+u (v-2)+(v-1)^2\right)}{v^3}\, .
}[Lrational]
It is instructive to compare our result to the ones known in the literature for the axio-dilaton currents in~\cite{Eden:2000bk, DHoker:1999pj,Drummond:2006by}. The axion $\CO_C$ and the dilaton $\CO_\phi$ are obtained as a  linear combination of the self-dual and anti-self dual SYM Lagrangian: $\CO_\phi \sim \CL+\bar{\CL}$ and $\CO_C \sim i(\CL-\bar{\CL})$. By using the fact that the four-point functions with only one $\CL$ or $\bar{\CL}$ vanish, it is straightforward to get
\eqna{
\langle \CO_\phi\CO_C\CO_\phi\CO_C  \rangle=-\frac{\CK_{4444}}{\pi^6} \left(\CG_{\CL}^{\mathrm{full}}(u,v)-u^4 \CG_{\CL}^{\mathrm{full}}\left(\frac{1}{u}, \frac{v}{u}\right)-\CG_{\CL}^{\mathrm{full}}\left(\frac{u}{v}, \frac{1}{v}\right)\right)\, .
}[]
Notice that in order to reproduce the correct result,  we have to consider the full correlator.  Then according to the analysis in~\cite{Drummond:2006by}, it should  be possible to write not only the anomalous part as the action of $\Delta^{(8)}$ on a scalar function of $u$ and $v$, as in~\eqref{LLb},  but  the protected part as well.  Indeed we find that we can rewrite~\eqref{Lrational} as
\eqna{
\CG_{\CL}^{\mathrm{full}}-\CG_{\CL}^{\mathrm{anom}} \equiv \CG_{\CL}^{\mathrm{rational}}=\Delta^{(8)} \left(\frac{u}{v} \right)\, .
}[]
Quite remarkably we find that the same seed function correctly reproduces the rational part also of the $\CN=2$ superpotential correlator in~\eqref{Wrational}
\eqna{
\CG_{\CW}^{\mathrm{full}}-\CG_{\CW}^{\mathrm{anom}} \equiv \CG_{\CW}^{\mathrm{rational}}= \frac{\mathtt{c}_t-\mathtt{c}_s}{3}\Delta^{(4)}\left( \frac{u}{v} \right)\, .
}[]
Unfortunately the same property does not apply to the other spinning components. 

\section{Towards double copy relations in position space}\label{Sec:doublecopy}

In this section we present preliminary results on the AdS generalization of the double copy relation \cite{Bern:2010ue}. In flat space, the double copy relation provides an interesting perspective on the nature of gravity, and gives an efficient way to construct gravitational amplitudes from the amplitudes in non-gravitational gauge theories (see \cite{Bern:2019prr} for a pedagogical review). As already mentioned in the introduction, a Mellin space double copy relation has been found in \cite{Zhou:2021gnu} at tree level for the four-point functions of the lowest components of the supermultiplets. Unfortunately,  Mellin amplitudes are difficult to define for spinning correlators \cite{Goncalves:2014rfa}, and therefore this formalism is not suitable for the discussion of AdS gluon and graviton amplitudes. In the absence of a better representation, in this section we search for such generalizations directly in position space. \footnote{The existence of a double copy for CFT correlators has also been explored in momentum space, see for instance~\cite{Farrow:2018yni,Armstrong:2020woi,Albayrak:2020fyp}.}
Although we did not succeed in identifying a  prescription that precisely gives the gravity correlators, we find a lot of evidence which suggests such a relation should exist. At the same time, it should be noted that where to look for the double copy relation is {\it a priori} unclear as there exist many possibilities. Therefore, it is important to clarify what we have considered. Before we proceed to enumerating the evidence, let us make two brief comments regarding this point and mention some of the subtleties. 

The first comment concerns the extra complexity coming from R-symmetry in supersymmetric theories.  If the double copy structure is extended by  superconformal symmetry to all the component correlators, it has to manifest itself in both the spacetime dependence and the R-symmetry dependence. But since here we are only taking the first step in finding such a relation, we will only focus on the spacetime part and avoid discussing how double copy acts on R-symmetry. We are allowed to do that because we will only look at correlators whose R-symmetry dependence can essentially be ``factored out''. For example, in the correlators of the bottom components the R-symmetry variables only appear in the superconformal factors multiplying the reduced correlators. As another example, in $\langle \mathcal{J}\mathcal{O}\mathcal{O}\mathcal{O} \rangle$ and $\langle \mathcal{T}\mathcal{O}\mathcal{O}\mathcal{O} \rangle$ of the gluon and graviton theories respectively, both $\mathcal{J}$ and $\mathcal{T}$ are R-symmetry neutral and there is a unique R-symmetry structure which appears as a three-point function. 

Another comment is about whether the AdS double copy relation should relate the whole correlators or only parts of them. There are two different perspectives and they lead to different expectations. From the superconformal symmetry perspective, all component correlators naturally split into the sum of a protected part and a dynamical anomalous part. It is possible that double copy exists only for the anomalous part. The reason to suspect such a scenario is that the protected part is a bit too ``simple'' to host such a structure. Especially when the theory admits a marginal coupling, the protected part is just the free theory correlator where the dual theory is highly nonlocal and far from YM theory or gravity. It is this possibility which we will investigate in the section. On the other hand, this splitting is quite artificial for the $\langle \mathcal{J}\mathcal{J}\mathcal{J}\mathcal{J} \rangle$ and $\langle \mathcal{T}\mathcal{T}\mathcal{T}\mathcal{T} \rangle$ correlators as they coincide with the correlators in the bosonic theories. In the bosonic theories no such distinctions exist. Since in flat space double copy holds for the whole amplitude, we might also expect that the AdS generalization is only present at the level of the full correlator as well. We will leave the exploration of this possibility for future work.

\subsection{Bottom components revisited}\label{sec:bottomDC}

Let us begin by writing the supergluon reduced correlators \FxLowest in the following form 
\begin{equation}
\Hgl \propto \mathtt{c}_s\lsp\BBN_sW_s + \mathtt{c}_t\lsp\BBN_tW_t + \mathtt{c}_u\lsp\BBN_sW_u\,.\label{HgluonNW}
\end{equation}
The functions 
\begin{equation}
W_s=x_{12}^{-2}D_{1122}\;,\quad W_t=x_{14}^{-2}D_{1221}\;,\quad W_u=x_{13}^{-2}D_{1212}\;,
\end{equation}
are the exchange Witten diagrams in the $s$-, $t$-, $u$-channels respectively where all external and internal dimensions are 2. As we will see, they correspond to the scalar propagators, $1/s$, {\it etc}., in flat space. The differential operators $\mathbb{N}_{s,t,u}$ are given by
\begin{equation}
\mathbb{N}_s=\mathbb{D}_t-\mathbb{D}_u\;,\quad \mathbb{N}_t=\mathbb{D}_u-\mathbb{D}_s\;,\quad \mathbb{N}_u=\mathbb{D}_s-\mathbb{D}_t\;,
\end{equation}
where
\begin{equation}
\mathbb{D}_{s}=\frac{1}{x_{12}^2}\frac{\partial}{\partial x_{34}^2}\;,\quad \mathbb{D}_{t}=\frac{1}{x_{14}^2}\frac{\partial}{\partial x_{23}^2}\;,\quad \mathbb{D}_{u}=\frac{1}{x_{13}^2}\frac{\partial}{\partial x_{24}^2}\;.
\end{equation} 
They play the role of the kinematic numerators and increase the conformal dimension of each external point by one. In deriving this representation we have used~\eqref{derOnD}.

Clearly, we have the following relation by construction
\begin{equation}
\mathbb{N}_s+\mathbb{N}_t+\mathbb{N}_u=0\;,
\end{equation}
which generalizes the color-kinematic duality in flat space. Note that this way of writing the reduced correlator might seem similar to the differential representation advocated in \cite{Diwakar:2021juk,Herderschee:2022ntr,Cheung:2022pdk}. However, an important difference is that our differential operators act on exchange Witten diagrams, instead of on contact Witten diagrams. If we replace $\mathbb{N}_{s,t,u}$ by $\mathtt{c}_{s,t,u}$ we get the four-point function of bi-adjoint scalars
\begin{equation}
\mathtt{c}_s^{\phantom{\prime}} \mathtt{c}'_s\lsp W_s+\mathtt{c}_t^{\phantom{\prime}} \mathtt{c}'_t \lsp W_t+\mathtt{c}_u^{\phantom{\prime}} \mathtt{c}'_u \lsp W_u\;.
\end{equation}
This agrees precisely with the Mellin space result of \cite{Zhou:2021gnu} where it showed that the zero copy gives conformally coupled bi-adjoint scalars on $\mathrm{AdS}_5\times S^1$.\footnote{For the correlator of the lowest KK mode, the internal space is invisible and the correlator coincide with that of the scalar theory on $\mathrm{AdS}_5$. The latter is a consistent truncation of the $\mathrm{AdS}_5\times S^1$ theory.} On the other hand, if we replace $\mathtt{c}_{s,t,u}$ by $\mathbb{N}_{s,t,u}$, we get a dimension 4 four-point function. Using the identities of $D$-functions in Appendix \ref{App:Dfunctions}, the result can be written as
\begin{equation}\label{DCscalar}
\mathbb{N}^2_s W_s+\mathbb{N}^2_tW_t+ \mathbb{N}^2_uW_u=\frac{9\pi^2N^2}{8}\Hgr+\mathtt{R}\;,
\end{equation}
which is almost the supergraviton reduced correlator except there is a rational term
\begin{equation}
\mathtt{R}=\frac{1}{(x_{12}^2x_{34}^2)^4}\frac{u^2(u+v+uv)}{2v^2}\;.
\end{equation}
Therefore, we find a double copy relation in position space which works up to terms with transcendental degree zero.

\subsection{Double copy structures for spinning correlators}

If we look at the expressions~\JJJJ and~\TTTT together with~\eqref{HgluonNW} and~\eqref{DCscalar} we immediately notice that both the double copy and the zero copy are trivially satisfied, provided that the ``square'' of the operators is defined as
\eqn{
(\diff_1\diff_2\diff_3\diff_4 \lsp \Lambda\lsp \BBN_{s,t,u})^2 \equiv \diff_1^2\diff_2^2\diff_3^2\diff_4^2 \lsp \Lambda^2\lsp \BBN_{s,t,u}^2\,.
}[]
The same would hold for the other insertions, as the reader can verify easily. However, the expression above does not hold as an identity for differential operators --- where the square literally means applying the operator twice and it is only true when the operators commute. Thus, defining the ``square'' in such a way would seem rather artificial. In this subsection we propose a way to make this a bit more precise and also more similar to the flat-space prescription. However, this comes at the expense of having to introduce the inverse of the operator ``$\diff_i$''. Taking an inverse of a differential operator is not necessarily a bad thing, for example $\square^{-1}$ can be easily defined. But we have to take care with the domain of $\diff_i$ and make sure that the space of functions we deal with does not belong to the kernel of $\diff_i$. In this preliminary exploration we will not trouble ourselves with these details and assume that such issues do not arise. Let us now analyze more carefully the cases of one and four insertions.

For the anomalous part of the correlators with one spinning insertion we have the following formulas
\twoseqn{
\langle\CJ^{I_1}\CO_2^{I_2}\CO_2^{I_3}\CO_2^{I_4}\rangle &= \diff_1\,\BBL_{234}^1\, \Hgl\,,
}[gluonOneInsertionDL]{
\langle\CT\CO_2\CO_2\CO_2\rangle &= \diff_1^2\,(\BBL_{234}^1)^2\, \Hgr\,,
}[][]
where the relevant definitions can be found around~\JOOO and~\TOOO. We are also dropping overall numerical factors such as powers of $2$, $C_{2,2,2}^2$ and $N^2$. As a trivial consequence of~(\ref{DCscalar}) we can obtain a double copy relation for the spinning correlator through the following replacement
\eqn{
\mathtt{c}_{s,t,u} \;\longrightarrow\; \diff_1^2\,(\BBL_{234}^1)^2\,\BBN_{s,t,u}\,(\diff_1\BBL_{234}^1)^{-1}\,.
}[DCone]
In performing this replacement we are considering a definite  prescription for the action of $\mathtt{c}_{s,t,u}$. More precisely, we assume to  first take all color factors to the left of every spacetime or polarization dependent quantity and then to perform the formal replacement above. In this way the new numerator acts, from the left, on the gluon correlator giving the  graviton one. The expression~\DCone is only formal for now because we have not given a concrete definition of the (nonlocal) operator $(\diff_1\BBL_{234}^1)^{-1}$. A more careful treatment is needed to make this rigorous and we hope to come back to it in a future work. Note that $\diff_1\BBL_{234}^1$ may be written as
\eqn{
\diff_1\BBL_{234}^1 = \etab_1 \partial_{1} \rmx_{12}\rmx_{23}\rmx_{34}\rmx_{41}\eta_1 - \etab_1 \partial_{1} \rmx_{14}\rmx_{43}\rmx_{32}\rmx_{21}\eta_1\,.
}[]
We can carry out the same exercise for two and four insertions. For brevity, let us show only the case of four insertions
\twoseqn{
\langle\CJ^{I_1}\CJ^{I_2}\CJ^{I_3}\CJ^{I_4}\rangle &= \diff_1\lsp\diff_2\lsp\diff_3\lsp\diff_4\,\Lambda(x, \eta)\, \Hgl\,,
}[gluonFourInsertionsDDDDLam]{
\langle\CT\CT\CT\CT\rangle &= \diff_1^2\lsp\diff_2^2\lsp\diff_3^2\lsp\diff_4^2\,\Lambda(x, \eta)^2\, \Hgr\,,
}[][]
with the definitions in~\JJJJ and~\TTTT. This, formally, leads to the double copy replacement
\eqn{
\mathtt{c}_{s,t,u} \;\longrightarrow\; \diff_1^2\lsp\diff_2^2\lsp\diff_3^2\lsp\diff_4^2\,\Lambda(x, \eta)^2\,\BBN_{s,t,u}\,(\diff_1\lsp\diff_2\lsp\diff_3\lsp\diff_4\,\Lambda(x, \eta))^{-1}\,.
}[DCfour]

Let us now address the case of the zero copy to a bi-adjoint scalar. Recall for the correlator of the bottom component, replacing the numerators $\BBN_{s,t,u}$ with the color factors $\mathtt{c}_{s,t,u}$ yielded the amplitude for bi-adjoint scalars. We can check if the same is true for spinning components. Formally, inverting the relations~\DCone and~\DCfour we have
\twoseqn{
\BBN_{s,t,u} &\;\longrightarrow\; \mathtt{c}'_{s,t,u}\, (\BBL_{234}^1)^{-2}\lsp\diff_1^{-1} \lsp\BBL_{234}^1\,,
}[ZCOne]{
\BBN_{s,t,u} &\;\longrightarrow\; \mathtt{c}'_{s,t,u}\, \Lambda(x,\eta)^{-2}(\diff_1^2\diff_2^2\diff_3^2\diff_4^2)^{-1} \diff_1\diff_2\diff_3\diff_4\, \Lambda(x,\eta)\,.
}[ZCFour][]
Of course, there is no obvious prescription on how to perform this replacement on an arbitrary function. We can however use the representations~\gluonOneInsertionDL and~\gluonFourInsertionsDDDDLam together with~\eqref{HgluonNW}. This leads to
\twoseqn{
\langle\CJ^{I_1}\CO_2^{I_2}\CO_2^{I_3}\CO_2^{I_4}\rangle\big|_{\ZCOne} &= \mathtt{c}^{\phantom{\prime}}_s\,\mathtt{c}_s'\, \diff_1\lsp(\BBL_{234}^1)^{-1} \diff_1^{-1}\lsp\BBL_{234}^1\,W_s + (s\to t,u)\,,
}[]{
\langle\CJ^{I_1}\CJ^{I_2}\CJ^{I_3}\CJ^{I_4}\rangle\big|_{\ZCFour} &= \mathtt{c}^{\phantom{\prime}}_s\,\mathtt{c}_s'\, \diff_1\diff_2\diff_3\diff_4\Lambda^{-1}(\diff_1\diff_2\diff_3\diff_4)^{-1}\Lambda\, W_s+ (s\to t,u)\,.
}[][ZCresult]
Given the result of~\eqref{DCscalar}, we do not expect the outcome to be the bi-adjoint scalar correlator on the nose, but we might have to allow for some rational terms as well, possibly different for each number of insertions. Notice that the operators in~\ZCresult are written in the form of a commutator ``$[\lnsp[a,b]\lnsp] \equiv aba^{-1}b^{-1}$.'' If the operators $\diff_i$ and $\BBL$ or $\Lambda$ commuted, then the zero copy relation would be trivially satisfied. This unfortunately  does not seem to be the case. However, we could content ourselves with a slightly weaker property, namely that the commutator acting on $W_{s,t,u}$ equals the identity up to rational terms
\eqn{
[\lnsp[\diff_1,\lsp\BBL^{-1} ]\lnsp]\, W_{s,t,u} \overset{?}{=} W_{s,t,u} + \mathtt{R}'\,,
}[]
and similarly for the four-insertions case. Unfortunately, we were not able to check this because we did not manage to find an explicit expression for the nonlocal inverse operators $\diff_i^{-1}$. We will leave the verification of this possibility for the future. 

Finally, let us consider the top components, namely the superpotential and the Lagrangian. Comparing the two four-point functions we see
\twoseqn{
\langle\CW\overbar\CW\CW\overbar\CW\rangle &= \frac{1}{(x_{12}^2x_{34}^2)^3} \Delta^{(4)} \, (x_{12}^2x_{34}^2)^2 x_{13}^2x_{24}^2  \, \Hgl\,,
}[]{
\langle\CL\bar\CL\CL\bar\CL\rangle &= \frac{1}{(x_{12}^2x_{34}^2)^4}\,\Delta^{(8)}\, (x_{12}^2x_{34}^2x_{13}^2x_{24}^2)^2\, \Hgr\,,
}[][]
with all relevant definitions given in~\WWb and~\LLb. A double copy relation for these operators readily follows
\eqn{
\mathtt{c}_{s,t,u} \;\longrightarrow\;
\frac{1}{(x_{12}^2x_{34}^2)^4}\,\Delta^{(8)}\, (x_{12}^2x_{34}^2x_{13}^2x_{24}^2)^2
\,\BBN_{s,t,u}\,
\left(
\frac{1}{(x_{12}^2x_{34}^2)^3} \Delta^{(4)} \, (x_{12}^2x_{34}^2)^2 x_{13}^2x_{24}^2
\right)^{-1}\,.
}[DCtop]
Also in this case the differential operator is just a formal expression because we do not know how to write the inverse of $\Delta^{(4)}$ in general. However, the situation here is a bit better because we can diagonalize $\Delta^{(4)}$ by simply expanding $\CHgl$ in the basis of superconformal blocks in the $u$ channel
\eqna{
\Delta^{(4)}\,\CHgl(z,\zb) &=\,\sum_{\Delta,\ell} a_{\Delta,\ell}\,  \Delta^{(4)}\,G_{\Delta,\ell}^u(z,\zb)  \\&=
\frac{1}{16}  \sum_{\Delta,\ell} a_{\Delta,\ell}\,(\Delta -\ell) (\Delta -\ell-2) (\Delta +\ell) (\Delta +\ell+2)\,g^u_{\Delta+2,\ell}(z,\zb)\,,
}[]
where the superconformal blocks in the $u$ channel of $\CH_{2222}^{\mathrm{gluon}}$ and the ordinary conformal blocks in the $u$ channel of $\CG_\CW$ are given by, respectively\footnote{The blocks in the $s$ channel are $u^{-1}g_{\Delta+2,\ell}(z,\zb)$. To obtain those in the $u$ channel we send $z\to1/z$ and multiply by  $u$ coming from the crossing of the kinematic factor. For the correlator of superpotentials the kinematic factor gives a $u^3$ instead.}
\twoseqn{
G_{\Delta,\ell}^u(z,\zb) &= u^2g_{\Delta+2,\ell}\mleft(\frac1z,\frac1\zb\mright)\,,
}[]{
g^u_{\Delta,\ell}(z,\zb) &= u^3 g_{\Delta,\ell}\mleft(\frac1z,\frac1\zb\mright)\,.
}[][]
Therefore, the color-kinematic dual in the basis of conformal blocks is given by\footnote{Note that the factor $(x_{12}^2x_{34}^2)^2 x_{13}^2x_{24}^2$ in front of $\Delta^{(4)}$ in~\DCtop is precisely the one that allows us to go from $\Hgl$ to $\CHgl$.}
\eqna{
\mathtt{c}_{s,t,u}|_{\Delta,\ell\;(u\text{-}\mathrm{chan.})} \;\longrightarrow\;&
\frac{1}{(x_{12}^2x_{34}^2)^4}\,\Delta^{(8)}\, (x_{12}^2x_{34}^2x_{13}^2x_{24}^2)^2
\,\BBN_{s,t,u}\,\\&
\frac{16\, (x_{12}^2x_{34}^2)^3}{(\Delta -\ell) (\Delta -\ell-2) (\Delta +\ell) (\Delta +\ell+2)}\,.
}[]

\section{Future directions}
In this paper, we performed a detailed superspace analysis for four dimensional SCFTs with $\mathcal{N}=2$ and $\mathcal{N}=4$ superconformal symmetry. We presented explicit formulas for how spinning component correlators are related to the super primary correlators and results of these correlators in the holographic limit. Among them, a particularly interesting result is the tree-level four gluon and four graviton scattering amplitudes in bosonic YM and Einstein gravity, which are otherwise difficult to obtain using diagrams. We also presented preliminary results for how gluon and graviton correlators are related, which provide evidence that an AdS extension of the double copy relation should exist. Our investigation leads to a number of natural future directions.

\begin{itemize}[leftmargin=1.2em]
\item {\bf AdS double copy\;} Although we did not manage to extract a precise double copy prescription in this paper, we obtained very explicit results for the spinning correlators in terms of $D$-functions. It will be important to analyze in detail the structures of these correlators in a future research and compare them with the structures in flat space. This will probably provide more hints for how to find an exact double copy relation. We should also investigate the other possibilities which were mentioned but not considered here.

\item {\bf Numerical bootstrap\;} In Sec.\ \ref{subsec:ortho}, we pointed out that a closed sector of the spinning correlators is obtained by restricting the polarizations to be in the orthogonal configuration. This sector resembles a scalar four-point function. A concrete research problem is to set up the numerical conformal bootstrap for this system and obtain bounds on various CFT data. Such a bootstrap problem will not exploit the full set of constraints coming from the spinning four-point function, but will still give rigorous bounds, albeit suboptimal. On the other hand, we computed all structures of $\langle \CJ\CJ\CJ\CJ\rangle$ explicitly for an $\CN=2$ holographic theory at tree level. To our knowledge, this is the only  nontrivial example of an explicitly known spinning four-point function. This result is therefore also useful to the full, four dimensional, bootstrap problem of currents as it can be used to put points in the exclusion plots and compare with the bounds.

\item {\bf Higher KK modes\;} In this paper, we restricted our attention to correlators of half-BPS multiplets with the lowest dimension. However, the superspace analysis can also be generalized to more general half-BPS multiplets which correspond to higher KK modes in AdS. Spinning correlators of these higher KK modes will be useful for bootstrapping higher-point functions involving general half-BPS operators.

\item {\bf Other dimensions\;} It would also be interesting to extend the superspace analysis to consider theories in other spacetime dimensions. In particular, the gluon and graviton four-point functions are also determined by the super primary correlators via supersymmetry and coincide with those in the bosonic theories at tree level. Using the results for super primary correlators \cite{Rastelli:2016nze,Rastelli:2017udc,Alday:2020lbp,Alday:2020dtb,Alday:2021odx}, we will be able to compute gluon and graviton amplitudes in different AdS backgrounds. It would be interesting to see how the behavior of these amplitudes changes with respect to spacetime dimensions. 
\end{itemize}

\acknowledgments
The work of A.B., G.F. and A.M. is supported by Knut and Alice Wallenberg Foundation under grant KAW 2016.0129 and by VR grant 2018-04438. The work of X.Z. is supported by funds from University of Chinese Academy of Sciences (UCAS), funds from the Kavli Institute for Theoretical Sciences (KITS), and also by the Fundamental Research Funds for the Central Universities.

\newpage
\appendices
\section{Superconformal algebra} \label{app:SuperconformalAlgebra}
Following~\cite{Dolan:2002zh}, we list of all non-vanishing (anti)commutators for the $\CN=2$ superconformal algebra in four dimensions $\mathrm{SU}(2,2|2)$ and in Lorentzian signature, $\eta^{\mu\nu}=\mathrm{diag}(-1,1,1,1)$ with spinor indices contracted with $\epsilon^{12}=\epsilon_{21}=1$. The conventions for Pauli matrices are those of~\cite{WessnBagger}. Furthermore, we denote
\eqn{
\rmx_{\alpha\alphad} \equiv \sigma^\mu_{\alpha\alphad}\lsp x_\mu\,,\qquad
\tilde\rmx^{\alphad\alpha} \equiv \sigmab^{\mu\,\alphad\alpha}\lsp x_\mu\,.
}[]
Similarly, we can express the Poincaré and special conformal generators in spinor notation as follows
\eqna{
\mathrm{P}_{\alpha \alphad}&=\sigma^\mu_{\alpha\alphad}P_{\mu}\, , \qquad &&\tilde{\mathrm{K}}^{\alphad\alpha}=\sigmab^{\mu\alphad\alpha}K_{\mu}\, ,\\
\mathrm{M}\indices{_\alpha^\beta}&=-\frac{1}{4}i (\sigma^{\mu}\bar{\sigma}^{\nu})\indices{_\alpha^\beta }M_{\mu\nu}\, , \qquad && \overbarUp{\mathrm{M}}\indices{^\alphad_\betad}=-\frac{1}{4}i (\bar{\sigma}^{\mu}\sigma^{\nu})\indices{^\alphad_\betad}M_{\mu\nu}\, .
}[]
Let us denote the supercharges as $Q^a_{\alpha},\, \Qb_{\alphad a}$ and $S^\alpha_a, \, \Sb^{\alphad a}$, with $a=1,2$. Then
\eqna{
&\{Q^a_\alpha,\Qb_{\alphad b}\}=2\delta\indices{^a_b}\mathrm{P}_{\alpha\alphad}\, , \quad &&\{Q^a_\alpha, Q^b_\beta\}=\{\Qb_{\alphad a},\Qb_{\betad b}\}=0\,, \\
&\{\Sb^{\alphad a},S_{b}^\alpha\}=2\delta\indices{^a_b}\tilde{\mathrm{K}}^{\alphad\alpha}\, , \quad &&\{S_a^\alpha, S_b^\beta\}=\{\Sb^{\alphad a},\Sb^{\betad b}\}=0\,, \\
&\{Q_\alpha^a, S_b^\beta \}=4\lsp\big(\delta^a_b\big(\mathrm{M}\indices{_\alpha^\beta}-\tfrac{i}{2} \delta\indices{_\alpha^\beta}D\big)-\delta\indices{_\alpha^\beta}R\indices{^a_b}\big)\, , \quad && \{Q^a_\alpha,\Sb^{\alphad b}\}=0\, ,\\
&\{\Sb^{\alphad a}, \Qb_{\betad b} \}=4\lsp\big(\delta^a_b\big(\mathrm{M}\indices{^\alphad_\betad}+\tfrac{i}{2} \delta\indices{^\alphad_\betad}D\big)-\delta\indices{^\alphad_\betad}R\indices{^a_b}\big)\, , \quad && \{S_a^\alpha,\Qb_{\alphad b}\}=0\, ,\\
& [\mathrm{M}\indices{_\alpha^\beta},\mathrm{M}\indices{_\gamma^\delta}]=\delta\indices{_\gamma^\beta}\mathrm{M}\indices{_\alpha^\delta}-\delta\indices{_\alpha^\delta}\mathrm{M}\indices{_\gamma^\beta}\, , \quad && [\overbarUp{\mathrm{M}}\indices{^\alphad_\betad},\overbarUp{\mathrm{M}}\indices{^\gammad_\deltad}]=-\delta\indices{^\alphad_\deltad}\overbarUp{\mathrm{M}}\indices{^\gammad_\betad}+\delta\indices{^\gammad_\betad}\overbarUp{\mathrm{M}}\indices{^\alphad_\deltad}\,,\\
&[\mathrm{M}\indices{_\alpha^\beta}, P_\mu] = (\sigma_{\mu\nu})\indices{_\alpha^\beta}\lsp P^\nu\,,
&&[\overbarUp{\mathrm{M}}\indices{^\alphad_\betad}, P_\mu] = (\sigmab_{\mu\nu})\indices{^\alphad_\betad}\lsp P^\nu\,,\\
&[\mathrm{M}\indices{_\alpha^\beta}, K_\mu] = (\sigma_{\mu\nu})\indices{_\alpha^\beta}\lsp K^\nu\,,
&&[\overbarUp{\mathrm{M}}\indices{^\alphad_\betad}, K_\mu] = (\sigmab_{\mu\nu})\indices{^\alphad_\betad}\lsp K^\nu\,,\\
&[\mathrm{M}\indices{_\alpha^\beta}, Q^a_\gamma]=\delta\indices{_\gamma^\beta}Q^a_\alpha-\tfrac{1}{2}\delta\indices{_\alpha^\beta}Q_\gamma^{a}\, , \quad && [\overbarUp{\mathrm{M}}\indices{^\alphad_\betad}, \Qb_{\gammad a}]=-\delta\indices{^\alphad _\gammad}\Qb_{a\lsp \betad}+\tfrac{1}{2}\delta\indices{^\alphad_\betad}\Qb_{\gammad a}\, ,\\
&[\mathrm{M}\indices{_\alpha^\beta}, S_a^\gamma]=-\delta\indices{_\alpha^\gamma}S_a^\beta+\tfrac{1}{2}\delta\indices{_\alpha^\beta}S^\gamma_{a}\, , \quad && [\overbarUp{\mathrm{M}}\indices{^\alphad_\betad}, \Sb^{\gammad a}]=\delta\indices{^\gammad _\betad}\Sb^{\alphad a}-\tfrac{1}{2}\delta\indices{^\alphad_\betad}\Sb^{\gammad a}\, ,\\
&[D, Q^a_\alpha]=\tfrac{i}{2}Q^a_\alpha \, , \quad && [D, \Qb_{\alphad a}]=\tfrac{i}{2}\Qb_{\alphad a}\, , \\
&[D, S^\alpha_a]=-\tfrac{i}{2}S^\alpha_a\, , \quad && [D, \Sb^{\alphad a}]=-\tfrac{i}{2}\Sb^{\alphad a}\, , \\
&[K_\mu, Q^a_\alpha]=-(\sigma_\mu)_{\alpha\alphad}\Sb^{\alphad a} \, , \quad && [K_\mu, \Qb_{\alphad a}]=S^a_\alpha(\sigma_\mu)_{\alpha\alphad}\, , \\
&[P_{\mu}, S^\alpha_a]=
\Qb_{\alphad a}(\bar{\sigma}_\mu)^{\alphad \alpha}\, , \quad && [P_{\mu}, \Sb^{\alphad a}]=-(\bar{\sigma}_\mu)^{\alphad \alpha}Q^a_\alpha\, .
}[]
Finally we have the relations involving the U(2) R-symmetry generators
\eqn{
[R\indices{^a_b},R\indices{^c_d}]=\delta\indices{^c_b}R\indices{^a_d}-\delta\indices{^a_d}R\indices{^c_b} .
}[]
Their action on the supercharges reads
\eqna{
&[R\indices{^a_b}, Q^c_\alpha]=\delta\indices{^c_b}Q^a_\alpha-\tfrac{1}{4}\delta\indices{^a_b} Q^c_\alpha\, , \qquad && [R\indices{^a_b}, \Qb_{\alphad c}]=-\delta\indices{^a_c}\Qb_{\alphad b}+\tfrac{1}{4}\delta\indices{^a_b} \Qb_{\alphad c}\, ,\\
& [R\indices{^a_b}, S^\alpha_c]=-\delta\indices{^a_c}S^\alpha_b+\tfrac{1}{4}\delta\indices{^a_b}S^\alpha_c \, , \qquad && [R\indices{^a_b}, \Sb^{\alphad c}]=\delta\indices{^c_b}\Sb^{\alphad a}-\tfrac{1}{4}\delta\indices{^a_b} \Sb^{\alphad c}\, .
}[]
One could also consider the $\SU(2)$ generators $T\indices{^a_b}$ defined as
\eqn{
T\indices{^a_b} = R\indices{^a_b} - \frac12 \delta\indices{^a_b} R\indices{^c_c}\,.
}[]
The trace part is the $\rmU(1)$ R-symmetry $r \equiv R\indices{^c_c}$, which assigns charge $1/2$ ($-1/2$) to $Q_\alpha$ ($\Qb_\alphad$). With these generators the above relations become (note the ``$\frac{1}{2}$'' instead of the ``$\frac{1}{4}$'')
\eqna{
&[T\indices{^a_b}, Q^c_\alpha]=\delta\indices{^c_b}Q^a_\alpha-\tfrac{1}{2}\delta\indices{^a_b} Q^c_\alpha\, , \qquad && [T\indices{^a_b}, \Qb_{\alphad c}]=-\delta\indices{^a_c}\Qb_{\alphad b}+\tfrac{1}{2}\delta\indices{^a_b} \Qb_{\alphad c}\, ,\\
& [T\indices{^a_b}, S^\alpha_c]=-\delta\indices{^a_c}S^\alpha_b+\tfrac{1}{2}\delta\indices{^a_b}S^\alpha_c \, , \qquad && [T\indices{^a_b}, \Sb^{\alphad c}]=\delta\indices{^c_b}\Sb{}^{\alphad a}-\tfrac{1}{2}\delta\indices{^a_b} \Sb^{\alphad c}\,,\\
&[r,Q^c_\alpha]  = \tfrac12 Q^c_\alpha\,, \qquad && [r,\Qb_{\alphad c}] = -\tfrac12 \Qb_{\alphad c}\,,\\
&[r,S_c^\alpha]  = -\tfrac12 S_c^\alpha\,, \qquad && [r,\Sb{}^{\alphad c}] = \tfrac12 \Sb{}^{\alphad c}\,,
}[]
Finally recall that we require all our generators to be Hermitian, namely
\eqn{
(Q^a_\alpha)^\dagger=\Qb_{\alphad a }\,,\quad (S^\alpha_a)^\dagger= \Sb^{\alphad a}\,, \quad( \mathrm{M}\indices{_\alpha^\beta})^\dagger= \overbarUp{\mathrm{M}}\indices{^\betad _\alphad}\, , \quad (R\indices{^a_b})^\dagger=R\indices{^b_a}\, .
}[]
When going to harmonic superspace it is cleaner to use the $\pm$ notation
\eqna{
Q_\alpha^1 &= Q_\alpha^+\,,&\qquad Q_\alpha^2 &= Q_\alpha^-\,,&\qquad \Qb_{\alphad1} &= \Qb^-_\alphad \,,&\qquad \Qb_{\alphad2} &= -\Qb_\alphad^+\,,\\
S^\alpha_1 &= S^{\alpha-}\,,&\qquad S^\alpha_2 &= -S^{\alpha+}\,,&\qquad \Sb^{\alphad1} &= \Sb^{\alphad+} \,,&\qquad \Sb^{\alphad2} &= \Sb^{\alphad-}\,,\\
T^0 &= \mathrlap{2\lsp T\indices{^1_1}=-2\lsp T\indices{^2_2}\,,}&&&\qquad T^{++} &= T\indices{^1_2}\,,&\qquad T^{--} &= T\indices{^2_1}\,.
}[]
The commutators in this notation read
\eqna{ 
[T^0,Q^+] &= Q^+\,,&\qquad [T^0,Q^-] &= -Q^-\,,\\
[T^{++},Q^+] &= 0\,,&\qquad [T^{++},Q^-] &= Q^+\,,\\
[T^{--},Q^+] &= Q^-\,,&\qquad [T^{--},Q^-] &= 0\,,\\
}[]
\eqna{
[T^0,S^+] &= S^+\,,&\qquad [T^0,S^-] &= -S^-\,,\\
[T^{++},S^+] &= 0\,,&\qquad [T^{++},S^-] &= S^+\,,\\
[T^{--},S^+] &= S^-\,,&\qquad [T^{--},S^-] &= 0\,,\\
}[]
\eqna{
\{Q^+_\alpha,\Qb{}^+_\alphad\} & =  \{Q^-_\alpha,\Qb{}^-_\alphad\} = 0\,,\\
\{Q_\alpha^+,\Qb{}^-_\alphad\} & = - \{Q_\alpha^-,\Qb{}^+_\alphad\} = 2 \lsp\mathrm{P}_{\alpha\alphad}\,.
}[]
\eqna{
\{S^{\alpha+},\Sb{}^{\alphad+}\} & =  \{S^{\alpha-},\Sb{}^{\alphad-}\} = 0\,,\\
\{S^{\alpha-},\Sb{}^{\alphad+}\} & = - \{S^{\alpha+},\Sb{}^{\alphad-}\} = 2 \lsp \tilde{\mathrm{K}}^{\alphad\alpha}\,.
}[]
\eqna{
\{Q^\pm_\alpha,\Sb{}^{\alphad\pm}\} & =  0\,,&\qquad \{S^{\alpha\pm},\Qb{}_\alphad^\pm\} & =  0\\
\{Q_\alpha^+,S^{\beta-}\} & = 4\lsp\big(\mathrm{M}\indices{_\alpha^\beta} - \tfrac12\delta\indices{_\alpha^\beta}\big(i D + T^0+r\big)\big)\,,&
\{Q_\alpha^+,S^{\beta+}\} & = 4\delta\indices{_\alpha^\beta}\lsp T^{++}\,,\\
\{Q_\alpha^-,S^{\beta+}\} & = 4\lsp\big(-\mathrm{M}\indices{_\alpha^\beta} + \tfrac12\delta\indices{_\alpha^\beta}\big(i D - T^0+r\big)\big)\,,&
\{Q_\alpha^-,S^{\beta-}\} & = -4\delta\indices{_\alpha^\beta}\lsp T^{--}\,,\\
\{\Sb^{\alphad+},\Qb_\betad^-\} & = 4\lsp\big(\overbarUp{\mathrm{M}}\indices{^\alphad_\betad} + \tfrac12\delta\indices{^\alphad_\betad}\big(i D - T^0-r\big)\big)\,,&
\{\Sb^{\alphad+},\Qb_\betad^+\} & = 4\delta\indices{^\alphad_\betad}\lsp T^{++}\,,\\
\{\Sb^{\alphad-},\Qb_\betad^+\} & = 4\lsp\big(-\overbarUp{\mathrm{M}}\indices{^\alphad_\betad} - \tfrac12\delta\indices{^\alphad_\betad}\big(i D + T^0-r\big)\big)\,,&
\{\Sb^{\alphad-},\Qb_\betad^-\} & = -4\delta\indices{^\alphad_\betad}\lsp T^{--}
\,.
}[]
\eqn{
[P_{\mu}, S^{\alpha\pm}]=
\Qb_\alphad^\pm\lsp\bar{\sigma}_\mu^{\alphad \alpha}\, , \qquad [P_{\mu}, \Sb^{\alphad \pm}]=-\bar{\sigma}_\mu^{\alphad \alpha}\lsp Q^\pm_\alpha\, .
}[]
\section{Differential representation}\label{app:covariantDeriv}

Below we write the covariant derivatives in the full superspace
\twoseqn{
D^a_\alpha &= \frac{\partial}{\partial \theta^\alpha_a} + i \lsp \sigma^\mu_{\alpha\alphad}\lsp \thetab^{\alphad a} \lsp \frac{\partial}{\partial x^\mu}\,,
}[Ddef]{
\Db_{\alphad a} &= -\frac{\partial}{\partial \thetab^{\alphad a}} - i\lsp  \theta^{\alpha}_a\lsp \sigma^\mu_{\alpha\alphad}\lsp\frac{\partial}{\partial x^\mu}\,.
}[Dbdef][DDbdef]
With the variables defined in~\eqref{thetapmWithxi} the covariant derivatives read
\twoseqn{
D^a_\alpha &= -\xi^a\frac{\partial}{\partial \theta^{+\alpha}} - 2i \lsp \xi^a \lsp \sigma^\mu_{\alpha\alphad}\lsp \thetab^{-\alphad} \lsp \frac{\partial}{\partial z^\mu} + \xib^a\frac{\partial}{\partial \theta^{-\alpha}}\,,
}[]{
\Db_{\alphad a} &= \phantom{+}\xi_a\frac{\partial}{\partial \thetab^{+\alphad}} + 2i\lsp \xi_a \lsp \theta^{-\alpha}\lsp \sigma^\mu_{\alpha\alphad}\lsp\frac{\partial}{\partial z^\mu} - \xib_a\frac{\partial}{\partial \thetab^{-\alphad}}\,.
}[][Dnewvar]

Now we show all the Poincaré and special conformal supercharges in the two half-BPS superspace formulations: harmonic and analytic. The ``$+$'' in $\theta^+$ and $\thetab^+$ is omitted.

In harmonic superspace one has
\twoseqn{
Q_\alpha^a &=- u^{+a}\frac{\partial}{\partial\theta^{\alpha}}
+2 i\lsp u^{-a}\lsp\thetab^{\alphad} \sigma^\mu_{\alpha\alphad}\frac{\partial}{\partial z^\mu} \,,
}[]{
\Qb_{\alphad a} &=u^+_a \frac{\partial}{\partial\thetab^{\alphad}}+2i\lsp u^-_a \theta^{\alpha}\sigma^\mu_{\alpha\alphad} \frac{\partial}{\partial z^\mu}\, .
}[][Qharmonic]
\twoseqn{
S^\alpha_a &=-4 u^+_a u^-_b\theta^{\alpha}\frac{\partial}{\partial u_b^+} +2u^-_a \theta^2\epsilon^{\alpha\beta}\frac{\partial}{\partial \theta^{\beta}} +i \lsp u^+_a \tilde{\mathrm{z}}^{\alphad\alpha} \frac{\partial}{\partial \thetab^{\alphad}}    -2u^-_a  \lsp \theta^{\beta} \sigma^\mu _{\beta\alphad}\tilde{\mathrm{z}}^{\alphad\alpha}\frac{\partial}{\partial z^\mu}\,,
}[]{
\Sb^{\alphad a}&=4u^{+a}u^-_b \thetab^{\alphad} \frac{\partial}{\partial u_b^+} +i \lsp u^{+a}\tilde{\mathrm{z}}^{\alphad\alpha} \frac{\partial}{\partial \theta^{\alpha}}+2 u^{-a} \thetab^2 \epsilon^{\alphad \betad}\frac{\partial}{\partial \thetab^{\betad}}+2u^{-a} \tilde{\mathrm{z}}^{\alphad\alpha}\sigma^{\mu}_{\alpha\betad}\thetab^{\betad}\frac{\partial}{\partial z^\mu}\,,
}[][Sharmonic]
with $\tilde{\mathrm{z}}^{\alphad\alpha} = \sigmab^{\mu\lsp\alphad\alpha} z_\mu$ and $\theta^2 = \theta^\alpha\theta_\alpha$, $\thetab^2 = \thetab_\alphad\thetab^\alphad$.

In analytic superspace instead the generators read
\twoseqn{
Q_\alpha^+ &= -y \frac{\partial}{\partial\theta^\alpha} + 2i \lsp \sigma_{\alpha\alphad}\thetab^\alphad\frac{\partial}{\partial z^\mu} \,,\qquad
Q_\alpha^- = \frac{\partial}{\partial\theta^\alpha}\,,
}[]{
\Qb_\alphad^+ &= -y \frac{\partial}{\partial\thetab^\alphad} - 2i \lsp \theta^\alpha\sigma_{\alpha\alphad}\frac{\partial}{\partial z^\mu} \,,\qquad
\Qb_\alphad^- = \frac{\partial}{\partial\thetab^\alphad}\,,
}[][Qanalytic]

\fourseqn{
S^{\alpha+} &= -2\lsp\theta^2\epsilon^{\alpha\beta}\frac{\partial}{\partial\theta^\beta}+4\llsp y\lsp\theta^\alpha\frac{\partial}{\partial y} -i \llsp y\lsp \tilde{\mathrm{z}}^{\alphad\alpha}\frac{\partial}{\partial\thetab^\alphad} + 2\llsp\theta^\beta\sigma^\mu_{\beta\alphad}\tilde{\mathrm{z}}^{\alphad\alpha}\frac{\partial}{\partial z^\mu}\,,
}[]{
S^{\alpha-}&=-4\lsp\theta^\alpha\frac{\partial}{\partial y}+i\lsp\tilde{\mathrm{z}}^{\alphad\alpha}\frac{\partial}{\partial\thetab^\alphad}\,,
}[]{
\Sb^{\alphad+} &= 2\lsp\thetab^2\epsilon^{\alphad\betad}\frac{\partial}{\partial\thetab^\betad}+4\llsp y\lsp\thetab^\alphad\frac{\partial}{\partial y} +i \llsp y\lsp \tilde{\mathrm{z}}^{\alphad\alpha}\frac{\partial}{\partial\theta^\alpha} + 2\llsp\tilde{\mathrm{z}}^{\alphad\alpha}\sigma^\mu_{\alpha\betad}\thetab^\betad\frac{\partial}{\partial z^\mu}\,,
}[]{
\Sb^{\alphad-}&=-4\lsp\thetab^\alphad\frac{\partial}{\partial y}-i\lsp\tilde{\mathrm{z}}^{\alphad\alpha}\frac{\partial}{\partial\theta^\alpha}\,.
}[][Sanalytic]

\section{\texorpdfstring{$\boldsymbol{\Db}$}{Db} functions}\label{App:Dfunctions}
In this appendix, we collect some useful properties of $D$-functions. A general $D$-function is defined as an integral over $z \in \mathrm{AdS}_{d+1}$
\eqna{
D_{\D{1}\D{2}\D{3}\D{4}}(x_1, x_2, x_3, x_4)=\int \frac{d z_0 d^d z}{z_{0}^{d+1}} \prod_{i=1}^4 \left(\frac{z_0}{z_0^2+(\vec{z}-\vec{x_i})^2}\right)^{\D{i}}\, .
}[]
$D$-function of higher external dimensions can be obtained from lower ones by acting with derivatives in $x_{ij}^2$ as follows
\eqna{
D_{\cdots( \D{i}+1)\cdots(\D{j}+1)\cdots}=\frac{d-2\Sigma}{2 \D{i}\D{j}} \frac{\partial}{\partial{x_{ij}^2}} D_{\cdots\D{i}\cdots\D{j}\cdots}\, ,
}[derOnD]
where we have introduced $\Sigma=\frac{1}{2}\sum_{i=1}^4\D{i}$.
We can write $D$-functions as functions of cross ratios by introducing the $\bar{D}$-functions
\eqna{
D_{\D{1}\D{2}\D{3}\D{4}}=\pi^2\frac{\Gamma\mleft(\Sigma-\frac{d}{2} \mright)}{2 \prod_{i}^4 \Gamma(\D{i})}\frac{(x_{14}^2)^{\Sigma-\D{1}-\D{4}}(x_{34}^2)^{\Sigma-\D{3}-\D{4}}}{(x_{13}^2)^{\Sigma-\D{4}}(x_{24}^2)^{\D{2}}}\Db_{\D{1}\D{2}\D{3}\D{4}}(u,v)\, .
}[]
In this paper, all $D$-functions can be related to the basic $D$-function
\begin{equation}
\Db_{1111}\equiv\Phi(z,\bar{z})=\frac{1}{z-\bar{z}}\left(2{\rm Li}_2(z)-2{\rm Li}_2(\bar{z})+\log(z\bar{z})\log\big(\frac{1-z}{1-\bar{z}}\big)\right)\;.
\end{equation}
From this expression, it is not difficult to verify the following useful relations 
\begin{equation}
\begin{split}
\partial_z\Phi(z,\bar{z})=&-\frac{\Phi(z,\bar{z})}{z-\bar{z}}+\frac{\log u}{(z-1)(z-\bar{z})}-\frac{\log v}{z(z-\zb)}\;,\\
\partial_{\bar{z}}\Phi(z,\bar{z})=&\frac{\Phi(z,\bar{z})}{z-\bar{z}}-\frac{\log u}{(\bar{z}-1)(z-\bar{z})}+\frac{\log v}{\bar{z}(z-\bar{z})}\;.
\end{split}
\end{equation}
These identities allow us to explicitly write $\Db$-functions as elementary functions of $z$ and $\bar{z}$. There also exist simple properties relating $\Db$-functions with permuted external dimensions
\eqna{
\Db_{\D{1}\D{2}\D{3}\D{4}}(u,v)&= v^{\Sigma-\D{2}-\D{3}}\Db_{\D{2}\D{1}\D{3}\D{4}}(u,v)\, 
=u^{\Sigma-\D{1}-\D{2}}\Db_{\D{4}\D{3}\D{2}\D{1}}(u,v)\\
&=\Db_{(\Sigma-\D{3})(\Sigma-\D{4})(\Sigma-\D{1})(\Sigma-\D{2})}(u,v)\, ,\\
&=\Db_{\D{3}\D{2}\D{1}\D{4}}(v, u)\, ,\\
&= v^{-\D{2}}\Db_{\D{1}\D{2}\D{4}\D{3}}\mleft(\frac{u}{v}, \frac{1}{v}\mright)
= v^{\D{4}-\Sigma}\Db_{\D{2}\D{1}\D{3}\D{4}}\mleft(\frac{u}{v}, \frac{1}{v}\mright)\, ,\\
&= u^{-\D{2}}\Db_{\D{4}\D{2}\D{3}\D{1}}\mleft(\frac{1}{u}, \frac{v}{u}\mright)
}[]
Finally, the Mellin space representation for these functions reads
\eqna{
\Db_{\D{1}\D{2}\D{3}\D{4}}(u,v)&=\int d\Ms\lsp d\Mt \lsp u^{\frac{\Ms}{2}}v^{\frac{\Mt}{2}}\,\Gamma \mleft(-\frac{\Ms}{2}\mright) \,\Gamma \mleft(\frac{1}{2} (-\D{1}-\D{2}+\D{3}+\D{4})-\frac{\Ms}{2}\mright) \\
&\,\quad\times \Gamma \mleft(-\frac{\Mt}{2}\mright) \,\Gamma \mleft(\frac{1}{2} (\D{1}-\D{2}-\D{3}+\D{4})-\frac{\Mt}{2}\mright)\\
&\,\quad\times \Gamma \mleft(\frac{\Ms+\Mt}{2}+\D{2}\mright)\, \Gamma \mleft(\frac{1}{2} (\D{1}+\D{2}+\D{3}-\D{4})+\frac{\Ms+\Mt}{2}\mright)\, .
}[DbMellin]

\section{Embedding space formalism}\label{app:embedding}
It is well known that a convenient way to rephrase $d$ dimensional conformal symmetry is by embedding it in a higher dimensional space, $\mathbb{M}^{d+2}$, where conformal transformations act as linear isometries. For this brief review we fix $d=4$.  Given a point $P^A \in \mathbb{M}^{6}$ in light-cone coordinates $(P^+, P^-, P^\mu)$, the four dimensional theory lives in a section of the null cone defined as
\eqna{
P \cdot P= -P^+ P^-+\delta_{\mu\nu}P^{\mu\nu}=0\, .
}[]
A particular parametrization of the coordinates give the so-called Poincaré section
\eqna{
P^{A}=(1, x^2, x^\mu)\, , \qquad x^\mu \in \mathbb{R}^4\,.
}[defP]
With this particular coordinate choice it is straightforward to see that
\eqna{
-2P_i \cdot P_j=x_{ij}^2\,.
}[Pcoord]
Every field in the original space can be uplifted to the null six dimensional cone by requiring homogeneity and transversality properties~\cite{Costa:2011mg}. Assuming the field to be a symmetric traceless tensor with dimension $\Delta$ and spin $\ell$, the homogeneity amounts to this scaling behavior
\eqn{
V_{A_1\cdots A_{\ell}}(\lambda P)=\lambda^{-\Delta} V_{A_1 \cdots A_\ell} (P)\, , \qquad \lambda>0\, ,
}[]
while transversality requires
\eqna{
P^A V_{A A_2 \cdots A_\ell}=0 \, . 
}[]
It is convenient to further contract the tensor indices with an additional polarization vector $Z$ such that a generic tensor $V_{A_1\cdots A_\ell}(P)$ is just a polynomials in these variables
\eqna{
V(P, Z)\equiv V_{A_1\cdots A_\ell}(P) Z^{A_1} \cdots Z^{A_\ell}\, ,  \qquad
V(P, \alpha Z+\beta P)=\alpha^\ell V(P,Z)\,.
}[]
To describe a symmetric traceless tensor, the polarization has to satisfy
\eqna{
Z \cdot P=0 \, , \qquad Z^2=0\, ,
}[Zcond]
so a convenient parametrization turns out to be
\eqna{
Z^A=(0, 2 h\cdot x, h^\mu)\, , \qquad
 h^2=0\,,
}[Zcoord]
where $x$ is the point on which $P$ depends according to~\eqref{defP}. As we have already seen in~\eqref{hDef},  the $4$d polarization $h^\mu$ can be expressed in terms of the auxiliary commuting spinor variables $\eta^\alpha$ and $\etab^\alphad$, we have been using, as $h^\mu=-\frac12\sigma^\mu_{\alpha\alphad}\etab_i^\alphad\eta_i^\alpha$.  Notice that this rewriting makes  the null requirement automatically realized.

It can be shown~\cite{Costa:2011mg} that, for parity invariant theories, all the possible tensor structures appearing in any $n$-point function can be built out of just two basic invariants,  $H_{ij}$ and $V_{i, jk}$ which we already encountered in Sec.~\ref{Sec:altpresentation}.  These building blocks in embedding space read
\eqna{
H_{ij}&=-2 \left( Z_i\cdot Z_j \; P_i \cdot P_j -Z_i \cdot P_j \; Z_j \cdot P_i \right)\, ,\\
V_{i, jk}&= \frac{Z_i \cdot P_j \; P_i \cdot P_k-Z_i \cdot P_k \; P_i \cdot P_j}{P_j \cdot P_k}\, .
}[]
The relations between these formulas and their four-dimensional analogue in~\eqref{HandVdefs} can be easily worked out using the explicit parametrizations in~\eqref{defP} and~\eqref{Zcoord} --- see also~\cite[Appendix A]{Karateev:2019pvw}. 

Let us conclude by mentioning how the orthogonal configuration we consider in Sec.~\ref{subsec:ortho} is translated in embedding space. The orthogonal condition is naturally realized by requiring that for any $i,j=1, \ldots, 4$ 
\eqna{
Z_i \cdot P_j=0 \quad\xRightarrow[]{\eqref{Pcoord},\eqref{Zcoord}}\quad -h_i \cdot x_{ij}=0\, .
}[]
\section{Tensor invariants} \label{app:tensors}
For reader's convenience, we collect various basic invariant tensor structures appearing in correlators involving spin-1 and spin-2 operators which were defined in Appendix D of \cite{Cuomo:2017wme}, and express them in terms of the auxiliary fermion variables $\eta^\alpha$ and $\etab^\alphad$.
\eqna{
&\hat{\mathbb{J}}^i_{jk}=\frac{\eta_i \rmx_{ik}\etab_i\; x_{ij}^2-\eta_i \rmx_{ij}\etab_i\; x_{ik}^2}{x_{jk}^2}\,, \qquad &&\hat{\mathbb{I}}_{ij}=\eta_j x_{ij}\etab_i\, , \\
&{\mathbb{K}}^{jk}_i=-\eta_i \rmx_{ij}\bar{\rmx}_{ik} \eta_k \, , \qquad && \hat{\mathbb{K}}^{jk}_i=\sqrt{\frac{x_{jk}^2}{x_{ij}^2x_{ik}^2}} {\mathbb{K}}^{jk}_i \, ,\\
&\mathbb{L}^i_{jkl}= \eta_i \rmx_{ij}\bar{\rmx}_{jk}\rmx_{kl}\bar{\rmx}_{il}\eta_i \, , \qquad && \hat{\rule{0.3em}{0em}\mathbb{L}}^i_{jkl}=\frac{1}{\sqrt{x_{jk}^2x_{jl}^2 x_{kl}^2}}\mathbb{L}^i_{jkl}\, .
}[defTensor]
In terms of these invariants, the tensors appearing in the correlator  of three currents in~\eqref{tensors3} are
\eqna{
\mathbb{T}_i^+&=\lbrace \hat{\mathbb{J}}^1_{23} \,  \hat{\mathbb{J}}^2_{13} \, \hat{\mathbb{J}}^3_{12}, \, \hat{\mathbb{J}}^1_{23} \hat{\mathbb{I}}_{23}\hat{\mathbb{I}}_{32}, \,\hat{\mathbb{J}}^2_{13} \hat{\mathbb{I}}_{13}\hat{\mathbb{I}}_{31}, \, \hat{\mathbb{J}}^3_{12} \hat{\mathbb{I}}_{12}\hat{\mathbb{I}}_{21}\rbrace\, ,\\
\mathbb{T}_1^-&= \hat{\mathbb{I}}_{12} \hat{\mathbb{I}}_{23}\hat{\mathbb{I}}_{31}+\hat{\mathbb{I}}_{13} \hat{\mathbb{I}}_{21}\hat{\mathbb{I}}_{32}\, .
}[tensorthreeE]
We will not report the tensors appearing in the four-point functions we have discussed since they can be easily obtained from~\cite{Cuomo:2017wme}.

\Bibliography[refs.bib]
\end{document}